\documentclass[11pt]{article}
\usepackage{amsmath}
\usepackage{epsfig}
\usepackage{amsfonts}
\usepackage{amssymb}
\usepackage{bbm,bm}

\usepackage{cite}
\allowdisplaybreaks[1]

\usepackage[table,dvipsnames]{xcolor}
\usepackage{multirow}
\usepackage{arydshln}
\usepackage{graphicx}

\usepackage{comment}

 \usepackage[
   colorlinks=true,
   linkcolor={red!50!black},
   citecolor={blue!50!black},
   filecolor=black,
   urlcolor=black,
   breaklinks=true
   ]{hyperref}
\usepackage{orcidlink}

\usepackage{array}
\usepackage{makecell}
\usepackage[table]{xcolor}
\usepackage{float}
\usepackage{tikz}
\usetikzlibrary{arrows.meta,calc,positioning}

\newcommand{\rmi}[1]{{\mbox{\scriptsize #1}}}
\newcommand{\rmii}[1]{{\mbox{\tiny\rm{#1}}}}
\newcommand{\smallIdx}[1]{{\mbox{\tiny\rm{#1}}}}

\newcommand{\gs}{g_\rmi{s}}

\newcommand{\tr}{\text{tr}}

\newcommand{\T}{\rmii{$T$}}

\renewcommand{\vec}[1]{{\bf #1}}
\newcommand{\nn}{\nonumber \\}

\newcommand{\lrD}{\overleftrightarrow{\!D}}

\newcommand{\sumint}[1]{{\hbox{$\sum$}\!\!\!\!\!\!\!\int\,}_{\!\!\!\!\raise-0.9ex\hbox{$\scriptstyle{#1}$}}}
\newcommand{\Tint}[1]{{\hbox{$\sum$}\!\!\!\!\!\!\!\int\,}_{\!\!\!\!\raise-0.9ex\hbox{$\scriptstyle{#1}$}}}
\newcommand{\Tinti}[1]{{{\Sigma}\!\!\!\!\raise0.3ex\hbox{$\int$}_\rmii{${#1}$}}}
\newcommand{\Tintip}[1]{{{\Sigma'}\!\!\!\!\!\raise0.3ex\hbox{$\int$}_\rmii{${#1}$}}}

\makeatletter \@addtoreset{equation}{section} \makeatother
\renewcommand{\theequation}{\arabic{section}.\arabic{equation}}
\newcounter{constraint}
\renewcommand{\theconstraint}{C\Roman{constraint}}
\makeatletter
\newcommand{\constraintsubsubsection}[3]{%
	\refstepcounter{constraint}%
	\let\constraint@seccntformat\@seccntformat
	\def\@seccntformat##1{}%
	\let\constraint@addcontentsline\addcontentsline
	\def\addcontentsline##1##2##3{}%
	\subsubsection[\theconstraint. #1]{%
		\theconstraint. #1:
		\textbf{#2}%
	}%
	\let\addcontentsline\constraint@addcontentsline
	\let\@seccntformat\constraint@seccntformat
	\protected@edef\@currentlabel{\theconstraint}%
	\label{#3}%
}
\makeatother
\makeatletter
\renewcommand\section{\@startsection{section}{1}{\z@}%
  {-5.5ex \@plus -1ex \@minus -.2ex}%
  {2.3ex \@plus.2ex}%
  {\normalfont\large\bfseries}}
\renewcommand\subsection{\@startsection{subsection}{2}{\z@}%
  {-3.25ex\@plus -1ex \@minus -.2ex}%
  {1.5ex \@plus .2ex}%
  {\normalfont\normalsize\bfseries}}
\renewcommand\thesection{\@arabic\c@section}
\renewcommand\thesubsection{\thesection.\@arabic\c@subsection}
\renewcommand{\@seccntformat}[1]{%
  \csname the#1\endcsname.\hspace{1.0em}}
\makeatother

\def\magenta{\color{magenta}}

\def\Green{\color{OliveGreen}}

\newcommand{\cellmag}{\cellcolor{magenta!20}}
\newcommand{\cellgreen}{\cellcolor{green!10}}

\allowdisplaybreaks

\begin{document}

\flushbottom

\begin{titlepage}

\begin{flushright}
	September 2026
\end{flushright}
\begin{centering}

\vfill

{\Large{\bf%
	Finite-temperature operator basis on
	$\mathbb{R}^3 \times S^1$
	for SMEFT
}}

\vspace{0.8cm}

\renewcommand{\thefootnote}{\fnsymbol{footnote}}
Joydeep Chakrabortty%
\orcidlink{0000-0001-8709-916X},%
$^{\rm a,}$%
\footnotemark[1]
Bruno Siqueira Eduardo%
\orcidlink{0000-0002-2523-8686},%
$^{\rm b,c,}$%
\footnotemark[2]
\\
Siddhartha Karmakar%
\orcidlink{0009-0003-0609-9689},%
$^{\rm a,}$%
\footnotemark[3]
and
Philipp Schicho%
\orcidlink{0000-0001-5869-7611}
$^{\rm c,}$%
\footnotemark[4]

\vspace{0.8cm}

$^\rmi{a}$%
{\em
  Indian Institute of Technology Kanpur, Kalyanpur,
  Kanpur 208016, Uttar Pradesh, India
}
\vspace{0.3cm}

$^\rmi{b}$%
{\em
 Instituto de Física, Universidade de São Paulo, C.P. 66.318, 05315-970 São Paulo, Brazil
}
\vspace{0.3cm}

$^\rmi{c}$%
{\em
  D\'epartement de Physique Th\'eorique, Universit\'e de Gen\`eve,\\
  24 quai Ernest Ansermet, CH-1211 Gen\`eve 4, Switzerland
}
\vspace{0.3cm}

\vspace*{0.8cm}

\mbox{\bf Abstract}

\end{centering}

\vspace*{0.3cm}

\noindent
We present the first complete non-redundant operator basis for
the Standard Model Effective Field Theory (SMEFT) at finite temperature,
using the imaginary-time formalism.
By employing the
Hilbert series method on
the space-time manifold $\mathbb{R}^3 \times S^1$,
we classify all effective operators up to dimension-six.
In constructing the basis, we consistently impose
integration-by-parts and equations-of-motion constraints along spatial directions.
We further analyze the impact of additional constraints,
including the vanishing of the curl of the electric and magnetic fields
and gauge choices for the temporal components on an operator basis.
We also realize them in terms of static
three-dimensional spatial and zero-temperature SMEFT operators.
At dimension five and six, we identify intrinsically thermal operators
that vanish with temperature. Our framework is fully general
and extends to arbitrary mass dimension
and compact connected internal symmetry groups.

\vfill
\renewcommand{\thefootnote}{\fnsymbol{footnote}}
\footnotetext[1]{joydeep@iitk.ac.in}
\footnotetext[2]{brunoseduardo2002@usp.br}
\footnotetext[3]{siddharthak@iitk.ac.in}
\footnotetext[4]{philipp.schicho@unige.ch}
\renewcommand{\thefootnote}{\arabic{footnote}}
\setcounter{footnote}{0}
\end{titlepage}

{\hypersetup{hidelinks}
\tableofcontents
}

\clearpage

%%%%%%%%%%%%%%%%%%%%%%%%% SECTION %%%%%%%%%%%%%%%%%%%%%%%%%%%%%%%%%%%%%%%%%
%
\section{Introduction}
\label{sec:intro}

Finite-temperature quantum field theory,
formulated in the imaginary-time (Matsubara) formalism~\cite{Matsubara:1955ws},
provides a natural framework for
describing systems in thermal equilibrium, and
its long-distance dynamics is
often captured by effective theories (EFTs)
in three dimensions~\cite{Ginsparg:1980ef,Appelquist:1981vg,Susskind:1979up,Weiss:1980rj}.
These dimensionally reduced EFTs (“3D”)
were originally motivated to study
thermodynamics and phase transitions in
non-Abelian gauge theories~\cite{Kajantie:1995dw,Bochkarev:1989kp}.
Recently, they have
been reinvigorated for precision studies of
cosmological phase transitions in
beyond the Standard Model (SM) theories~\cite{Laine:2017hdk,Croon:2020cgk,Ekstedt:2022bff,Ekstedt:2024etx,Bernardo:2025vkz,Fuentes-Martin:2026bhr}.
The same framework also captures soft contributions to
real-time lightcone observables~\cite{Caron-Huot:2008zna}.
Representative examples include
the transverse collision kernel relevant for jet quenching
in a hot QCD medium~\cite{Panero:2013pla,DOnofrio:2014mld},
the soft sectors of thermal photon and dilepton emission rates~\cite{Ghiglieri:2013gia,Ghiglieri:2014kma},
and neutrino interaction rates in the electroweak plasma~\cite{Ghiglieri:2016xye}.

In these applications,
dimensional reduction assumes a systematic separation of scales.
The validity of the resulting high-temperature EFT expansion is controlled by
the hierarchy of scales generated at finite temperature. Once this hierarchy is
established, a complete set of allowed higher-dimensional operators at each mass
dimension is required to perform a systematic computation at fixed order.
Although the validity of scale separation
has been rigorously investigated in the context
of phase transitions~\cite{%
	Kajantie:1995dw,Laine:1996nz,Laine:1999rv,Chala:2024xll,
	Chala:2025aiz,Bernardo:2025vkz,Chala:2025cya,Bernardo:2026whs,Bernardo:2026nyq
},
the construction of
the finite-temperature higher-dimensional operator basis from first principles
is still typically performed by directly adapting
a zero-temperature operator basis,
such as a Warsaw-type basis~\cite{%
	Weinberg:1979sa,Buchmuller:1985jz,Grzadkowski:2010es};
see~\cite{Aebischer:2025qhh} for a review of zero-temperature
SM effective theory (SMEFT).
At the same time,
finite-temperature analyses have already identified genuinely thermal structures
with no zero-temperature counterpart~\cite{Kajantie:1997ky,Chala:2025aiz,Bernardo:2026nyq}.
This top-down generation of dimension-six operators
was performed in several theories
using correlator matching,
including
QCD~\cite{Laine:2018lgj},
the Abelian Higgs model~\cite{Bernardo:2025vkz},
Higgs-Yukawa theories~\cite{Chala:2024xll},
SMEFT~\cite{Kajantie:1995dw,Moore:1995jv,Chala:2025aiz,Chala:2025gif}, and
generalized theories~\cite{Bernardo:2026nyq}.

At finite temperature,
such a top-down procedure
does not allow for a UV-generic complete basis construction.
Since the thermal background selects a preferred timelike direction and, therefore, reduces
Lorentz symmetry,%
\footnote{%
	\label{footnote:brokenLorentz}
	While the underlying theory remains Lorentz invariant,
	the thermal ensemble introduces a preferred timelike four-velocity $u^\mu$;
	see~\cite{Weldon:1982aq} for a fully covariant treatment.
	It is convenient to decompose field components with respect to
	$u^\mu$, typically working in the rest frame $u^\mu = (1,\vec{0})$.
}
the electric and magnetic components of the field strengths
can be organized as independent building blocks.
As a result,
additional operator structures can appear at finite temperature
that are absent or redundant at zero temperature.
Moreover,
for a fixed UV completion and matching order,
top-down matching populates only the subset of operators generated in that setup.
A direct construction
by expanding thermal functional determinants
using the Heat Kernel expansion in
Schwinger time~\cite{Schwinger:1951nm,DeWitt:1975ys}
was carried out in the context of
QCD~\cite{Dyakonov:1984,Chapman:1994vk,Megias:2003ui}
and the Abelian Higgs model~\cite{Bandyopadhyay:2026nrv},
and was recently generalized
to generic thermal effective actions~\cite{Chakrabortty:2024wto,Balui:2025yvd,Balui:2025kat}.
In this approach,
the space-time symmetry in the imaginary-time formalism is
automatically encoded.
However,
the resulting thermal effective action is not automatically written
in a non-redundant basis and
can still contain operators related by
integration by parts (IBP) and
equations of motion (EOMs).
This reduction step removes on-shell-redundant directions and projects the
operator set onto a minimal basis in a fixed convention.
Moreover,
medium-induced operators with no zero-temperature counterpart,
including those relevant for
CP- or
CPT-odd effects~\cite{
	Kajantie:1997ky,Gynther:2003za%
	},
can arise and require a dedicated operator analysis.

Consequently, a first-principles construction is needed.
In this work,
we employ the
Hilbert series (HS) method~\cite{%
	Lehman:2015via,Henning:2015daa,Henning:2015alf,Henning:2017fpj
}
to derive the non-redundant finite-temperature SMEFT operator basis directly
from symmetry and redundancy constraints.
This construction goes beyond operator counting.
We identify the allowed non-redundant operator classes in the thermal EFT.
This provides a systematic and model-independent classification of operators
in the thermal theory and enables
precision applications to thermal electroweak and QCD phenomena.
While in the strict static limit
top-down functional matching and Hilbert-series constructions provide
useful mutual cross-checks,
currently available functional implementations are not yet fully general for
capturing Polyakov-loop-related thermal structures at the basis level.

This paper is organized as follows.
Sec.~\ref{sec:finiteTmanifold}
discusses the characteristics of the space-time manifold,
including the boundary conditions of fields
at finite temperature introduced through the imaginary-time formalism.
We comment on the non-perturbative effects,
e.g., the Polyakov loop.
In sec.~\ref{sec:HSonR3S1},
we introduce the finite-temperature
Hilbert series,
where we detail the formalism to compute
the Haar measures and bosonic characters for space-time and
internal gauge symmetries.
We employ these results to compute the non-redundant operator basis
of SMEFT at finite temperature in sec.~\ref{sec:HS-SMEFT}.
We tabulate the complete set of
dimension-five and dimension-six SMEFT operators in the presence of
different constraints in sec.~\ref{sec:SMEFT}.
This section first discusses the covariant form of
the operators and then shows how they are reduced to
the ones that are suitable for a static 3D thermal EFT.
In sec.~\ref{sec:classification_operators},
we classify the effective operators in light of different constraints on the electric and magnetic fields,
identify those that appear only at finite temperature and vanish as $T\to 0$,
and discuss the role of the gauge choice on the operator set.
In sec.~\ref{sec:dim-reduc}, we perform a comparative analysis between our operator basis and those
discussed in the literature, and systematically bridge them.
We briefly highlight specific operators, such as those even and odd under P and CP,
and their roles in possible observables.
We then conclude in sec.~\ref{sec:conclusions}.
Appendix~\ref{app:HS:details}
provides the raw Hilbert series output for reference and
appendix~\ref{app:OperatorGrowth} displays
the growth in the number of operators up to mass dimension ten.

%%%%%%%%%%%%%%%%%%%%%%%%% SECTION %%%%%%%%%%%%%%%%%%%%%%%%%%%%%%%%%%%%%%%%%
%%%%%%%%%%%%%%%%%%%%%%%%% SECTION %%%%%%%%%%%%%%%%%%%%%%%%%%%%%%%%%%%%%%%%%
%
%
\section{%
		The static finite-temperature manifold
    }
\label{sec:finiteTmanifold}

We construct the non-redundant operator basis for
the SMEFT at finite temperature in
the imaginary-time formalism~\cite{Matsubara:1955ws}
by extending
the Hilbert series calculation of operators from
zero-temperature QFT to
finite-temperature QFT.
The set of effective operators at finite temperature
is known as 3-dimensional static
effective field theory (3D-EFT), defined on $\mathbb{R}^3$~\cite{Ginsparg:1980ef,Appelquist:1981vg}.
By introducing the temperature $(T)$ in QFT as a geometric entity
via the imaginary time $(it\equiv \tau)$ formalism,
the temporal direction is compactified on
a circle $S^1$ with radius $\beta = 1/T$.
As a consequence,
$D = d+1$ dimensional finite-temperature QFT
with $d=3$
is defined on the modified manifold
\begin{equation}
\label{eq:finiteTmanifold}
	\mathcal{M}_\T=\mathbb{R}^3\times S^1
	\,,
\end{equation}
instead of zero-temperature $\mathcal{M}_{\T=0}$;
see~\eqref{eq:zeroTmanifold}.%
\footnote{%
	One could attempt a Hilbert series construction directly in static $\mathbb{R}^3$
	treating the temporal gauge-field component $X_0$ as an extra scalar.
	This way, additional identifications
	(e.g.\ $(D_i X_0)\equiv X_{i0}$ and $X_{0i}\equiv[D_0,D_i]$)
	are needed to avoid overcounting.
	Working on $\mathbb{R}^3\times S^1$ instead encodes these structures covariantly
	and tracks operators with multiple temporal derivatives before reduction to the static 3D basis.
}

Due to the thermal-trace in
the partition function,
bosonic fields $\phi$ satisfy periodic and
fermionic fields $\psi$ anti-periodic boundary conditions,
{\em viz.}
\begin{align}
\label{eq:BCs}
    \phi (\tau, \vec{x}) &=  \phi (\beta+\tau, \vec{x})
	\,,& 
    \psi (\tau, \vec{x}) &=  -\psi (\beta+\tau, \vec{x})
	\,.
\end{align}

To construct the 3D-EFT,
we dimensionally reduce the theory by integrating out all
non-zero Matsubara modes~\cite{Matsubara:1955ws}.
The remaining infrared (IR) degrees of freedom are therefore the zero modes.
Since fermions satisfy anti-periodic boundary conditions, they do not admit
zero Matsubara modes, while the non-zero modes acquire
effective masses of
$m_\rmi{eff} \sim \mathcal{O}(\pi T)$.
Consequently, all fermionic modes are integrated out, and
the resulting $3D$-EFT contains only bosonic zero-modes.

%%%%%%%%%%%%%%%%%%%%%%%%%%%%%%%%%%%%%%%%%%%%%%%%%%%%%%%%%%%%%%%%%%%%%%%%%%%%%%%%%%%%%%%

To construct a
finite-temperature QFT
on the $\mathbb{R}^3$ manifold,
one can consider
the Euclidean group ${\rm E}(3)$ or
a double cover of the Euclidean group $\widetilde{{\rm E}}(3)$,
\begin{align}
	{\rm E}(3) &= SO(3) \ltimes \mathbb{T}_3
	\,,\\
	\label{eq:doublecover}
	\widetilde{{\rm E}}(3) &= SU(2)_\T \ltimes \mathbb{T}_3
	\,,
\end{align}
where
$SU(2)_\T$ generates rotations and
$\mathbb{T}_3$ translations in $\mathbb{R}^3$.
Since $SO(3)$, the group of spatial rotations in three dimensions,
has fundamental group $\pi_1(SO(3))=\mathbb{Z}_2$ and is therefore not simply connected,
it is convenient to work with its simply connected double cover
$SU(2)_\T$,
for which $\pi_1(SU(2)_\T)=0$.
Consequently, we formulate the theory using
the covering
$\widetilde{{\rm E}}(3)$ of eq.~\eqref{eq:doublecover} rather than
${\rm E}(3)$.
This choice is natural in the HS construction, both in relativistic~\cite{Lehman:2015via,Lehman:2015coa,Anisha:2019nzx,Henning:2015alf,Henning:2017fpj,Ruhdorfer:2019qmk, Delgado:2022bho,Grojean:2023tsd,Alonso:2024usj,Banerjee:2020bym,Marinissen:2020jmb,Melia:2020pzd,Kondo:2022wcw,Delgado:2023ivp,Sun:2025zuk} and non-relativistic cases~\cite{Kobach:2018nmt,Li:2026eym},
where the number of invariant operators is obtained by integrating characters of the field representations over the symmetry group.
This procedure requires the
characters to be single-valued functions of the group elements,
which is ensured when all fields transform in representations
that are well-defined on the group.
In particular, for
fields with non-zero spin
it is relevant to work with simply connected groups~\cite{%
	Beckers:1981yq,Sitenko:2023rex,aguilar1999,Bacry1970,Douglas_2011}.

In contrast, the compact thermal circle $S^1$ carries an $SO(2)$ symmetry,
which is not simply connected with $\pi_1(SO(2))=\mathbb{Z}$.
However, unlike for $SO(3)$,
this does not prevent a direct HS construction, since all relevant representations of $SO(2)$ are already single-valued.
This non-trivial topology reflects the presence of winding configurations,
such as the untraced Polyakov loop~\cite{Polyakov:1975rs}
\begin{equation}
\label{eq:polyakov}
   \Omega (\vec{x}) =
   	\mathbb{T}\, \Big[\exp \Big(-\int_{0}^{\beta}\! {\rm d}\tau \;X_0(\vec{x}) \Big) \Big]
		\stackrel{X_0(\vec{x})=X_0}{=}
		e^{-\beta X_0}
	\,,
\end{equation}
which is non-local in imaginary time.
Here,
$X_0$ is the temporal component of a generic gauge field,
$\mathbb{T}$ denotes path ordering with respect to Euclidean time $\tau$, and
the final equality is maintained in the gauge $X_0=\mathrm{const}$.
The Polyakov loop is equivalent to
an imaginary chemical potential~\cite{Popov:1988fdi,Prokofev:2011pof},
which shifts integer Matsubara modes by
a non-integer real number~\cite{Popov:1988fdi,Megias:2003ui,Moral-Gamez:2011wcb,Chakrabortty:2024wto}.
It can also be viewed as an
order parameter for confinement~\cite{Susskind:1979up,Weiss:1980rj,Gross:1980br,Pisarski:2000eq,Pisarski:2002ji}.
Its effects are encoded in
the thermal Wilson coefficients and
can affect the thermodynamics of
the phase transition~\cite{%
	Kajantie:1998yc,
	Vuorinen:2006nz,Kurkela:2007dh,Langelage:2010yr,
	Bergner:2013qaa,
	Fukushima:2017csk,Ghiglieri:2020dpq,
	Chakrabortty:2024wto}.
A local expansion of eq.~\eqref{eq:polyakov}
in powers of
the dimensionless $\beta X_0$
recovers a basis in terms of higher $X_0$ powers,
at the price of a tower of higher-dimensional operators
whose Wilson coefficients absorb
the non-perturbative effects~\cite{%
	Chapman:1994vk,Megias:2003ui,Laine:2018lgj,Bernardo:2026nyq}.
The boundary conditions along $S^1$ differ for bosons and fermions,
as given in eq.~\eqref{eq:BCs}.

It is now evident that,
 in Euclidean finite-temperature,
with space-time manifold $\mathcal{M}_\T$ of eq.~\eqref{eq:finiteTmanifold},
the continuous space-time symmetry is reduced compared to $T=0$.
The full finite-temperature symmetry group is
\begin{equation}
\label{eq:symmetryGroup:finiteT}
	\widetilde{{\rm E}}(3)\times SO(2)
	\;\simeq\;
	SU(2)_\T\ltimes \mathbb{T}_3 \times SO(2)
	\,.
\end{equation}

\begin{figure}[t]
    \centering
\begin{tikzpicture}[
	font=\small,
	>=Latex,
	line width=0.5pt,
	bluebox/.style={
		draw=blue!55!black,
		rectangle,
		fill=white,
		align=center,
		inner sep=8pt
	},
	redbox/.style={
		draw=red!70!black,
		rectangle,
		fill=white,
		align=center,
		inner sep=8pt
	},
	blackbox/.style={
		draw=black,
		rectangle,
		fill=white,
		align=center,
		inner sep=8pt
	},
	blueoval/.style={
		draw=blue!55!black,
		rectangle,
		fill=white,
		align=center,
		inner sep=8pt
	},
	redoval/.style={
		draw=red!70!black,
		rectangle,
		fill=white,
		align=center,
		inner sep=8pt
	},
	blackoval/.style={
		draw=black,
		rectangle,
		fill=white,
		align=center,
		inner sep=8pt
	},
	bluearrow/.style={->,draw=blue!60!black},
	redarrow/.style={->,draw=red!70!black},
	blackarrow/.style={->,draw=black}
]
	\def\xleftcol{4.0}
	\def\xrightcol{12.0}
	\def\xhaarT{2.1}
	\def\xcharT{6.5}
	\def\xhaarG{10.5}
	\def\xcharG{14.0}

	\node[blueoval,minimum width=4.8cm,minimum height=1.35cm,text=blue!60!black]
		(space) at (\xleftcol,10.2)
		{Space-time
		[eq.~\eqref{eq:finiteTmanifold}]
		\\
		$\mathcal{M}_\T = \mathbb{R}^3 \times S^1$,
	};

	\node[bluebox,minimum width=5.8cm,minimum height=2.15cm,text=blue!60!black]
		(group) at (\xleftcol,7.8)
		{Space-time symmetry [eq.~\eqref{eq:symmetryGroup:finiteT}]
		\\[2pt]
		$\widetilde E(3)\times SO(2)
		= SU(2)_\T\ltimes \mathbb{T}_3\times SO(2)$};

	\node[redbox,minimum width=5.8cm,minimum height=2.15cm,text=red!70!black]
		(internal) at (\xrightcol,7.8)
		{Internal symmetry [eq.~\eqref{eq:SMgauge}]
		\\[2pt]
		$\prod_j G_j =
			\underbrace{
				SU(3)_c \times
				SU(2)_\rmii{$L$} \times
				U(1)_\rmii{$Y$}}_{\text{SM gauge}}
		$
		};

	\node[blueoval,minimum width=2.9cm,minimum height=1.35cm,inner sep=5pt,
		text=blue!60!black]
		(haarT) at (\xhaarT,4.9)
		{$SU(2)_\T$ Haar measure\\[2pt]
		$\int_\T {\rm d}\mu_\T$};

	\node[bluebox,minimum width=2.9cm,minimum height=1.35cm,inner sep=5pt,
		text=blue!60!black]
		(charT) at (\xcharT,4.9)
		{Character under $SU(2)_\T$
		\\[2pt]
		$\widetilde\chi$
		(EOM in $D_i$)
		};

	\node[redoval,minimum width=2.9cm,minimum height=1.35cm,inner sep=5pt,
		text=red!70!black]
		(haarG) at (\xhaarG,4.9)
		{$G$ Haar measure\\[2pt]
		$\int_{G_j} {\rm d}\mu_j$};

	\node[redbox,minimum width=2.9cm,minimum height=1.35cm,inner sep=5pt,
		text=red!70!black]
		(charG) at (\xcharG,4.9)
		{Character under $G$\\[2pt]
		$\prod_j \chi_j$};

	\node[blackoval,minimum width=5.8cm,minimum height=1.7cm]
		(totalhaar) at (\xleftcol,2.2)
		{Total Haar measure
		\\[2pt]
		$\int {\rm d}\mu = \int_\T {\rm d}\mu_\T \prod_j\int_{G_j} {\rm d}\mu_j$};

	\node[blackbox,minimum width=5.8cm,minimum height=1.7cm]
		(pe) at (\xrightcol,2.2)
		{Plethystic exponential
		for integer spin
		\\[2mm]
		$
			\mathrm{PE}[\phi]
		$,  [eq.~\eqref{eq:PE:integerSpin}]
		};

	\node[blackbox,minimum width=5.8cm,minimum height=1.7cm]
		(momentum) at (\xleftcol,-1.05)
		{Momentum gen.\ function [eq.~\eqref{eq:P3:momentumGeneratingFunction}]
		\\[2pt]
		$P^{(3)}(D_i,\alpha)$,
		(IBP in $D_i$)
	};

	\node[blackbox,minimum width=5.8cm,minimum height=1.7cm]
		(hs) at (\xrightcol,-1.05)
		{Hilbert series\\[1mm]
		$
		\begin{aligned}
		\mathcal{H}^\text{spatial} &=
        \Bigl(
		\int_\T {\rm d}\mu_\T
		\prod_j \int_{G_j}{\rm d}\mu_j
		\Bigr)\,
		\frac{Z}{P^{(3)}(D_i,\alpha)}
		\nn &
		+ \Delta \mathcal{H}^\text{spatial}
		\end{aligned}
		$
		};

	\coordinate (groupsplit) at ($(group.south)+(0,-0.45)$);
	\coordinate (internalsplit) at ($(internal.south |- groupsplit)$);
	\coordinate (haarTturn) at ($(haarT.north)+(0,0.55)$);
	\coordinate (charTturn) at ($(charT.north)+(0,0.55)$);
	\coordinate (haarGturn) at ($(haarG.north)+(0,0.55)$);
	\coordinate (charGturn) at ($(charG.north)+(0,0.55)$);
	\coordinate (haarTsplit) at ($(groupsplit -| haarTturn)$);
	\coordinate (charTsplit) at ($(groupsplit -| charTturn)$);
	\coordinate (haarGsplit) at ($(internalsplit -| haarGturn)$);
	\coordinate (charGsplit) at ($(internalsplit -| charGturn)$);
	\coordinate (totalhaarturn) at ($(totalhaar.north)+(0,0.55)$);
	\coordinate (peturn) at ($(pe.north)+(0,0.80)$);
	\coordinate (totalhaarsplit) at ($(haarT.south |- totalhaarturn)$);
	\coordinate (totalhaargsplit) at ($(haarG.south |- totalhaarturn)$);
	\coordinate (pesplitT) at ($(charT.south |- peturn)$);
	\coordinate (pesplitG) at ($(charG.south |- peturn)$);
	\coordinate (hsfrommeasure) at ($(totalhaar.south)+(0,-0.55)$);
	\coordinate (hsmiddlesplit) at ($(hs.north |- hsfrommeasure)$);
	\coordinate (hsfromabove) at ($(hs.north)+(0,0.55)$);

	\draw[bluearrow] (space.south) -- (group.north);
	\draw[draw=blue!60!black] (group.south) -- (groupsplit);
	\draw[bluearrow] (groupsplit) -- (haarTsplit) -- (haarTturn) -- (haarT.north);
	\draw[bluearrow] (groupsplit) -- (charTsplit) -- (charTturn) -- (charT.north);
	\draw[draw=red!70!black] (internal.south) -- (internalsplit);
	\draw[redarrow] (internalsplit) -- (haarGsplit) -- (haarGturn) -- (haarG.north);
	\draw[redarrow] (internalsplit) -- (charGsplit) -- (charGturn) -- (charG.north);
	\draw[blackarrow] (haarT.south) -- (totalhaarsplit) -- (totalhaarturn) -- (totalhaar.north);
	\draw[blackarrow] (haarG.south) -- (totalhaargsplit) -- (totalhaarturn) -- (totalhaar.north);
	\draw[blackarrow] (charT.south) -- (pesplitT) -- (peturn) -- (pe.north);
	\draw[blackarrow] (charG.south) -- (pesplitG) -- (peturn) -- (pe.north);
	\draw[blackarrow] (totalhaar.south) -- (hsfrommeasure) -- (hsmiddlesplit) -- (hsfromabove) -- (hs.north);
	\draw[blackarrow] (pe.south) -- (hs.north);
	\draw[blackarrow] (momentum.east) -- (hs.west);
\end{tikzpicture}%
  \caption{%
		Schematic workflow of the finite-temperature
		Hilbert series construction.
		}
  \label{fig:HSFT_flowchart}
\end{figure}
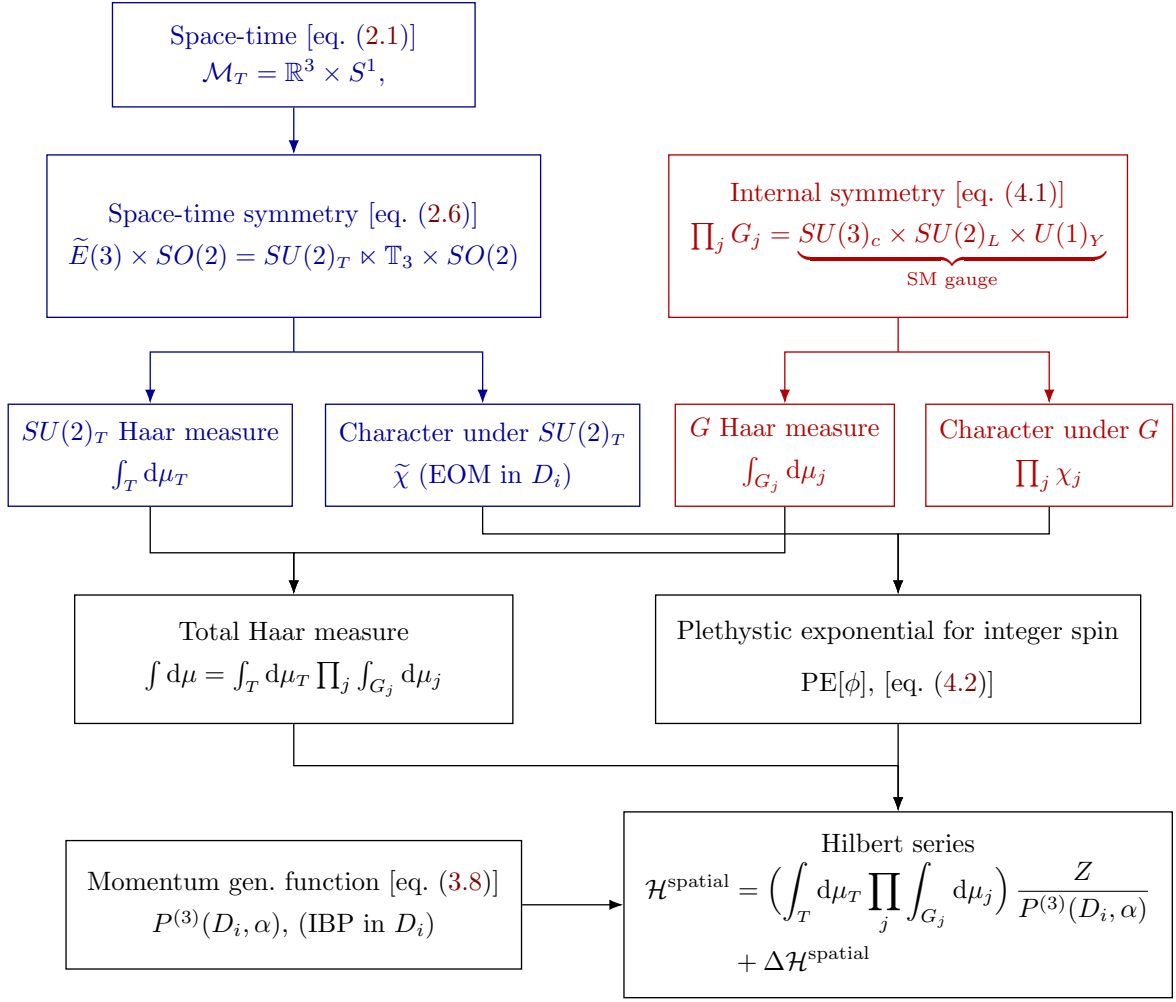

As a result, and as indicated
schematically in fig.~\ref{fig:HSFT_flowchart},
derivative structures are organized
according to the finite-temperature symmetry group in eq.~\eqref{eq:symmetryGroup:finiteT}.
In the next section,
we implement the symmetry group eq.~\eqref{eq:symmetryGroup:finiteT},
via its characters and Haar measures.
%%%%%%%%%%%%%
%%%%%%%%%%%%%%%%%%%%%%%%

%%%%%%%%%%%%%%%%%%%%%%%%% SECTION %%%%%%%%%%%%%%%%%%%%%%%%%%%%%%%%%%%%%%%%%
%
\section{%
    Finite-temperature Hilbert series on
    $\mathbb{R}^3 \times S^1$}
\label{sec:HSonR3S1}

We first highlight the symmetry of the space-time manifold
relevant to defining the zero-temperature Hilbert series.
The flat $(3+1)$-dimensional space-time manifold
with $\mathcal{X} \equiv (t, x^i)$
of zero-temperature QFT is
\begin{equation}
\label{eq:zeroTmanifold}
	\mathcal{M}_{\T = 0} = \mathbb{R}^3 \times \mathbb{R}^1
	\,,
\end{equation}
where
$\mathbb{R}^3$ is the three-dimensional non-compact spatial manifold.
The corresponding space-time symmetry group is
the Lorentz group, $SO(3,1)$.
This group is non-compact, and therefore lacks finite-dimensional unitary representations.
As a result, the standard HS construction,
which relies on finite-dimensional characters and Haar integration over
compact groups, cannot be applied directly.

When constructing the effective Lagrangian,
using the HS method,
we are interested in those operators that are invariant
under space-time and internal symmetries.
This method has been successfully developed for
compact connected groups~\cite{%
Hanany:2014dia,Lehman:2015via,Henning:2015daa,Henning:2017fpj,Banerjee:2020bym,Anisha:2019nzx
}; see~\cite{Grinstein:2023njq,Grinstein:2024jqt,Grinstein:2024iyf} for other applications involving internal symmetries.
To treat the non-compact Lorentz symmetry, it is convenient to work in the
Weyl (chiral) representation, in which fields are decomposed into
left- and right-chiral components.
At the level of algebra, this corresponds to expressing the Lorentz
generators in terms of two commuting $\mathfrak{su}(2)$ algebras.
For the HS construction, one then works with the associated
compact group
$SO(3,1) \simeq SU(2)_\rmii{$L$} \times SU(2)_\rmii{$R$}$.
In this formulation, fields are assigned to representations of
$SU(2)_\rmii{$L$} \times SU(2)_\rmii{$R$}$, and, for example,
the field-strength tensor $F^{\mu\nu}$ is decomposed into its
self-dual and anti-self-dual components
$F^{\mu\nu}_{\rmii{$L$},\rmii{$R$}}$.
This allows one to define Weyl characters for fields of different spin,
together with the corresponding Haar measures for these compact groups,
expressed in terms of coordinates on their maximal tori
to which the groups are mapped;
see~\cite{%
	Henning:2015alf,Henning:2015daa,Henning:2017fpj,Anisha:2019nzx,
	Banerjee:2020bym}.

In this zero-temperature construction,
the space-time manifold is non-compact, and
the fields only satisfy the standard boundary conditions (BCs)
such that
$\Psi(\mathcal{X}) \to 0$ sufficiently fast as $\mathcal{X} \to \infty$.
Consequently, total-derivative terms do not contribute, and
operators that differ only by a total derivative are identified as equivalent.
For this reason, integration by parts (IBP) acts as a redundancy relation in
the operator construction, identifying operators that differ only by a total
derivative.
A complementary source of redundancy arises from
the equations of motion (EOM).
Upon imposing the EOM (i.e.\ working on-shell),
additional linear relations among operators are obtained, allowing
the operator basis to be reduced further.
The construction of the non-redundant operator bases using HS
is detailed in~\cite{Henning:2015daa,Henning:2017fpj,Banerjee:2020bym,Anisha:2019nzx,Alonso:2024usj}. 

%%%%%%%%%%%%%%%%%%%%%%%%% SECTION %%%%%%%%%%%%%%%%%%%%%%%%%%%%%%%%%%%%%%%%%
%
\subsection{Space-time induced characters of bosonic infrared modes}
\label{sec:finiteTcharacters-ST}

With the deformation of the space-time manifold
from $\mathcal{M}_{\T=0}$ to $\mathcal{M}_\T$ at finite temperature,
also the basic building blocks of operator construction change.
At finite temperature, the non-trivial boundary
conditions eq.~\eqref{eq:BCs} are employed only along
the compact temporal direction ($S^1$).
Here, the spatial directions ($\mathbb{R}^3$) remain unchanged, and
the EOMs are associated with the spatial derivatives,
$D_i;\; \forall i=1,2,3$.
In contrast,%
\footnote{%
	The distinction between temporal and spatial derivatives in
	the Hilbert series has also proven necessary in
	non-relativistic QFTs~\cite{Kobach:2017xkw,Kobach:2018nmt,Li:2026eym}, though by distinct physical reasons.
}
the temporal derivative $D_0$ is tied to the Matsubara modes.
Thus, while constructing the
dimensionally reduced 3D-EFT, we refrain from imposing an EOM driven by $D_0$.
The constraints owing to the selective EOM are encoded explicitly in
the finite-temperature characters of bosonic fields. 
We compute them in the following sections based on their respective
single particle modules (SPMs)~\cite{Henning:2017fpj}.

In tab.~\ref{tab:ch},
we list the characters of the relevant fields based on
the different constraints imposed subsequently.
\begin{table}[t]
    \centering
    \renewcommand{\arraystretch}{1.5}
	\begin{tabular}{|c|r|l|}
	\hline
     \textbf{Field} &
     \multicolumn{1}{c|}{\textbf{Constraint}} &
     \multicolumn{1}{c|}{\textbf{Character under $SU(2)_\T$}}
	\\
	\hline
    \hline
    \textbf{scalar} &
		$D_i D^i \phi = 0$,
		eq.~\eqref{eq: scalar EOM}
		&
		$P^{(0,3)}(D_0,D_i,\alpha)(1-D_i^2)$,
        eq.~\eqref{eq:chi_scalar_1}
    \\
    \hline
    \multirow{2}{*}{\textbf{vector}} &
		$D_i V_i = 0$,
		eq.~\eqref{eq:GaussBianchi}
		& 
		$P^{(0,3)}(D_0,D_i,\alpha) \left[\chi_1(\alpha)-D_i\right]$,
    eq.~\eqref{eq:chi_spin1_div}
    \\
    \cline{2-3} &

		\makecell[r]{
		$\epsilon_{ijk}D_i U_j = 0$,
		\hphantom{eq.~\eqref{eq: spin-1 full}}\\
		$D_i U_i = 0$,
		eq.~\eqref{eq: spin-1 full}}
		&
		$P^{(0,3)}(D_0,D_i,\alpha)(1-D_i)[\chi_1(\alpha)-D_i]$,
    eq.~\eqref{eq:chi_spin1_div+curl}
	\\
	\hline
    \end{tabular}
    \caption{%
		Characters of bosonic fields at
		finite temperature given different constraints.
	}
    \label{tab:ch}
\end{table}

%%%%%%%%%%%%%%%%%%%%%%%%% SECTION %%%%%%%%%%%%%%%%%%%%%%%%%%%%%%%%%%%%%%%%%
%
\subsubsection[Scalar fields: spin-0]{%
    Scalar fields: spin-0
}
 
Scalar fields, singlets under space-time symmetry, are not affected by the deformation of the
space-time manifold from $\mathcal{M}_{\T=0}$ to $\mathcal{M}_\T$,
except that they satisfy periodic boundary conditions along $S^1$.
The SPM of a scalar field $\phi$ at finite temperature can be written as
\begin{align}
\label{eq: scalar module}
    D_0^n D_{\{ i_1} \cdots D_{i_p \}} \phi &
	\,,&
	\text{where} \;\; n,p &\in \mathbb{Z}_{\geq 0}
	\,.
\end{align}
Here, $\{\cdot\}$ denotes the traceless symmetric part, 
while the antisymmetric part, i.e., $[D_i,D_j]$, is related to the magnetic components of the gauge field tensors.
Thus, this part needs to be removed by suitably defining the SPMs of
the field tensors (see spin-1 section).
The removal of traces is equivalent to imposing constraints due to
the spatial EOM that reads
\begin{equation}
	\label{eq: scalar EOM}
    D_i D^i \phi = 0
	\,.
\end{equation}
It is worth mentioning that
it is sufficient to choose one particular position of the $n$-temporal derivatives for each $p$ in eq.~\eqref{eq: scalar module}, as their arbitrary distribution will be related to $[D_0,D_j]$,
which is equivalent to the electric component of the field tensor, and
therefore already taken care of by its own SPM.

Since we work with the covering group $\widetilde{E}(3)$, see eq.~\eqref{eq:doublecover},
we express the character of the traceless symmetric representation of $SO(3)$ with $p$-spatial vector indices as
the character of the spin-$p$ representation of $SU(2)_\T$,
\begin{equation}
\label{eq: SU(2) ch}
    \chi_p (\alpha) = \sum_{k=-p}^p \alpha^{2k}
	\,,
\end{equation}
where $\alpha$ is the unimodular complex number, the coordinate of
the maximal torus $\mathbb{T}^1$ of $SU(2)_\T$.
Employing constraints from
eqs.~\eqref{eq: scalar module} and~\eqref{eq: SU(2) ch},
the character of the scalar SPM at finite temperature is
\begin{align}
\label{eq:chi_scalar}
    \chi_{\text{scalar}}^{\T} &=
		\sum_{n,p=0}^{+\infty} D_0^n D_i^p \chi_p (\alpha) =
		\frac{1+D_i}{(1-D_0)(1-D_i \alpha^{-2})(1-D_i \alpha^2)}
	\,,
\end{align} 
where $D_0$ and $D_i$ are spurions.

This character computation can also be performed in an alternative way.
First, we isolate the contribution from the spatial derivatives as follows
\begin{align}
\label{eq: deriv module}
    D_{( i_1} \cdots D_{i_p )}&
	\,,&
	p& \in \mathbb{Z}_{\geq 0}
	\,,
\end{align}
where $(\cdot)$ denotes the symmetric part; see the discussion around eq.~\eqref{eq: scalar module},
the antisymmetric part has to be removed. Here, $D_i$ transforms as a vector of $SU(2)_\T$. Thus, the character in association with eq.~\eqref{eq: deriv module} is the 3-momentum (descendant) generating function $P^{(3)}(D_i,\alpha)$,
\begin{align}
     P^{(3)}(D_i,\alpha) &=
 	\sum_{p=0}^{\infty} D_i^{\,p}\;\chi_{\mathrm{sym}^p_{(1)}}(\alpha) =
	\sum_{p=0}^{\infty} D_i^{\,p}\;\mathrm{sym}^p[\chi_1(\alpha)]
	\,,
\end{align}
where $\mathrm{sym}^p_{(1)}$ is the $p$-th symmetric tensor of
an $SU(2)_\T$ vector, and
$\mathrm{sym}^p[\cdot]$  denotes the $p$-th symmetric product of
a function~\cite{Henning:2017fpj}.%
\footnote{%
    The $n$-th symmetric product of $f(x_1,...,x_r)$, denoted as $\mathrm{sym}^n[f(x_1,...,x_r)]$,
		is defined through the plethystic exponential
    $\mathrm{PE}[w f(x_1,...,x_r)] \equiv \exp\bigl\{\sum_{m=1}^\infty \frac{w^{m}}{m}f(x_1^m,...,x_r^m)\bigr\}$ as its coefficient in $w^n$, where $w$ is a generic bookkeeping parameter.
		In contrast, the $n$-th antisymmetric product of $f(x_1,...,x_r)$, denoted as $\wedge^n[f(x_1,...,x_r)]$, is defined through the antisymmetric plethystic exponential $\mathrm{PE}_A[w f(x_1,...,x_r)] \equiv \exp\bigl\{-\sum_{m=1}^\infty \frac{(-w)^{m}}{m}f(x_1^m,...,x_r^m)\bigr\}$ as its coefficient in $w^n$.
}
Using properties of the bosonic plethystic exponential,
we can write
\begin{align}
\label{eq:P3:momentumGeneratingFunction}
    P^{(3)}(D_i,\alpha) &= \exp\left[\sum_{m=1}^\infty \frac{D_i^{\,m}}{m}\chi_1(\alpha^m)\right]
		= \exp\left[\sum_{m=1}^\infty \frac{D_i^{\,m}}{m}\left(\frac{1}{\alpha^{2m}}+1+\alpha^{2m}\right)\right] \notag
    \\
    &= \frac{1}{(1-D_i)(1-D_i \alpha^{-2})(1-D_i \alpha^2)}
	\,.
\end{align}
Equivalently, because $J_z$ has eigenvalues $m_j=\{+1,0,-1\}$,
\begin{align}
\label{eq:P3}
 P^{(3)}(D_i,\alpha) & =
	\sum_{m_{-1},\, m_0,\, m_{+1} = 0}^{\infty}
	D_i^{\,m_{-1}+m_0+m_{+1}}\,
	\alpha^{2[(+1)m_{+1}+(0)m_{0}+(-1)m_{-1}]}
	\nn & =
	\Biggl[\sum_{m_{-1}=0}^{\infty}(D_i\,\alpha^{-2})^{m_{-1}}\Biggr]
	\Biggl[\sum_{m_0=0}^{\infty} D_i^{\,m_0}\Biggr]
	\Biggl[\sum_{m_{+1}=0}^{\infty}(D_i\,\alpha^{2})^{m_{+1}}\Biggr]
	\nn[2mm] &=
	\frac{1}{(1-D_i)\,(1-D_i/\alpha^{2})\,(1-D_i\alpha^{2})}
	\,.
\end{align}

Together with the contribution of the temporal derivatives,
$P^{(0)}(D_0)$,
we can write the full momentum generating function
\begin{align}
	\label{eq:P0}
    P^{(0)}(D_0) &= \sum_{n=0}^\infty D_0^n = \frac{1}{1-D_0}
	\,,
	\\[2mm]
	\label{eq:P30}
	P^{(0,3)}(D_0,D_i,\alpha) & =
	P^{(0)}(D_0)
	\times
	P^{(3)}(D_i,\alpha)
	\,,
\end{align}
which appears in eq.~\eqref{eq:chi_scalar} as
\begin{equation}
	\label{eq:chi_scalar_1}
    \chi_{\text{scalar}}^{\T} = P^{(0,3)}(D_0,D_i,\alpha)(1-D_i^2)\,,
\end{equation}
highlighting the removal of the relevant descendants.
These descendants are proportional to the spatial EOM of
massless scalars, see eq.~\eqref{eq: scalar EOM}, and are also encapsulated in the form of 
$D_0^n D_iD^i \phi$ for $n>0$.
The purely spatial part of
the character,
$\chi_{\text{scalar}}^{\T}/P^{(0)}(D_0)$, is a short conformal character of
a free scalar field~\cite{Dolan:2005wy,Henning:2017fpj}.

%%%%%%%%%%%%%%%%%%%%%%%%% SECTION %%%%%%%%%%%%%%%%%%%%%%%%%%%%%%%%%%%%%%%%%
%
\subsubsection[Vector fields: spin-1]{%
    Vector fields: spin-1
}

The bosonic fields with non-zero spin transform non-trivially under the thermal symmetry group in
eq.~\eqref{eq:doublecover} of
the manifold $\mathcal{M}_\T$~\cite{%
	Beckers:1981yq, Sitenko:2023rex, aguilar1999, Bacry1970, Douglas_2011}.
At finite temperature, in the imaginary-time formalism, the thermal state
selects a preferred rest frame, so that full Lorentz invariance is no longer
realized on physical observables;
see footnote~\ref{footnote:brokenLorentz}.
Accordingly, the gauge-field tensor $X_{\mu\nu}$ decomposes
naturally into
electric ($X_{\rmii{E}\,i}$) and magnetic ($X_{\rmii{M}\,i}$) components,
\begin{align}
	\label{eq:EB}
	X_{\rmii{E}\,i} &= X_{0i}
	\,,&
	X_{\rmii{M}\,i} &= \frac{1}{2}\epsilon_{ijk} X_{jk}
	\,. 
\end{align}
Both $\mathbf{X_{\rmii{E}}}$ and $\mathbf{X_{\rmii{M}}}$ transform as vectors under
$SO(3)$, or equivalently as spin-$1$ fields under $SU(2)_\T$, while both fields are invariant under translations $\mathbb{T}_3$.%
\footnote{%
    Under three-dimensional rotations, $\mathbf{X_{\rmii{E}}}$ transforms as a
    vector whereas $\mathbf{X_{\rmii{M}}}$ transforms as an axial vector. Consequently,
    under parity (P) one has
    $\mathbf{X_{\rmii{E}}} \to -\mathbf{X_{\rmii{E}}}$ and
    $\mathbf{X_{\rmii{M}}} \to \mathbf{X_{\rmii{M}}}$, while under
    time reversal (T) along the thermal
    $S^1$ circle,
    $\mathbf{X_{\rmii{E}}} \to \mathbf{X_{\rmii{E}}}$ and
    $\mathbf{X_{\rmii{M}}} \to -\mathbf{X_{\rmii{M}}}$.
}

%%%%%%%%%%%%%%%%%%%%%%%%% SECTION %%%%%%%%%%%%%%%%%%%%%%%%%%%%%%%%%%%%%%%%%
%
\constraintsubsubsection
	{Constraint~I}
	{$\nabla\cdot\mathbf{X_{\rmii{E}}}=\nabla\cdot\mathbf{X_{\rmii{M}}}=0$}
	{sec:constraint-I:def}

If subjected to only the source-free Gauss' law and Bianchi identity, respectively,
\begin{align}
\label{eq:GaussBianchi}
  \nabla\cdot\mathbf{X_{\rmii{E}}} \equiv  D_i X_{\rmii{E}\,i} &= 0
	\,,&
 \nabla\cdot\mathbf{X_{\rmii{M}}} \equiv  D_i X_{\rmii{M}\,i}  &= 0
	\,,
\end{align}
the SPM of the electric or magnetic components will be
\begin{align}
    D_0^n D_{( i_1} \cdots D_{i_p )} V_{i_0} &
	\,,&
	\text{where }n,p & \in \mathbb{Z}_{\geq 0}
	\,,
	&
	\text{with}\;\;i_0 &\neq i_1,\dots,i_0 \neq i_p
	\,,
\end{align}
where $V_i\in\{X_{\rmii{E}\,i},X_{\rmii{M}\,i}\}$ depicts the vector field that satisfies eq.~\eqref{eq:GaussBianchi}. The distinguished $V$-index guarantees that terms proportional to
$D_iV_i$ are removed from the module.
The symmetrization of the spatial derivatives follows
from eq.~\eqref{eq: deriv module}, and
the contribution of the temporal derivatives is analogous to
the case of the scalar field SPM in eq.~\eqref{eq: scalar module}.

Thus,
the character of the {\em divergence-free} vector SPM at finite temperature reads
\begin{align}
\label{eq:chi_spin1_div}
	\chi^{\T,\rmii{\ref{sec:constraint-I:def}}}_{\text{vector}} &=
	\sum_{n,p=0}^{+\infty}
		D_0^n D_i^p\;\chi_{\mathrm{sym}^p_{(1)}}(\alpha)
		\bigl[
			\chi_1(\alpha)-D_i
		\bigr]
    =
	\frac{1}{1-D_0}\sum_{p=0}^{+\infty} D_i^p\;\chi_{\mathrm{sym}^p_{(1)}}(\alpha)
		\bigl[
			\chi_1(\alpha)-D_i
		\bigr]
	\nn[2mm]
	&= 
	P^{(0,3)}(D_0,D_i,\alpha) \bigl[\chi_1(\alpha)-D_i\bigr]
	\,,
\end{align}
where
the superscript~\ref{sec:constraint-I:def}
denotes the divergence constraint in eq.~\eqref{eq:GaussBianchi}.
The interpretation of the character follows from the fact that
the ideal generated by $D_i V_i$ is a scalar at grade $D_i^1$, 
freely acted on by derivatives.
Thus, $P^{(0,3)}(D_0,D_i,\alpha)D_i$ is the character that should be removed to
be consistent with the divergence constraint. 
It is worth mentioning that the purely spatial part of
the character $\chi^{\T,\rmii{\ref{sec:constraint-I:def}}}_{\text{vector}}/P^{(0)}(D_0)$ is
a short conformal character of
a free vector current~\cite{Dolan:2005wy,Henning:2017fpj}.

%%%%%%%%%%%%%%%%%%%%%%%%% SECTION %%%%%%%%%%%%%%%%%%%%%%%%%%%%%%%%%%%%%%%%%
%
\constraintsubsubsection
	{Constraint~II}
	{$\nabla\cdot\mathbf{X_{\rmii{E}}}=\nabla\cdot\mathbf{X_{\rmii{M}}}=0\;\oplus\;\nabla\times\mathbf{X_{\rmii{E}}}=\nabla\times\mathbf{X_{\rmii{M}}}=0$}
	{sec:constraint-II:def}

At finite temperature, constraints involving spatial derivatives $D_i$
are considered to be on the same footing as covariant constraints involving $D_\mu$
at zero temperature.
For example,
the zero-temperature scalar EOM is
$D_\mu D^\mu \phi = 0$,
whereas its finite-temperature counterpart is given in
eq.~\eqref{eq: scalar EOM}.
For the gauge-field strength,
the zero-temperature EOM and Bianchi identity are
$D^\mu F_{\mu\nu} = 0$ and
$D^\mu \Tilde{F}_{\mu\nu} = 0$, respectively, with
$\Tilde{F}_{\mu\nu} \equiv \epsilon_{\mu\nu\alpha\beta}F^{\alpha\beta}$.
At finite temperature,
these reduce to
\begin{align}
	\label{eq: spin-1 full}
    D^i F_{i\nu} &= 0
	\,,&
	D^i \Tilde{F}_{i\nu} &= 0
	\,.
\end{align}
The choice $\nu = 0$ reproduces
eq.~\eqref{eq:GaussBianchi},
which is already accounted for in
$\chi^{\T,\rmii{\ref{sec:constraint-I:def}}}_{\text{vector}}$.
The spatial choice $\nu = j$ in
eq.~\eqref{eq: spin-1 full} yields
\begin{align}
	\label{eq:curl} 
  (\nabla \times \mathbf{X_{\rmii{M}}})_j \equiv  \epsilon_{jik} D_i X_{\rmii{M}\,k} &= 0
	\,,&
  (\nabla \times \mathbf{X_{\rmii{E}}})_j \equiv	\epsilon_{jik} D_iX_{\rmii{E}\,k} &= 0
	\,,
\end{align}
which emerge as different constraints.
The SPM
\begin{align}
    D_0^n D_{\{ i_1} \cdots D_{i_p} U_{i_0\}}&
	\,,&
	n,p& \in \mathbb{Z}_{\geq 0}\,, 
\end{align}
automatically mod out curls and divergences by removing
the antisymmetric part and the traces, respectively.
Here, $U_i \in \{X_{\rmii{E}\,i}, X_{\rmii{M}\,i} \}$ denotes a vector field that satisfies
both eq.~\eqref{eq:GaussBianchi} and eq.~\eqref{eq:curl}.
Consequently,
the spatial part of the SPM transforms in the spin-$(p+1)$
representation of $SU(2)_\T$.
This gives the character of the
{\em divergence-free and curl-free} vector SPM at finite temperature,
\begin{align}
	\label{eq:chi_spin1_div+curl}
    \chi^{\T,\rmii{\ref{sec:constraint-II:def}}}_{\text{vector}} &=
	\sum_{n,p=0}^{+\infty} D_0^n D_i^p \chi_{p+1} (\alpha) =
	P^{(0,3)}(D_0,D_i,\alpha)(1-D_i)\bigl[\chi_1(\alpha)-D_i\bigr]
	\,,
\end{align}
where
the superscript~\ref{sec:constraint-II:def}
denotes the divergence and curl constraints in
eqs.~\eqref{eq:GaussBianchi} and~\eqref{eq:curl}, respectively.

By expanding the last two factors, 
${(1-D_i)[\chi_1(\alpha)-D_i] =
	\chi_1(\alpha)-D_i-D_i\chi_1(\alpha)+D_i^2\,,}$
we can interpret the characters as follows:
\begin{itemize}
	\item[(i)]
		Since $U_i$ transforms as a vector,
		the primary contributes with $[\chi_1(\alpha)]$; 
	\item[(ii)] $[-D_i]$ removes the divergences as in eq.~\eqref{eq:chi_spin1_div};
	\item[(iii)] $[\epsilon_{ijk}D_i U_j]$ transforms as a vector and contains one spatial derivative, so
	$[-D_i\chi_1(\alpha)]$ removes the curls;
	\item[(iv)]
		the divergence-free and curl-free conditions remove
		$[\epsilon_{ijk} D_i D_j U_k]$.
		Since this object carries two spatial derivatives and transforms as a scalar,
		a $[+D_i^2]$ contribution must be added back to ensure correct counting.
\end{itemize}
It is worth mentioning that the purely spatial part of the character
$\chi^{\T,\rmii{\ref{sec:constraint-II:def}}}_{\text{vector}}/P^{(0)}(D_0)$ no longer corresponds to a conformal character.
Instead, it gives the character of a harmonic 1-form~\cite{Henning:2017fpj}.

We discuss in sec.~\ref{sec:classification_operators}
how the different constraints and their corresponding characters
affect the operator bases.
This, in turn, helps to identify the most suitable operator structures for the thermal field theory of interest.
We employ the equivalent constraints uniformly
in both the Abelian and non-Abelian gauge sectors.

%%%%%%%%%%%%%%%%%%%%%%%%% SECTION %%%%%%%%%%%%%%%%%%%%%%%%%%%%%%%%%%%%%%%%%
%
\subsection{%
		Internal gauge symmetry induced characters of bosonic infrared modes}
\label{sec:finiteTcharacters-IGS}

The fields also transform under the SM gauge symmetry, which is connected and compact.
The representations of the SM fields under the internal gauge symmetry
remain unchanged in the presence of temperature. 
We briefly discuss the computation of the characters of
the fields under the SM gauge group. 

For the Abelian group $U(1)$,
the character is completely determined by
the $U(1)$ charge $q$ of the field,
\begin{align}
    \chi_{U(1)} (z)=z^q
	\,.
\end{align}

For a general $SU(N)$ factor, let $R$ denote an irreducible representation labeled by
$r_1>r_2>\cdots>r_{N-1}>0$, with $r_N\equiv 0$.
The Weyl character evaluated on a maximal-torus element
$M(\epsilon)=\mathrm{diag}(\epsilon_1,\ldots,\epsilon_N)$ is given by \cite{Balantekin:2001id,Plymen:1976,koike1987,littlewood1977theory,rossmann2006lie}
\begin{align}
	\chi_R\big(M(\epsilon)\big)
	=
	\frac{\det \mathcal{N}(\epsilon;r)}{\det \Delta(\epsilon)}
	=
	\frac{\det(\epsilon_a^{r_b})}{\prod_{1\le a<b\le N}(\epsilon_a-\epsilon_b)}
	\,,
\end{align}
where
\begin{align}
\label{eq:Vandermonde}
	\mathcal{N}(\epsilon;r) &=
	\begin{pmatrix}
		\epsilon_1^{r_1} & \epsilon_1^{r_2} & \cdots & \epsilon_1^{r_{N-1}} & 1\\
		\epsilon_2^{r_1} & \epsilon_2^{r_2} & \cdots & \epsilon_2^{r_{N-1}} & 1\\
		\vdots & \vdots & \ddots & \vdots & \vdots\\
		\epsilon_N^{r_1} & \epsilon_N^{r_2} & \cdots & \epsilon_N^{r_{N-1}} & 1
	\end{pmatrix}
	\,,&
	\Delta(\epsilon) &=
	\begin{pmatrix}
		\epsilon_1^{N-1} & \epsilon_1^{N-2} & \cdots & \epsilon_1 & 1\\
		\epsilon_2^{N-1} & \epsilon_2^{N-2} & \cdots & \epsilon_2 & 1\\
		\vdots & \vdots & \ddots & \vdots & \vdots\\
		\epsilon_N^{N-1} & \epsilon_N^{N-2} & \cdots & \epsilon_N & 1
	\end{pmatrix}
	\,.
\end{align}
The denominator is the Vandermonde determinant,
\begin{align}
	\det\Delta(\epsilon)=\prod_{1\le a<b\le N}(\epsilon_a-\epsilon_b)
	\,.
\end{align}

To explicitly evaluate the character, the torus eigenvalues $\epsilon_a$ must satisfy
the $SU(N)$ constraint: $\prod_{a=1}^N \epsilon_a=1$.
A convenient parameterization introduces $N-1$ independent phases,
$z_i=e^{i\theta_i}$ for $i=1,\dots,N-1$, and expresses $\epsilon_a$ as monomials in these variables.
A standard choice is
\begin{align}
	\epsilon_1&=z_1
	\,,&
	\epsilon_2&=z_1^{-1}z_2
	\,,&
	\epsilon_k&=z_{k-1}^{-1}z_k \qquad (3\le k\le N-1)
	\,,&
	\epsilon_N&=z_{N-1}^{-1}
	\,.
\end{align}
This automatically enforces $\prod_{a=1}^N\epsilon_a=1$ and $|\epsilon_a|=1$, and expresses
both $\mathcal{N}(\epsilon;r)$ and $\Delta(\epsilon)$ as Laurent-polynomial matrices in variables $z_i$.

The remaining input is the set $\{r_i\}$ characterizing the chosen irreducible representation.
For an $SU(N)$ representation with Dynkin labels $(a_1,a_2,\ldots,a_{N-1})$, one may write
\begin{align}
    r_i&=\lambda_i+\rho_i
	\,,&
	\text{with}&&
	\rho_i &= N-i
	\,,&
	\lambda_i &=
	a_i+a_{i+1}+\cdots+a_{N-1}
	\,.
\end{align}

As an explicit example,
consider the adjoint representation of $SU(3)$,
with $R = {\bf 8}$ and the Dynkin label $\{1,1\}$.
This gives rise to
\begin{align}
    \Bigl\{
	\lambda_1=2
	\,,\;
	\lambda_2=1
	\,,\;
    \rho_1=2
	\,,\;
	\rho_2=1
	\Bigr\}
	&&\implies&&
	\Bigl\{
    r_1=4
	\,,\;
	r_2=2
	\Bigr\}
	\,.
\end{align}
The resulting character is
\begin{align}
    \chi_{SU(3)_{\bf 8}}(z_2,z_3)
    &=
    z_2 z_3 + \frac{1}{z_2 z_3} + 2
    + \frac{z_2}{z_3^2}
    + \frac{z_2^2}{z_3}
    + \frac{z_3}{z_2^2}
    + \frac{z_3^2}{z_2}
	\,.
\end{align}
In the present finite-temperature construction, the $SU(3)_c$ adjoint representation
is carried by the gauge-field components
$G_\rmii{E}$ and $G_\rmii{M}$.

Similarly,
for $SU(2)_\rmii{$L$}$ the relevant characters are those of
the fundamental representation ($R={\bf 2}$), associated with $H$, and of
the adjoint representation ($R={\bf 3}$),
associated with $W_\rmii{E}$ and $W_\rmii{M}$
\begin{align}
    \chi_{SU(2)_{\bf 2}}(z_1)
    &= z_1+\frac{1}{z_1}
	\,,\\
    \chi_{SU(2)_{\bf 3}}(z_1)
    &= z_1^2+1+\frac{1}{z_1^2}
	\,.
\end{align}
Here, we consider only bosonic fields as input to the Hilbert series construction,
while the fermionic characters for the same representations
have the same form as the bosonic ones.

The Haar measure for $SU(N)$ may be written as \cite{Gray:2008yu,Hanany:2014dia,Henning:2017fpj,Banerjee:2020bym,Anisha:2019nzx}
\begin{align}
\label{eq:haar}
	\int_{SU(N)} {\rm d}\mu_{SU(N)}
	=
	\frac{1}{(2\pi i)^{N-1}N!}
	\oint_{|z_l|=1}
	\prod_{l=1}^{N-1}\frac{{\rm d}z_l}{z_l}\,
	\Delta(\epsilon)\Delta(\epsilon^{-1})
	\,.
\end{align}
The explicit measures needed here are those for $SU(2)$ and $SU(3)$:
\begin{align}
    {\rm d}\mu_{SU(2)}
    &=
	\frac{1}{2(2\pi i)}
	\frac{{\rm d}z_1}{z_1}
	\left(1-z_1^2\right)\left(1-\frac{1}{z_1^2}\right)
	\,,\\[2mm]
	{\rm d}\mu_{SU(3)}
	&=
	\frac{1}{6(2\pi i)^2}
	\frac{{\rm d}z_2}{z_2}
	\frac{{\rm d}z_3}{z_3}
	\left(1-z_2z_3\right)
	\nn &
	\hphantom{=\frac{1}{6(2\pi i)^2}}
	\times
	\left(1-\frac{1}{z_2z_3}\right)
	\left(1-\frac{z_2}{z_3^2}\right)
	\left(1-\frac{z_3^2}{z_2}\right)
	\left(1-\frac{z_3}{z_2^2}\right)
	\left(1-\frac{z_2^2}{z_3}\right)
	\,.
\end{align}
The full projection measure then factorizes into
a thermal space-time part and an internal-group part,
\begin{equation}
	\int {\rm d}\mu
	=
	\int_\T {\rm d}\mu_\T\;
	\prod_j \int_{G_j}{\rm d}\mu_j
	\,,
\end{equation}
so that the Hilbert series is obtained by integrating over all space-time and internal group variables.

%%%%%%%%%%%%%%%%%%%%%%%%%%%%%%%%%%%%%%%%%%%%%%%%%%%%%%%%%%%%%%%%%
\subsection{%
    Constraints due to integration by parts
    on $\mathbb{R}^3$
}
%%%%%%%%%%%%%%%%%%%%%%%%%%%%%%%%%%%%%%%%%%%%%%%%%%%%%%%%%%%%%%%%%
The spatial manifold $\mathbb{R}^3$ is non-compact,
so the fields do not satisfy any non-trivial boundary conditions.
As a result, IBP involving $D_i$ in finite-temperature QFT is treated
on the same footing as $D_\mu$ in zero-temperature QFT.
The Hilbert series encoding the constraint from spatial IBP is
\begin{align}
\label{eq:HSspatial}
    \mathcal{H}^\text{spatial} &= \mathcal{H}_0^\text{spatial} + \Delta\mathcal{H}^\text{spatial}
    \,,
	\\[2mm]
    \mathcal{H}_0^\text{spatial}
	&=
	\biggl(
		\int_\T {\rm d}\mu_\T
		\prod_j \int_{G_j}{\rm d}\mu_j
	\biggr)
	\;
	\frac{Z}{P^{(3)}(D_i,\alpha)}
	\,.
\end{align}
Here,
$\mathcal{H}_0^\text{spatial}$
offers the relevant contribution of the HS, and
$P^{(3)}(D_i,\alpha)$ accounts for the removal of
spatial IBP-redundant operators.
$\Delta\mathcal{H}^\text{spatial}$ is
an \textit{addendum} to the series that accurately accounts for
co-closed but not co-exact forms.
We collectively represent the spurions for respective fields satisfying
eqs.~\eqref{eq: scalar module},~\eqref{eq:GaussBianchi},
and~\eqref{eq: spin-1 full}, by the sets
${S_\text{scalar}^{ }}$,
$S_\text{vector}^{\rmii{\ref{sec:constraint-I:def}}}$, and
$S_\text{vector}^{\rmii{\ref{sec:constraint-II:def}}}$, respectively.%
\footnote{%
    Now we simplify the notation, omitting the spatial vector indices, since here they are only spurions. We still write the spatial derivative spurion as $D_i$ though, as we still distinguish it from the temporal derivative spurion $D_0$.
}
The $\Delta\mathcal{H}^\text{spatial}$ in three spatial dimensions
can be written as~\cite{Henning:2017fpj}:
\begin{align}
\label{eq:deltaH}
  \Delta\mathcal{H}&^\text{spatial} =
			D_i^3
    \nn[2mm] &
		+ \biggl(\prod_j \int_{G_j} {\rm d}\mu_j\biggr)
		\biggl[
				D_i^2 \sum_{\phi\in S_\text{scalar}} \phi (1-D_0)^{-1}\chi_{\rmii{$G$},\phi}
    	+ D_i^2 \sum_{\phi\in S_\text{scalar}} \phi^2 \wedge^2 \bigl((1-D_0)^{-1}\chi_{\rmii{$G$},\phi}\bigr)
    \nn[1mm] &
		\hphantom{{} \biggl(\prod_j \int_{G_j} {\rm d}\mu_j\biggr)\biggl[}
			+ D_i^2 \sum_{\{\phi,\Tilde{\phi}\} \subseteq S_\text{scalar}}\! \phi\Tilde{\phi} (1-D_0)^{-2}\chi_{\rmii{$G$},\phi}\chi_{\rmii{$G$},\Tilde{\phi}}
	    	+ D_i \sum_{V\in S_\text{vector}^{\rmii{\ref{sec:constraint-I:def}}}}\!\! V (1-D_0)^{-1}\chi_{\rmii{$G$},\rmii{$V$}}
    \nn[1mm] &
		\hphantom{{} \biggl(\prod_j \int_{G_j} {\rm d}\mu_j\biggr)\biggl[}
			+ D_i (1-D_i) \sum_{U \in S_\text{vector}^{\rmii{\ref{sec:constraint-II:def}}}} U (1-D_0)^{-1}\chi_{\rmii{$G$},\rmii{$U$}}
    \nn &
		\hphantom{{} \biggl(\prod_j \int_{G_j} {\rm d}\mu_j\biggr)\biggl[}
	    	+ D_i \sum_{U\in S_\text{vector}^{\rmii{\ref{sec:constraint-II:def}}}} U^2 \wedge^2 \bigl((1-D_0)^{-1}\chi_{\rmii{$G$},\rmii{$U$}}\bigr)
    \nn[1mm] &
		\hphantom{{} \biggl(\prod_j \int_{G_j} {\rm d}\mu_j\biggr)\biggl[}
			+ D_i \sum_{\{U,\Tilde{U}\} \subseteq S_\text{vector}^{\rmii{\ref{sec:constraint-II:def}}}}\! U\Tilde{U} (1-D_0)^{-2}\chi_{\rmii{$G$},\rmii{$U$}}\chi_{\rmii{$G$},\rmii{$\tilde{U}$}} 
		\biggr]
		\,,
\end{align}
where $\chi_{G,\eta}$, with
$\eta \in S_\text{scalar}
\cup S_\text{vector}^{\rmii{\ref{sec:constraint-I:def}}}
\cup S_\text{vector}^{\rmii{\ref{sec:constraint-II:def}}}$,
denotes the character of the representation of
the spurion $\eta$ under the internal symmetry group $G$.
Here, $\{\phi,\Tilde{\phi}\} \subseteq S_\text{scalar}$ ensures that
we select every possible distinct unordered pair in $S_\text{scalar}$.%
\footnote{%
	Naturally,
	$\sum_{\{\phi,\Tilde{\phi}\} \subseteq S_\text{scalar}}$ is only present when $|S_\text{scalar}| \geq 2$.
	The same applies to
	$S_\text{vector}^{\rmii{\ref{sec:constraint-II:def}}} $.
}
It is worth mentioning that,
as opposed to the \textit{addenda} $\Delta\mathcal{H}$ computed so far in
the literature~\cite{%
	Henning:2017fpj,Henning:2015alf,Ruhdorfer:2019qmk,Graf:2020yxt,
	Dujava:2022vqz,Graf:2022rco,Bijnens:2022zqo,Delgado:2022bho,Grojean:2023tsd,
	Alonso:2024usj},
which contribute only up to a certain mass dimension,
we see that when expanded in powers of $D_0$, it actually contributes
in each order of the series, and thus cannot be ignored.

%%%%%%%%%%%%%%%%%%%%%%%%% SECTION %%%%%%%%%%%%%%%%%%%%%%%%%%%%%%%%%%%%%%%%%
%
\section{Hilbert series for SMEFT at finite temperature}
\label{sec:HS-SMEFT}

\begin{table}[t]
\centering
\renewcommand{\arraystretch}{1.15}
\begin{tabular}{|c|c|c|c|c|}
\hline
Field & $SU(2)_\T$ rep & 
$SU(3)_c$ rep &
$SU(2)_\rmii{$L$}$ rep &
$U(1)_\rmii{$Y$}$ rep \\
\hline
\hline
$H$      & $1$          & $1$ & ${\bf 2}$ & $1/2$ \\
\hline
$B_\rmii{E}$    & ${\bf 3}$           & $1$ & $1$ & $0$ \\
$B_\rmii{M}$    & ${\bf 3}$           & $1$ & $1$ & $0$ \\
$W_\rmii{E}$    & ${\bf 3}$         & $1$ & ${\bf 3}$ & $0$ \\
$W_\rmii{M}$    & ${\bf 3}$          & $1$ & ${\bf 3}$ & $0$ \\
$G_\rmii{E}$    & ${\bf 3}$         & ${\bf 8}$ & $1$ & $0$ \\
$G_\rmii{M}$    & ${\bf 3}$         & ${\bf 8}$ & $1$ & $0$ \\
\hline
\end{tabular}
\caption{%
    Standard Model field content and representations used as input in the Hilbert series calculations.}
\label{tab:SMinput}
\end{table}

We construct the finite-temperature
dimension-five and dimension-six effective operators for a theory whose particle content and internal gauge symmetries are the same as in the Standard Model. The fields used in the
Hilbert series input are listed in tab.~\ref{tab:SMinput}.
Throughout this section,
``dimension-five'' and ``dimension-six'' refer to the canonical 4D
mass-dimension assignment.
After reduction to the static 3D EFT,
the corresponding operators inherit different 3D canonical dimensions.
We keep the 4D labels to facilitate comparison with zero-temperature SMEFT.

Accordingly, the space-time part of the integrand is built from two ingredients:
\begin{itemize}
	\item[(i)]
	the character $\widetilde{\chi}(D_0,D_i,\alpha)$ of the relevant
	$SU(2)_\T$ representation together with its $SO(2)$ charge, and for each field it can be one of the options displayed in the third column of
	tab.~\ref{tab:ch};
	\item[(ii)]
	the Haar measure on the maximal torus of $SU(2)_\T$,
	denoted by $\int_\T{\rm d}\mu_\T$,
	as summarized in fig.~\ref{fig:HSFT_flowchart}.
\end{itemize}

The internal symmetry group remains the same as in the Standard Model
\begin{equation}
\label{eq:SMgauge}
	G =
	\prod_j G_j \equiv
	\underbrace{
	SU(3)_c \times
	SU(2)_\rmii{$L$} \times
	U(1)_\rmii{$Y$}}_{\text{SM gauge}}
	\,.
\end{equation}
For each factor $G_j$, we introduce the corresponding character $\chi_j(z_i)$ and Haar measure
$\int_{G_j}{\rm d}\mu_j$.
Thus, the effect of finite temperature appears only in the space-time sector, while the internal-group
characters and projection measures are the same as in
the usual zero-temperature  Hilbert series
construction~\cite{Hanany:2014dia,Henning:2015daa,Banerjee:2020bym}.

The plethystic exponential for a bosonic spurion $\phi$,
is defined as%
\footnote{%
    Here, $\phi$ refers to any of the species presented in the second column of tab.~\ref{tab:ch}, not only the scalar field.
}
\cite{Henning:2015alf,Henning:2015daa,Henning:2017fpj,Anisha:2019nzx,Banerjee:2020bym}
\begin{align}
	\label{eq:PE:integerSpin}
	\mathrm{PE}[\phi]
	&=
	\exp\!\Biggl[
		\sum_{r=1}^{\infty}\frac{1}{r}
		\phi^r
		\Big(\prod_j \chi_j(z_i^{\,r})\Big)\,
		\widetilde{\chi}(D_0^{\,r},D_i^{\,r},\alpha^r)
	\Biggr]
	\,.
\end{align}
Here,
$\mathrm{PE}[\phi]$ collects all structures that can be formed from $\phi$
together with their derivative dressings.
The character generating function
\begin{equation}
	\label{eq:Z:PE}
    Z = \prod_{\{\phi\}} \mathrm{PE}[\phi]
		\,,
\end{equation}
collects all composite structures that can be formed from fundamental spurions $\{\phi\}$.
Only a subset of these combinations is invariant under the full symmetry group.
The subsequent integration in group space projects onto these invariant terms, and the
Hilbert series leads to the generation of independent EFT operators~\cite{Henning:2015alf, Anisha:2019nzx, Banerjee:2020bym, Banerjee:2020jun}.

For later use,
we list the ingredients of the plethystic exponential of each SMEFT field,
\begin{align}
	H&\rightarrow\frac{1}{r} H^r \, \chi^{\T}_{\rmi{scalar}}\left(D_0^r,D_i^r,\alpha^r\right) \, \chi_{({SU(2)})_{\bf 2}}(z_1^r) \, z^{r/2}
	\,,\nn
	H^{\dagger}&\rightarrow \frac{1}{r} (H^{\dagger})^r \, \chi^{\T}_{\rmi{scalar}}\left(D_0^r,D_i^r,\alpha^r\right) \, \chi_{({SU(2)})_{\bf 2}}(z_1^r) \, z^{-r/2}
	\,,\nn
	B_\rmii{E/M}&\rightarrow\frac{1}{r} B_\rmii{E/M}^r \, \chi^{\T}_{\rmi{vector}}\left(D_0^r,D_i^r,\alpha^r\right)
	\,,\nn
	W_\rmii{E/M}&\rightarrow\frac{1}{r} W_\rmii{E/M}^r \, \chi^{\T}_{\rmi{vector}}\left(D_0^r,D_i^r,\alpha^r\right) \, \chi_{({SU(2)})_{\bf 3}}(z_1^r)
	\,,\nn
	G_\rmii{E/M}&\rightarrow\frac{1}{r} G_\rmii{E/M}^r \, \chi^{\T}_{\rmi{vector}}\left(D_0^r,D_i^r,\alpha^r\right) \, \chi_{({SU(3)})_{\bf 8}}(z_2^r,z_3^r)
	\,,
\end{align}
where $\chi^{\T}_{\rmi{vector}}$ can be either
$\chi^{\T,\rmii{\ref{sec:constraint-I:def}}}_{\rmi{vector}}$ or
$\chi^{\T,\rmii{\ref{sec:constraint-II:def}}}_{\rmi{vector}}$ depending on the constraints being imposed.

At finite temperature, it is convenient to decompose the gauge-field strengths into
their electric and magnetic components for all gauge groups.
Throughout, we use the notation of eq.~\eqref{eq:EB},
where $X_{\rmii{E}\,i}$ and $X_{\rmii{M}\,i}$ denote
the electric and magnetic components of
the field-strength tensor for $X \in \{B, W, G\}$,
corresponding to
$U(1)_\rmii{$Y$}$,
$SU(2)_\rmii{$L$}$, and
$SU(3)_c$.

The final form of the
Hilbert series also encodes the IBP relations.
In the following sections,
we discuss how such IBP relations are suitably blended in the HS constructions.
In finite-temperature QFT,
only the $\mathbb{R}^3$ manifold is non-compact, where $S^1$ is compact.
Thus, the IBP relations are allowed for spatial derivatives $D_i$ only.%
\footnote{%
    This condition should be relaxed if, for example,
		one is interested in scenarios where the fields do not vanish at infinity,
		such as where the divergence of the Poynting vector does not vanish.
		In those cases,
		the Hilbert series would be simply $\mathcal{H} = \int{\rm d}\mu \;Z$,
		where $Z$ is defined in eq.~\eqref{eq:Z:PE}.
}
This allows for an arbitrary number of temporal derivatives $D_0$ in an operator.

%%%%%%%%%%%%%%%%%%%%%%%%% SECTION %%%%%%%%%%%%%%%%%%%%%%%%%%%%%%%%%%%%%%%%%
%
\subsection{%
    Role of derivatives: spatial vs.\ temporal}
\label{sec:dimensional:reduction}

In the context of 3D-SMEFT at finite temperature,
the constraints imposed on the fields of tab.~\ref{tab:SMinput} are
\begin{align}
    D_i D^i H &= 0\,,
		&
		D_i D^i H^\dagger &= 0
		\,,
\end{align}
for the scalar fields, with mass dimension $[H]=[H^\dagger]=1$,
and
\begin{align}
    D_i X_{\rmii{E}\,i} &= 0\,,&
		D_i X_{\rmii{M}\,i} &= 0\,,
\end{align}
for vector fields
$X\in \{B,W,G\}$, with mass dimension $[X_\rmii{E}] = [X_\rmii{M}] = 2$.
Since no additional constraints will be considered on the SPM of the scalar fields, their spurions are always dressed with the field character in eq.~\eqref{eq:chi_scalar_1}.
Once we employ the spatial IBP relation, we first note the polynomial of the operators up to mass dimension four as
\begin{align}
\label{eq:dim4}
  \mathcal{LP}_4 &=
		M^4 + M^2\cdot H H^\dagger + M^1\cdot 2 H H^\dagger D_0 
	\nn &
	+ M^0\cdot\Big[
		B_\rmii{E}^2
		+ B_\rmii{M}^2
		+ G_\rmii{E}^2
		+ G_\rmii{M}^2
		+ W_\rmii{E}^2
		+ W_\rmii{M}^2
	\nn  &\hphantom{{}M^0\cdot\Big[}
		+ B_\rmii{E} B_\rmii{M}
		+ G_\rmii{E} G_\rmii{M}
		+ W_\rmii{E} W_\rmii{M}
		+ 3 H H^\dagger D_0^2
        + H^2(H^\dagger)^2
	\Big]
	\,,
\end{align}
where
$X_\rmii{E}^2 \equiv X_{\rmii{E}\,i}X_{\rmii{E}\,i}$ and $M$ is some mass scale. 
However, an important feature
is that at finite temperature
the temporal derivative $D_0$
is treated on a different footing from
the spatial derivatives $D_i$.
This is already reflected in the scalar kinetic sector:
there are three independent operators
of the form $H H^\dagger D_0^2$,
while there is only one operator
of the form $H H^\dagger D_i^2$.%
\footnote{%
    Due to the EOM, this term does not appear in the HS output,
		but it is certainly present when we write the Lagrangian.
}
In zero-temperature QFT, the scalar kinetic term can be written either as
$(D_\mu H^\dagger)(D^\mu H)$ or as
$H^\dagger \Box H$, since the two differ only by
the total derivative $D_\mu (H^\dagger D^\mu H)$,
which vanishes at the asymptotic boundary of
a non-compact space-time manifold.
However, in finite-temperature QFT, temporal and spatial derivatives are no longer on the same footing.
Because the temporal direction $S^1$ is compact,
there are three independent operators,
$(D_0 H^\dagger)(D^0 H)$,
$(D_0^2 H^\dagger)H$, and
$H^\dagger (D_0^2H)$,
while along the non-compact spatial directions $\mathbb{R}^3$
we have only
$(D_i H^\dagger)(D^i H)\sim H^\dagger D_i^2 H$.
This special role of the temporal derivative
persists for higher-dimensional operators as well,
which is why operators involving $D_0$
are more numerous than those involving $D_i$. 

%%%%%%%%%%%%%%%%%%%%%%%%% SECTION %%%%%%%%%%%%%%%%%%%%%%%%%%%%%%%%%%%%%%%%%
%
\section{SMEFT operators at finite temperature}
\label{sec:SMEFT}

In this work,
we compute the non-redundant operators within two frameworks,
relying on the roles of different constraints:
\begin{itemize}
	\label{item:constraints}
    \item
		{\textbf{
			\hyperref[sec:constraint-I:def]{Constraint~I}$\;\oplus\;$Spatial IBP:}}
    
			No other constraints are imposed in addition to these fundamental constraints,
			which already encode the IBP relations and EOM on the non-compact $\mathbb{R}^3$ manifold.
			In this scenario, the vector spurions should be dressed with
			the field character in eq.~\eqref{eq:chi_spin1_div}.

    \item
		{\textbf{
			\hyperref[sec:constraint-II:def]{Constraint~II}$\;\oplus\;$Spatial IBP:}}

			In addition to the former constraints, we also note the role of other possible constraints, e.g.,
			the null condition of the curl of the electric and magnetic components of the field tensors,
			which reads 
\begin{align}
\label{eq:curlSM}
    \epsilon_{ijk}D_i X_{\rmii{E}\,j} &= 0
		\,,&
		\epsilon_{ijk}D_iX_{\rmii{M}\,j} &= 0
		\,,
\end{align}
for $X\in \{B,W,G\}$.
Thus, the vector spurions must be dressed with
the field character in eq.~\eqref{eq:chi_spin1_div+curl}.

\end{itemize}
The HS output relying on these constraints is displayed in
Appendices~\ref{app:HS:details} and~\ref{app:OperatorGrowth}.

%%%%%%%%%%%%%%%%%%%%%%%%% SECTION %%%%%%%%%%%%%%%%%%%%%%%%%%%%%%%%%%%%%%%%%
%
\subsection{%
    Interplay between
    temporal derivatives and gauge fields
}\label{sec:temporal-deriv-gf}

When constructing the Hilbert series,
temporal covariant derivatives, $D_0$,
appear explicitly inside of invariant operators.
But, while discussing their structures for the static 3D case,
we extract the temporal gauge fields masked inside $D_0$ as follows:
\begin{align}
	\label{eq:temporal-and-gauge}
    D_0 H &= (ig \tau^\smallIdx{$I$} W_0^\smallIdx{$I$} + ig^\prime Y_\rmii{$H$} B_0)\,H
		\,,\\
    D_0 B_{0i} &= g^\prime B_0 B_{0i}
		\,,&
		D_{0}B_{ij} &= g^\prime B_{0}B_{ij}
		\,,\\
    D_0 W_{0i} & = ig[W_0, W_{0i}]
		\,,&
		D_0 W_{ij} &= ig[W_0, W_{ij}]
		\,,\\
    D_0 G_{0i} & = i\gs [G_0, G_{0i}]
		\,,&
		D_0 G_{ij} &= i\gs [G_0, G_{ij}]
		\,,
\end{align}
where
$\tau^\rmii{$I$}$ are the Pauli matrices, and
$Y_\rmii{$H$}$ is the hypercharge of the Higgs field.
Here, the couplings
$g$,
$g^\prime$, and
$\gs$, as well as
parameters of the fundamental $4D$ theory
are included within the Wilson coefficients.

%%%%%%%%%%%%%%%%%%%%%%%%% tables %%%%%%%%%%%%%%%%%%%%%%%%%%%%%%%%%%%%%%%%%
%
\newcommand{\classtablescale}{0.77}

\begin{table}[t]
\centering
\scalebox{\classtablescale}{%
\begin{tabular}{|c| p{0.35\textwidth} | p{0.35\textwidth}|}
	\hline
	\multicolumn{3}{|c|}{\textbf{Class: $H^4D$}} \\
	\hline
	\multicolumn{1}{|c|}{HS ($T\neq 0$)} &
	\multicolumn{1}{c|}{Inv.\ operators ($T\neq 0$)} &
	\multicolumn{1}{c|}{Static 3D form} \\
	\hline
	\hline
	\multirow{2}{*}{$2\,H^2(H^\dagger)^2D_0$}
	& $(H^\dagger H)\,D_0(H^\dagger H)$
	& $B_0(H^\dagger H)\,(H^\dagger H)$, \\ 
    & $(H^\dagger H)\,(H^\dagger D_0 H)$
	& \makebox[0pt][l]{\hypertarget{op:dim5-h4d-w0-h4}{}}$W_0^\smallIdx{$I$}(H^\dagger \tau^\smallIdx{$I$} H)\,(H^\dagger H)$\\
	\hline
	\hline
	\multicolumn{1}{|l|}{
		Operators:
	} &
	\multicolumn{1}{l|}{
		2
	} & 2
	\\
	\hline
\end{tabular}
}
	\caption{%
		Dimension-five: class $H^4D$.
	}
	\label{tab:dim5H4D}
\end{table}

\begin{table}[t]
\centering
\scalebox{\classtablescale}{%
\begin{tabular}{|c| p{0.35\textwidth} | p{0.35\textwidth}|}
	\hline
	\multicolumn{3}{|c|}{\textbf{Class: $H^2D^3$}} \\
	\hline
	\multicolumn{1}{|c|}{HS ($T\neq 0$)} &
	\multicolumn{1}{c|}{Inv.\ operators ($T\neq 0$)} &
	\multicolumn{1}{c|}{Static 3D form} \\
	\hline
	\hline
	\makebox[0pt][l]{\hypertarget{op:dim5-h2d3}{}}
	\multirow{4}{*}{$4 HH^\dagger D_0^3$}
	& $D_0^3(H^\dagger \,H)$
	& $B_0^3 H^\dagger \,H$, 
	\\
	& $D_0^2(H^\dagger D_0 \,H)$ &$B_0W_0^\smallIdx{$I$}W_0^\smallIdx{$I$}\,H^\dagger \,H$\\
	& $D_0(H^\dagger D_0^2\,H)$& $B_0^2 W_0^\smallIdx{$I$}\,H^\dagger \tau^\smallIdx{$I$}\,H$ \\
    & $(H^\dagger D_0^3 \,H)$&\\%$\epsilon^\smallIdx{$IJK$}W_0^\smallIdx{$I$}W_0^\smallIdx{$J$}W_0^\smallIdx{$K$}\,H^\dagger \,H$\\

	\hline
	\hline
	\multicolumn{1}{|l|}{
		Operators:
	} &
	\multicolumn{1}{l|}{
		4
	} & 3
	\\
	\hline
\end{tabular}
}
	\caption{%
		Dimension-five: class $H^2D^3$.
	}
	\label{tab:dim5H2D3}
\end{table}

\begin{table}[t]
\centering
\scalebox{\classtablescale}{%
\begin{tabular}{|c| p{0.35\textwidth} | p{0.35\textwidth}|}
	\hline
	\multicolumn{3}{|c|}{\textbf{Class: $H^2XD$}} \\
	\hline
	\multicolumn{1}{|c|}{HS ($T\neq 0$)} &
	\multicolumn{1}{c|}{Inv.\ operators ($T\neq 0$)} &
	\multicolumn{1}{c|}{Static 3D form} \\
	\hline
	\hline
	$HH^\dagger B_\rmii{E} D_i$ &
	$(H^\dagger \lrD_{\!\!i} H)\,B_{0i}$&
	\cellmag{$(H^\dagger \lrD_{\!\!i} H)\,[D_i, B_{0}]$}
	\\

	\hdashline[1pt/2pt]
	\makebox[0pt][l]{\hypertarget{op:dim5-h2xd-bm-di}{}}$HH^\dagger B_\rmii{M} D_i$
	& $(H^\dagger \lrD_{\!\!i} H)\,\epsilon_{ijk}B_{jk}$
	& $(H^\dagger \lrD_{\!\!i} H)\,\epsilon_{ijk}B_{jk}$
	\\

	\hline
	\makebox[0pt][l]{\hypertarget{op:dim5-h2xd-we-di}{}}$HH^\dagger W_\rmii{E} D_i$
	& $\big(H^\dagger \tau^\smallIdx{$I$}  i \lrD_{\!\!i} H\big)\,W^\smallIdx{$I$}_{0i}$
	& \cellmag{$\big(H^\dagger \tau^\smallIdx{$I$}  i \lrD_{\!\!i} H\big)\,[D_i, W^\smallIdx{$I$}_{0}]$}
	\\

	\hdashline[1pt/2pt]
	\makebox[0pt][l]{\hypertarget{op:dim5-h2xd-wm-di}{}}$HH^\dagger W_\rmii{M} D_i$ 
	& $(H^\dagger \tau^\smallIdx{$I$} \lrD_{\!\!i} H)\,\epsilon_{ijk}W^\smallIdx{$I$}_{jk}$
	& $(H^\dagger \tau^\smallIdx{$I$} \lrD_{\!\!i} H)\,\epsilon_{ijk}W^\smallIdx{$I$}_{jk}$
	\\

	\hline
	\hline
	\multicolumn{1}{|l|}{
		Operators:
	}
		&
	\multicolumn{1}{l|}{
		4
	} & 2 + {\magenta 2}
	\\
	\hline
\end{tabular}
}
\caption{%
		Dimension-five: class $H^2XD$.
		Magenta-shaded entries are present when temporal gauge fields are
		spatially varying,
		and vanish in the strict static limit with constant temporal gauge fields
		($X_{0i}\equiv -X_{\rmii{E}i}=0$).
	}
	\label{tab:dim5H2XD}
\end{table}

\begin{table}[t]
\centering
\scalebox{\classtablescale}{%
\begin{tabular}{|c| p{0.35\linewidth} | p{0.35\linewidth}|}
	\hline
	\multicolumn{3}{|c|}{\textbf{Class: $X^2D$}} \\
	\hline
	\multicolumn{1}{|c|}{HS ($T\neq 0$)} &
	\multicolumn{1}{c|}{Inv.\ operators ($T\neq 0$)} &
	\multicolumn{1}{c|}{Static 3D form} \\
	\hline
	\hline
	\multirow{2}{*}{$2\,B_\rmii{E}B_\rmii{M}D_0$}
	& $B_{0i}\,\epsilon_{ijk} \big(D_0B_{jk}\big)$ \newline $\big( D_0 B_{0i}\big)\,\epsilon_{ijk} B_{jk}$
	& \multirow{2}{*}{\cellmag{$B_0\big([D_i, B_{0}]\,\epsilon_{ijk}B_{jk}\big)$}}
	\\
    
	\hdashline[1pt/2pt]
	$B_\rmii{E}B_\rmii{M}D_i$&
	\cellgreen{$B_{0m} \big( \epsilon_{ijm} D_i\,\epsilon_{jkl}B_{kl}\big)$}
	&\cellmag{$[D_m,B_{0}] \big( \epsilon_{ijm} D_i\,\epsilon_{jkl}B_{kl}\big)$}
	\\
	
	\hline
	\multirow{2}{*}{$2\,G_\rmii{E}G_\rmii{M}D_0$}
	& $G^\smallIdx{$A$}_{0i}\,\epsilon_{ijk} \big( D_0 G^\smallIdx{$A$}_{jk}\big)$ \newline $\big( D_0 G^\smallIdx{$A$}_{0i}\big)\,\epsilon_{ijk}G^\smallIdx{$A$}_{jk}$& \multirow{2}{*}{\cellmag{$G_0 \big([D_i, G^\smallIdx{$A$}_{0}]\,\epsilon_{ijk}G^\smallIdx{$A$}_{jk}\big)$}}
	\\

    \hdashline[1pt/2pt]
	$G_\rmii{E}G_\rmii{M}D_i$ &
	\cellgreen{$ G^\smallIdx{$A$}_{0m}\big( \epsilon_{ijm}\,D_i\,\epsilon_{jkl}G^\smallIdx{$A$}_{kl}\big)$}
	& \cellmag{$[D_m, G^\smallIdx{$A$}_{0}]\big( \epsilon_{ijm}D_i\,\epsilon_{jkl}G^\smallIdx{$A$}_{kl}\big)$}
	\\

	\hline
	\makebox[0pt][l]{\hypertarget{op:dim5-x2d-wewm-d0}{}}
	\multirow{2}{*}{$2\,W_\rmii{E}W_\rmii{M}D_0$}
	& $W^\smallIdx{$I$}_{0i}\,\epsilon_{ijk}\big(D_0 W^\smallIdx{$I$}_{jk}\big)$ \newline $\big(D_0 W^\smallIdx{$I$}_{0i}\big)\,\epsilon_{ijk}W^\smallIdx{$I$}_{jk}$
	& \multirow{2}{*}{\cellmag{
		$\epsilon^\smallIdx{$IJK$}W_0^\smallIdx{$I$} \big([D_i, W^\smallIdx{$J$}_{0}]\,\epsilon_{ijk}W^\smallIdx{$K$}_{jk}\big)$}}
	\\
    \hdashline[1pt/2pt]
	$W_\rmii{E}W_\rmii{M}D_i$ &
	\cellgreen{$W^\smallIdx{$I$}_{0m} \big( \epsilon_{ijm} \,D_i\epsilon_{jkl}W^\smallIdx{$I$}_{kl}\big)$}
	&\cellmag{$[D_m, W^\smallIdx{$I$}_{0}] \big( \epsilon_{ijm} D_i\,\epsilon_{jkl}W^\smallIdx{$I$}_{kl}\big)$}
	\\
	\hline
	$B_\rmii{E}^2 D_0$ &  $D_0 (B_{0i} B_{0i})$
	& \cellmag{$B_0 ([D_i, B_{0}] [D_i,B_{0}])$}
	\\
	$B_\rmii{M}^2 D_0$ & $D_0 (\epsilon_{ijk}\epsilon_{ilm}B_{jk} B_{lm})$ & $B_0 (\epsilon_{ijk}\epsilon_{ilm}B_{jk} B_{lm})$
	\\
    \hdashline[1pt/2pt]
    $B_\rmii{E}^2 D_i$ & \cellgreen $B_{0i} (\epsilon_{ijk}D_jB_{0k})$ & \cellmag $[D_i,B_0](\epsilon_{ijk}D_j[D_k,B_0])$
	\\    
	\makebox[0pt][l]{\hypertarget{op:dim5-x2d-bm2-di}{}}$B_\rmii{M}^2 D_i$ & \cellgreen $ \epsilon_{ijk}B_{jk}(\epsilon_{ipq}D_{p}\epsilon_{qlm} B_{lm})$ & $ \epsilon_{ijk}B_{jk}(\epsilon_{ipq}D_{p}\epsilon_{qlm} B_{lm})$
	\\
    \hline
	$W_\rmii{E}^2 D_0$ & $D_0 (W^\smallIdx{$I$}_{0i} W^\smallIdx{$I$}_{0i})$ &  
	\\
	$W_\rmii{M}^2 D_0$ & $D_0 (\epsilon_{ijk}\epsilon_{ilm}W^\smallIdx{$I$}_{jk} W^\smallIdx{$I$}_{lm})$
	& $\epsilon^\smallIdx{$IJK$}\,W_0^\smallIdx{$I$} (\epsilon_{ijk}\epsilon_{ilm}W^\smallIdx{$J$}_{jk} W^\smallIdx{$K$}_{lm})$
	\\
    \hdashline[1pt/2pt]
    $W_\rmii{E}^2 D_i$ & \cellgreen $ W^\smallIdx{$I$}_{0i}(\epsilon_{ijk} D_j W^\smallIdx{$I$}_{0k})$ & \cellmag $ [D_i, W^\smallIdx{$I$}_{0}](\epsilon_{ijk} D_j [D_k, W^\smallIdx{$I$}_{0}])$  
	\\
	\makebox[0pt][l]{\hypertarget{op:dim5-x2d-wm2-di}{}}$W_\rmii{M}^2 D_i$ & \cellgreen $\epsilon_{ijk}W^\smallIdx{$I$}_{jk}(\epsilon_{ipq} D_p \epsilon_{qlm}   W^\smallIdx{$I$}_{lm})$ & $ \epsilon_{ijk}W^\smallIdx{$I$}_{jk}(\epsilon_{ipq}D_{p}\epsilon_{qlm} W^\smallIdx{$I$}_{lm})$
	\\
  \hline
	$G_\rmii{E}^2 D_0$ & $D_0 (G_{0i} G_{0i})$
	& \cellmag{$d^\smallIdx{$ABC$}G_0^\smallIdx{$A$}[D_i,G^\smallIdx{$B$}_0][D_i, G^\smallIdx{$C$}_0]$}
	\\
	$G_\rmii{M}^2 D_0$
	& $D_0 (\epsilon_{ijk}\epsilon_{ilm}G_{jk} G_{lm})$
	& $f^\smallIdx{$ABC$}\,G^\smallIdx{$A$}_0 (\epsilon_{ijk}\epsilon_{ilm}G^\smallIdx{$B$}_{jk} G^\smallIdx{$C$}_{lm})$
	\\
  \hdashline[1pt/2pt]
  	$G_\rmii{E}^2 D_i$ & \cellgreen $ G^\smallIdx{$A$}_{0i}(\epsilon_{ijk} D_j G^\smallIdx{$A$}_{0k})$
	& \cellmag{$ [D_i, G^\smallIdx{$A$}_{0}](\epsilon_{ijk} D_j [D_k, G^\smallIdx{$A$}_{0}])$} 
	\\
  $G_\rmii{M}^2 D_i$
	& \cellgreen{$\epsilon_{ijk}G^\smallIdx{$A$}_{jk} (\epsilon_{ipq} D_p \epsilon_{qlm} G^\smallIdx{$A$}_{lm})$}
	&
    $\epsilon_{ijk}G^\smallIdx{$A$}_{jk} (\epsilon_{ipq} D_p \epsilon_{qlm} G^\smallIdx{$A$}_{lm})$
	\\
	\hline
	\hline
	\multicolumn{1}{|l|}{
		Operators:
	} &
	\multicolumn{1}{l|}{
		12 + {\Green 9} 
         }
         &
         6 + {\magenta 11}
	\\
	\hline
\end{tabular}
	}
	\caption{%
		Dimension-five: class $X^2D$.
		Green-shaded entries vanish when
		\hyperref[sec:constraint-II:def]{Constraint~II} is imposed.
		Magenta-shaded entries are defined in tab.~\ref{tab:dim5H2XD}.
	}
	\label{tab:dim5X2D}
\end{table} 

\begin{table}[t]
	\centering
	\scalebox{\classtablescale}{%
	\begin{tabular}{| c | p{0.35\textwidth}|p{0.40\textwidth}|p{0.25\textwidth}|}
		\hline
		\multicolumn{4}{|c|}{\textbf{Class} $X^3$}\\
		\hline
		HS ($T\neq 0$) &
		\multicolumn{1}{c|}{Inv.\ operators ($T\neq 0$)} &
		\multicolumn{1}{c|}{Static 3D form} &
		\multicolumn{1}{c|}{Inv.\ operators ($T=0$)} \\
		\hline
		\hline
		
		$G_\rmii{E}^{3}$ &
		$f^\smallIdx{$ABC$}\epsilon_{ijk}G^\smallIdx{$A$}_{0i}G^\smallIdx{$B$}_{0j}G^\smallIdx{$C$}_{0k}$
		& \cellmag$f^\smallIdx{$ABC$}\epsilon_{ijk}
			\bigl[D_i, G^\smallIdx{$A$}_{0}\bigr]
			\bigl[D_j, G^\smallIdx{$B$}_{0}\bigr]
			\bigl[D_k, G^\smallIdx{$C$}_{0}\bigr]$ &
		\multirow{2}{*}{$f^\smallIdx{$ABC$}G_{\mu}^{\smallIdx{$A$}\nu}G_{\nu}^{\smallIdx{$B$}\rho}G_{\rho}^{\smallIdx{$C$}\mu}$}
		\\
		$G_\rmii{M}^{3}$
		& $f^\smallIdx{$ABC$}\epsilon_{ijk}\epsilon_{ilm}\epsilon_{jpq}\epsilon_{krs}G^\smallIdx{$A$}_{lm}G^\smallIdx{$B$}_{pq}G^\smallIdx{$C$}_{rs}$
		& $f^\smallIdx{$ABC$}\epsilon_{ijk}\epsilon_{ilm}\epsilon_{jpq}\epsilon_{krs}G^\smallIdx{$A$}_{lm}G^\smallIdx{$B$}_{pq}G^\smallIdx{$C$}_{rs}$ &
		\\
		\hdashline[1pt/2pt]
	
		\makebox[0pt][l]{\hypertarget{op:dim6-x3-we3}{}}$W_\rmii{E}^{3}$
		& $\epsilon^\smallIdx{$IJK$}\epsilon_{ijk}\,W^\smallIdx{$I$}_{0i}W^\smallIdx{$J$}_{0j}W^\smallIdx{$K$}_{0k}$
		& \cellmag $\epsilon^\smallIdx{$IJK$}\epsilon_{ijk}\,
			\bigl[D_i, W^\smallIdx{$I$}_{0}\bigr]
			\bigl[D_j, W^\smallIdx{$J$}_{0}\bigr]
			\bigl[D_k, W^\smallIdx{$K$}_{0}\bigr]$ &
		\multirow{2}{*}{$\epsilon^\smallIdx{$IJK$}W_{\mu}^{\smallIdx{$I$}\nu}W_{\nu}^{\smallIdx{$J$}\rho}W_{\rho}^{\smallIdx{$K$}\mu}$} \\
		
		$W_\rmii{M}^{3}$
		& $\epsilon^\smallIdx{$IJK$}\epsilon_{ijk}\epsilon_{ilm}\epsilon_{jpq}\epsilon_{krs}\,W^\smallIdx{$I$}_{lm}W^\smallIdx{$J$}_{pq}W^\smallIdx{$K$}_{rs}$
		& $\epsilon^\smallIdx{$IJK$}\epsilon_{ijk}\epsilon_{ilm}\epsilon_{jpq}\epsilon_{krs}\,W^\smallIdx{$I$}_{lm}W^\smallIdx{$J$}_{pq}W^\smallIdx{$K$}_{rs}$ &
		\\
		\hdashline[1pt/2pt]
		$G_\rmii{E}\,G_\rmii{M}^{2}$
		& $f^\smallIdx{$ABC$}\epsilon_{ijk}\epsilon_{jpq}\epsilon_{krs}G^\smallIdx{$A$}_{0i}G^\smallIdx{$B$}_{pq}G^\smallIdx{$C$}_{rs}$
		& \cellmag $
			f^\smallIdx{$ABC$}\epsilon_{ijk}\epsilon_{jpq}\epsilon_{krs}
			\bigl[D_i, G^\smallIdx{$A$}_{0}\bigr]
			G^\smallIdx{$B$}_{pq}G^\smallIdx{$C$}_{rs}$ &
		\multirow{2}{*}{$f^\smallIdx{$ABC$}\widetilde{G}_{\mu}^{\smallIdx{$A$}\nu}G_{\nu}^{\smallIdx{$B$}\rho}G_{\rho}^{\smallIdx{$C$}\mu}$} \\
		% --- corrected cubic gauge-mix terms ---
		
		$G_\rmii{E}^{2}G_\rmii{M}$
		& $f^\smallIdx{$ABC$}\,\epsilon_{ijk}\,G^\smallIdx{$A$}_{0i}G^\smallIdx{$B$}_{0j}\big(\epsilon_{k l m}G^\smallIdx{$C$}_{l m}\big)$
		& \cellmag $
			f^\smallIdx{$ABC$}\,\epsilon_{ijk}\,
			\bigl[D_i, G^\smallIdx{$A$}_{0}\bigr]
			\bigl[D_j, G^\smallIdx{$B$}_{0}\bigr]
			\big(\epsilon_{k l m}G^\smallIdx{$C$}_{l m}\big)$ &
		\\
		\hdashline[1pt/2pt]
		
		\makebox[0pt][l]{\hypertarget{op:dim6-x3-wewm2}{}}$W_\rmii{E}\,W_\rmii{M}^{2}$
		& $\epsilon^\smallIdx{$IJK$}\epsilon_{ijk}\epsilon_{jpq}\epsilon_{krs}\,W^\smallIdx{$I$}_{0i}W^\smallIdx{$J$}_{pq}W^\smallIdx{$K$}_{rs}$
		& \cellmag $\epsilon^\smallIdx{$IJK$}\epsilon_{ijk}\epsilon_{jpq}\epsilon_{krs}\,
		\bigl[D_i, W^\smallIdx{$I$}_{0}\bigr]
		W^\smallIdx{$J$}_{pq}W^\smallIdx{$K$}_{rs}$
		&
		\multirow{2}{*}{%
			$\epsilon^\smallIdx{$IJK$}\widetilde{W}_{\mu}^{\smallIdx{$I$}\nu}W_{\nu}^{\smallIdx{$J$}\rho}W_{\rho}^{\smallIdx{$K$}\mu}$} \\
		
		\makebox[0pt][l]{\hypertarget{op:dim6-x3-we2wm}{}}$W_\rmii{E}^{2}W_\rmii{M}$
		& $\epsilon^\smallIdx{$IJK$}\,\epsilon_{ijk}\,W^\smallIdx{$I$}_{0i}W^\smallIdx{$J$}_{0j}\big(\epsilon_{k l m}W^\smallIdx{$K$}_{l m}\big)$
		& %$\epsilon^\smallIdx{$IJK$}\,\epsilon_{ijk}\,W^\smallIdx{$I$}_{0i}W^\smallIdx{$J$}_{0j}\big(\epsilon_{k l m}W^\smallIdx{$K$}_{l m}\big)$
        \cellmag $\epsilon^\smallIdx{$IJK$}\,\epsilon_{ijk}\,[D_i,W^\smallIdx{$I$}_{0}][D_j,W^\smallIdx{$J$}_{0}]\big(\epsilon_{k l m}W^\smallIdx{$K$}_{l m}\big)$&
		\\
		\hline
		\hline
		\multicolumn{1}{|l|}{
				Operators:
		} &
		8 & 2 + {\magenta 6} &
		4 \\
		\hline
	\end{tabular}%
	}
	\caption{%
		Dimension-six: class $X^3$.
		See tabs.~\ref{tab:dim5H2XD} and~\ref{tab:dim5X2D}
		for the color coding.
		Among the operators of this class,
		those containing $\widetilde{X}$-type ($X_\rmii{E}\times X_\rmii{M}$) structures
		are CP-odd.
		The remaining ones with zero-temperature counterparts are CP-even.
		The purely finite-temperature operators of this class
		are listed separately in tab.~\ref{tab:dim6X3:finite_only_2}.
	}
    \label{tab:dim6X3}
\end{table}

\begin{table}[t]
	\centering
	\scalebox{\classtablescale}{%
	\begin{tabular}{| c | p{0.35\textwidth}|p{0.35\textwidth}|p{0.25\textwidth}|}
		\hline
		\multicolumn{4}{|c|}{\textbf{Operator class $H^6$ }}\\
		\hline
		HS ($T\neq 0$) &
		\multicolumn{1}{c|}{Inv.\ operators ($T\neq 0$)} &
		\multicolumn{1}{c|}{Static 3D form} &
		\multicolumn{1}{c|}{Inv.\ operators ($T=0$)} \\
		\hline
		\hline
		
		$H^{3}(H^{\dagger})^{3}$ &
		$(H^{\dagger}H)^{3}$ & $(H^{\dagger}H)^{3}$ &
		$(H^{\dagger}H)^{3}$ \\
		\hline
		\hline
		\multicolumn{1}{|c|}{
			Operators:
		} & 
		\multicolumn{1}{l|}{1} &
		\multicolumn{1}{l|}{1} &
		1\\
		\hline
	\end{tabular}%
	}\caption{Dimension-six: class $H^6$.}
    \label{tab:dim6H6}
\end{table}

\begin{table}[t]
	\centering
	\scalebox{\classtablescale}{%
	\begin{tabular}{| c | p{0.35\textwidth}|p{0.35\textwidth}|p{0.25\textwidth}|}
		\hline
		\multicolumn{4}{|c|}{\textbf{Operator class $H^2 D^4$ }}\\
		\hline
		HS ($T\neq 0$) &
		\multicolumn{1}{c|}{Inv.\ operators ($T\neq 0$)} &
		\multicolumn{1}{c|}{Static 3D form} &
		\multicolumn{1}{c|}{Inv.\ operators ($T=0$)} \\
		\hline
		\hline

		\multirow{5}{*}{$5H\,D_0^{4}H^{\dagger}$}
		& $D_0^4(H^\dagger H)$
		& $B_0^4 H^\dagger H$
		& \multirow{5}{*}{$(D_\mu D^\mu H)^\dagger (D_\nu D^\nu H) $}
		\\
		&$D_0^3(H^\dagger D_0 H)$
		& \((W_0^\smallIdx{$I$} W_0^\smallIdx{$I$} H^\dagger)(W_0^\smallIdx{$J$} W_0^\smallIdx{$J$} H)\)
		& \\
		& $D_0^2(H^\dagger D_0^2 H)$
		& \( (W_0^\smallIdx{$I$} W_0^\smallIdx{$I$} H^\dagger)(B_0 B_0 H)\)
		&\\
		&$D_0(H^\dagger D_0^3 H)$
		& $B_0^3 W_0^\smallIdx{$I$} H^\dagger \tau^\smallIdx{$I$} H$
		&
		\\
		&$(H^\dagger D_0^4 H)$
		& $ B_0 W_0^\smallIdx{$I$} W_0^\smallIdx{$I$} (W_0^\smallIdx{$J$} H^\dagger \tau^\smallIdx{$J$} H)\)&
		\\
		\hline
		\hline
		\multicolumn{1}{|c|}{
        Operators:
		} &
		\multicolumn{1}{l|}{
			5
		} & 5 &
		1\\
		\hline
	\end{tabular}%
	}
	\caption{%
		Dimension-six: class $H^2 D^4$.
	}
	\label{tab:dim6H2D4}
\end{table}

\begin{table}[t]
	\centering
	\scalebox{\classtablescale}{%
	\begin{tabular}{| c | p{0.35\textwidth}|p{0.35\textwidth}|p{0.25\textwidth}|}
		\hline
		\multicolumn{4}{|c|}{\textbf{Operator class $H^4 D^2$ }}\\
		\hline
		HS ($T\neq 0$) &
		\multicolumn{1}{c|}{Inv.\ operators ($T\neq 0$)} &
		\multicolumn{1}{c|}{Static 3D form} &
		\multicolumn{1}{c|}{Inv.\ operators ($T=0$)} \\
		\hline
		\hline
		
		\multirow{6}{*}{$6\,H^{2}D_0^{2}(H^{\dagger})^{2}$}
		& $( H^\dagger H)D_0^2(H^\dagger H)$
		& \( B_0^2 (H^\dagger H)^2\)
		& \multirow{8}{*}{\makecell[c]{
			$(H^{\dagger}H)\square(H^{\dagger}H)$ \\
			$(H^{\dagger}D^{\mu}H)^{\dagger}(H^{\dagger}D_{\mu}H)$ \\
			$(H^\dagger H)( D^\mu H)^\dagger (D_\mu H)$}}
		\\
		
		& $(H^\dagger H) D_0 (H^\dagger D_0 H)$
		& \((W^\smallIdx{$I$}_0 H^\dagger \tau^\smallIdx{$I$} H)(W_0^\smallIdx{$J$} H^\dagger \tau^\smallIdx{$J$} H)\)
		&
		\\
		
		& $( H^\dagger H) (D_0 H^\dagger) (D_0 H)$
		&\((H^\dagger W^\smallIdx{$I$}_0 H)(H W^\smallIdx{$I$}_0 H^\dagger)\)
		&
		\\
        
		& $[H^\dagger (D_0 H)]^\dagger [H^\dagger (D_0 H)]$
		& $B_0 W_0^\smallIdx{$I$}(H^\dagger \tau^\smallIdx{$I$} H)(H^\dagger H)$
		& 
		\\
        &%$[(D_0 H^\dagger) H]^\dagger [(D_0 H^\dagger) H]$
        {$[D_0(H^\dagger \, H) ] [D_0(H^\dagger\, H)]$}
		& &\\
        
		& $[H^\dagger (D_0 H)]^\dagger [(D_0\,H^\dagger) H]$
		& &\\
        
		\cdashline{1-3}[1pt/2pt]
		
		\multirow{2}{*}{$2\,H^{2}D_i^{2}(H^{\dagger})^{2}$}
		& $[(D_iH^\dagger) H]^\dagger[(D_i H^\dagger) H]$
		%$[H^\dagger (D_i H)]^\dagger[H^\dagger (D_i H)]$
        {$[H^\dagger H][(D_i H)^\dagger (D_i H)]$}
		& $[(D_iH^\dagger) H]^\dagger[(D_i H^\dagger) H]$
		%$[H^\dagger (D_i H)]^\dagger[H^\dagger (D_i H)]$
        {$[H^\dagger H][(D_i H)^\dagger (D_i H)]$}
		& 
		\\
		\hline
		\hline
		\multicolumn{1}{|c|}{
            Operators:
		} &
		\multicolumn{1}{l|}{
        8
		} & 6 &
		3\\
		\hline
	\end{tabular}%
	}
	\caption{%
		Dimension-six: class $H^4 D^2$.
	}
  \label{tab:dim6H4D2}
\end{table}
%%%%%%%%%%%%%%%%%%%%%%%%%%%%%%%%%%%%%%%%%%%%%%%%%%%%

\begin{table}[t]
	\centering
	\scalebox{\classtablescale}{%
	\begin{tabular}{| c | p{0.35\textwidth}|p{0.35\textwidth}|p{0.25\textwidth}|}
		\hline
		\multicolumn{4}{|c|}{\textbf{Operator class $H^2 X^2$ }}\\
		\hline
		HS ($T\neq 0$) &
		\multicolumn{1}{c|}{Inv.\ operators ($T\neq 0$)} &
		\multicolumn{1}{c|}{Static 3D form} &
		\multicolumn{1}{c|}{Inv.\ operators ($T=0$)} \\
		\hline
		\hline
	
		$H\,W_\rmii{E}^{2}H^{\dagger}$
		& $H^\dagger H\, W^\smallIdx{$I$}_{0i} W^\smallIdx{$I$}_{0i}$
		& \cellmag $H^\dagger H\, \bigl[D_i, W^\smallIdx{$I$}_{0}\bigr] \bigl[D_i, W^\smallIdx{$I$}_{0}\bigr]$ 
		& \multirow{2}{*}{$ H^\dag H\, W^\smallIdx{$I$}_{\mu\nu} W^{\smallIdx{$I$}\mu\nu}$}
		\\
		$H\,W_\rmii{M}^{2}H^{\dagger}$
		& $H^\dagger H\, \epsilon_{ijk}\epsilon_{ilm}\,W^\smallIdx{$I$}_{jk} W^\smallIdx{$I$}_{lm}$
		& $H^\dagger H\, \epsilon_{ijk}\epsilon_{ilm}\,W^\smallIdx{$I$}_{jk} W^\smallIdx{$I$}_{lm}$
		&
		\\
      \noalign{\vskip -.3pt}
				\cdashline{1-4}[1pt/2pt]
      \noalign{\vskip .3pt}
		\multirow{2}{*}{$2\,H\,W_\rmii{E}W_\rmii{M}\,H^{\dagger}$}
		& $H^\dagger  H\, \epsilon_{ijk}\,W^\smallIdx{$I$}_{0i}W^\smallIdx{$I$}_{jk}$
		& \cellmag $H^\dagger  H\, \epsilon_{ijk}\,\bigl[D_i, W^\smallIdx{$I$}_{0}\bigr]W^\smallIdx{$I$}_{jk}$
		& \multirow{2}{*}{$ H^\dag H\, \widetilde W^\smallIdx{$I$}_{\mu\nu} W^{\smallIdx{$I$}\mu\nu}$}
		\\
		& \makebox[0pt][l]{\hypertarget{op:dim6-h2x2-hwewm-tau}{}}
			$H^\dagger \tau^\smallIdx{$I$} H\, \epsilon^\smallIdx{$IJK$}\epsilon_{ijk}\,W^\smallIdx{$J$}_{0i}W^\smallIdx{$K$}_{jk}$
		& \cellmag $H^\dagger \tau^\smallIdx{$I$} H\, \epsilon^\smallIdx{$IJK$}\epsilon_{ijk}\,\bigl[D_i, W^\smallIdx{$J$}_{0}\bigr]W^\smallIdx{$K$}_{jk}$
		&
		\\
		\hline

		$B_\rmii{E}^{2}\,H\,H^{\dagger}$
		& $H^\dagger H\, B_{0i}B_{0i}$
		& \cellmag $H^\dagger H\, \bigl[D_i, B_{0}\bigr]\bigl[D_i, B_{0}\bigr]$
		& \multirow{2}{*}{$H^\dag H\, B_{\mu\nu} B^{\mu\nu}$}
		\\
		$B_\rmii{M}^{2}\,H\,H^{\dagger}$
		& $H^\dagger H\, \epsilon_{ijk}\epsilon_{ilm}\,B_{jk}B_{lm}$
		& $H^\dagger H\, \epsilon_{ijk}\epsilon_{ilm}\,B_{jk}B_{lm}$
		&
		\\
		\hdashline[1pt/2pt]
		
		$B_\rmii{E}B_\rmii{M}\,H\,H^{\dagger}$
		&  $H^\dagger H\, \epsilon_{ijk}\,B_{0i}B_{jk}$
		& \cellmag
			$H^\dagger H\, \epsilon_{ijk}\,\bigl[D_i, B_{0}\bigr]B_{jk}$
		& $H^\dagger H\, \widetilde B_{\mu\nu} B^{\mu\nu}$
		\\
		\hline
		
		$B_\rmii{E}\,H\,W_\rmii{E}\,H^{\dagger}$
		& $H^\dagger \tau^\smallIdx{$I$} H\, W^\smallIdx{$I$}_{0i}B_{0i}$
		& \cellmag
			$H^\dagger \tau^\smallIdx{$I$} H\, [D_i, W^\smallIdx{$I$}_{0}] \bigl[D_i, B_{0}\bigr]$
		& \multirow{2}{*}{$H^\dag \tau^\smallIdx{$I$} H\, W^\smallIdx{$I$}_{\mu\nu} B^{\mu\nu}$}
		\\
		$B_\rmii{M}\,H\,W_\rmii{M}\,H^{\dagger}$
		& $H^\dagger \tau^\smallIdx{$I$} H\, \epsilon_{ijk}\epsilon_{ilm}\,W^\smallIdx{$I$}_{jk}B_{lm}$
		& $H^\dagger \tau^\smallIdx{$I$} H\, \epsilon_{ijk}\epsilon_{ilm}\,W^\smallIdx{$I$}_{jk}B_{lm}$
		&
		\\
		\hdashline[1pt/2pt]
		
		$B_\rmii{M}\,H\,W_\rmii{E}\,H^{\dagger}$
		& $H^\dagger \tau^\smallIdx{$I$} H\, \epsilon_{ijk}\,W^\smallIdx{$I$}_{0i}B_{jk}$
		& \cellmag
			$H^\dagger \tau^\smallIdx{$I$} H\, \epsilon_{ijk}\,\bigl[D_i, W^\smallIdx{$I$}_{0}\bigr] B_{jk}$
		& \multirow{2}{*}{$H^\dag \tau^\smallIdx{$I$} H\, \widetilde W^\smallIdx{$I$}_{\mu\nu} B^{\mu\nu}$}
		\\
		$B_\rmii{E}\,H\,W_\rmii{M}\,H^{\dagger}$
		& $H^\dagger \tau^\smallIdx{$I$} H\, \epsilon_{ijk}\,B_{0i}W^\smallIdx{$I$}_{jk}$
		& %$H^\dagger \tau^\smallIdx{$I$} H\, \epsilon_{ijk}\,B_{0i}W^\smallIdx{$I$}_{jk}$
        \cellmag $H^\dagger \tau^\smallIdx{$I$} H\, \epsilon_{ijk}\,[D_i, B_{0}]W^\smallIdx{$I$}_{jk}$
        &
		\\
		\hline
		$G_\rmii{E}^{2}\,H\,H^{\dagger}$
		& $H^\dagger H\, G^\smallIdx{$A$}_{0i}G^\smallIdx{$A$}_{0i}$ 
		& \cellmag
			$H^\dagger H\, \bigl[D_i, G^\smallIdx{$A$}_{0}\bigr]\bigl[D_i, G^\smallIdx{$A$}_{0}\bigr]$
		&  \multirow{2}{*}{$H^\dag H\, G^\smallIdx{$A$}_{\mu\nu} G^{\smallIdx{$A$}\mu\nu}$}
		\\
		$G_\rmii{M}^{2}\,H\,H^{\dagger}$
		& $H^\dagger H\, \epsilon_{ijk}\epsilon_{ilm}\,G^\smallIdx{$A$}_{jk}G^\smallIdx{$A$}_{lm}$
		& $H^\dagger H\, \epsilon_{ijk}\epsilon_{ilm}\,G^\smallIdx{$A$}_{jk}G^\smallIdx{$A$}_{lm}$
		&
		\\
		
		\hdashline[1pt/2pt]
		
		$G_\rmii{E}G_\rmii{M}\,H\,H^{\dagger}$
		& $H^\dagger H\, \epsilon_{ijk}\,G^\smallIdx{$A$}_{0i}G^\smallIdx{$A$}_{jk}$
		& \cellmag
			$H^\dagger H\, \epsilon_{ijk}\,\bigl[D_i, G^\smallIdx{$A$}_{0}\bigr]G^\smallIdx{$A$}_{jk}$
		& $H^\dag H\, \widetilde G^\smallIdx{$A$}_{\mu\nu} G^{\smallIdx{$A$}\mu\nu}$
		\\
		
		\hline
		\hline
		\multicolumn{1}{|c|}{
			Operators:
		} &
		\multicolumn{1}{l|}{
			14
		 } & 4 + {\magenta 10} &
		\multicolumn{1}{l|}{
			8
		}
		\\
		\hline
	\end{tabular}}
	\caption{%
		Dimension-six: class $H^2 X^2$.
		See tabs.~\ref{tab:dim5H2XD} and~\ref{tab:dim5X2D}
		for the color coding.
	}
	\label{tab:dim6H2X2}
\end{table}

%%%%%%%%%%%%%%%%%%%%%%%%%%%%%%%%%%%%%%%%%%%%%%%%%
\begin{table}[t]
	\centering
    \resizebox{\textwidth}{!}{
	\begin{tabular}{|
			c|
			l|
			l|
			l|}
		\hline
		\multicolumn{4}{|c|}{\textbf{Operator class $H^2 X D^2$ }}\\
		\hline
		HS ($T\neq 0$) &
		Inv.\ operators ($T\neq 0$) &
		Static 3D form &
		Inv.\ operators ($T=0$)\\
        
		\hline
		\hline
	
		$B_\rmii{E} H D_i^{2}H^{\dagger}$
		& \cellgreen{$\epsilon_{ijk}\,D_j B_{0k}\,(H^\dagger i\,\lrD_{\!\!i} H)$}
		& \cellmag{$\epsilon_{ijk}\,D_j \bigl[D_k, B_{0}\bigr]\,(H^\dagger i\,\lrD_{\!\!i} H)$}
		& \multirow{8}{*}{$D_\nu B^{\mu\nu}(H^\dagger iD_\mu H)$}
		\\
        \noalign{\vskip -.3pt}
        \cdashline{1-3}[1pt/2pt]
        \noalign{\vskip .3pt}
		$B_\rmii{M} H D_i^{2}H^{\dagger}$
		& \cellgreen{$\epsilon_{ilm}\epsilon_{ijk}\,D_l B_{jk}\,(H^\dagger i\;\lrD_{\!\!m} H)$}
		& $\epsilon_{ilm}\epsilon_{ijk}\,D_l B_{jk}\,(H^\dagger i\;\lrD_{\!\!m} H)$
		&
		\\
        \noalign{\vskip -.3pt}
        \cdashline{1-3}[1pt/2pt]
        \noalign{\vskip .3pt}
		\multirow{3}{*}{$3\,B_\rmii{E} H D_0 D_i H^{\dagger}$}
		& $B_{0i}\,(D_i H^\dagger D_0 H + D_0 H^\dagger D_i H)$ 
		& \cellmag{$[D_i,B_{0}]\,(D_i H^\dagger B_0 H + B_0 H^\dagger D_i H)$}
		&\\
		
		& $B_{0i}\,(H^\dagger D_0 D_i H + D_0 D_i H^\dagger H)$
		& \cellmag{$[D_i, B_{0}]\,(H^\dagger B_0 D_i H + H\,B_0 D_i H^\dagger)$}
		&\\
		
		& $(D_0 B_{0i})\,(H^\dagger i\lrD_{\!\!i} H)$
		& \cellmag{$\bigl(B_0 \bigl[D_i, B_{0}\bigr]\bigr)\,(H^\dagger i\lrD_{\!\!i} H)$}
		&
		\\
        \noalign{\vskip -.3pt}
        \cdashline{1-3}[1pt/2pt]
        \noalign{\vskip .3pt}
		\multirow{3}{*}{$3\,B_\rmii{M} H D_0 D_i H^{\dagger}$}
		& $\epsilon_{ijk}B_{jk} (D_i H^\dagger D_0 H + D_0 H^\dagger D_i H)$ 
		& $\epsilon_{ijk}B_{jk} (D_i H^\dagger B_0 H + B_0 H^\dagger D_i H)$
		&\\
		
		& $\epsilon_{ijk}B_{jk} (H^\dagger D_0 D_i H + D_0 D_i H^\dagger H)$
		& $\epsilon_{ijk}B_{jk} (H^\dagger B_0 D_i H + H B_0 D_i H^\dagger)$
		&\\
		&
		$\epsilon_{ijk}(D_0 B_{jk}) (H^\dagger i\lrD_{\!\!i} H)$
		& $\epsilon_{ijk}(B_0 B_{jk}) (H^\dagger i\lrD_{\!\!i} H)$
		& \\
		
		\hline
		
		$H W_\rmii{E} D_i^{2} H^{\dagger}$ &
		\cellgreen{$\epsilon_{ijk}\,D_j W^\smallIdx{$I$}_{0k}\,(H^\dagger \tau^\smallIdx{$I$} D_i H)$}
		& \cellmag{$\epsilon_{ijk}\,D_j \bigl[D_k, W^\smallIdx{$I$}_{0}\bigr]\,(H^\dagger \tau^\smallIdx{$I$} D_i H)$}
		&
		\multirow{13}{*}{$D_\nu W^{\smallIdx{$I$}\mu\nu}(H^\dagger iD^\smallIdx{$I$}_\mu H)$}
		\\
        \noalign{\vskip -.3pt}
        \cdashline{1-3}[1pt/2pt]
        \noalign{\vskip .3pt}
		$H W_\rmii{M} D_i^{2} H^{\dagger}$ &
		\cellgreen{$\epsilon_{ilm}\epsilon_{ijk}\,D_l W^\smallIdx{$I$}_{jk}\,(H^\dagger \tau^\smallIdx{$I$} D_m H)$}
		& $\epsilon_{ilm}\epsilon_{ijk}\,D_l W^\smallIdx{$I$}_{jk}\,(H^\dagger \tau^\smallIdx{$I$} D_m H)$
		&
		\\
        \noalign{\vskip -.3pt}
        \cdashline{1-3}[1pt/2pt]
        \noalign{\vskip .3pt}
		\multirow{5}{*}{$3\,H W_\rmii{E} D_0 D_i H^{\dagger}$}
		& $W_{0i} (D_i H^\dagger D_0 H + D_0 H^\dagger D_i H)$ 
		& \cellmag{$[D_i, W_{0}] (D_i H^\dagger B_0 H + B_0 H^\dagger D_i H)$}
		&\\
		&
		& \cellmag{$[D_i, W_{0}] (D_i H^\dagger W_0^\smallIdx{$I$} \tau^\smallIdx{$I$} H + W_0^\smallIdx{$I$} H^\dagger \tau^\smallIdx{$I$} D_i H)$}
		&\\
		& $W_{0i} (H^\dagger D_0 D_i H + D_0 D_i H^\dagger H)$
		& \cellmag{$[D_i, W_{0}^\smallIdx{$I$}] (H^\dagger W_0^\smallIdx{$I$} D_i H + H W_0^\smallIdx{$I$} D_i H^\dagger)$}
		&\\
		&
		& \cellmag{$[D_i, W_{0}^\smallIdx{$I$}] (H^\dagger B_0 \tau^\smallIdx{$I$} D_i H + H B_0 \tau^\smallIdx{$I$} D_i H^\dagger)$}
		&\\
		
		& $(D_0 W_{0i})\,(H^\dagger i\lrD_{\!\!i} H)$
		& \cellmag{$(W_0^\smallIdx{$I$} [D_i, W_{0}^\smallIdx{$I$}])\,(H^\dagger i\lrD_{\!\!i} H)$}
		&
		\\
        \noalign{\vskip -.3pt}
        \cdashline{1-3}[1pt/2pt]
        \noalign{\vskip .3pt}
		
		\makebox[0pt][l]{\hypertarget{op:dim6-h2xd2-hwm-d0di}{}}
		\multirow{5}{*}{$3\,H W_\rmii{M} D_0 D_i H^{\dagger}$}
		& $\epsilon_{ijk}W_{jk} (D_i H^\dagger D_0 H + D_0 H^\dagger D_i H)$ 
		& $\epsilon_{ijk}W^\smallIdx{$I$}_{jk} (D_i H^\dagger W^\smallIdx{$I$}_0 H + W^\smallIdx{$I$}_0 H^\dagger D_i H)$,
		&\\
		& & $\epsilon_{ijk}W^\smallIdx{$I$}_{jk} (D_i H^\dagger B_0 \tau^\smallIdx{$I$} H + B_0 \tau^\smallIdx{$I$} H^\dagger D_i H)$
		&\\
        
		& $\epsilon_{ijk}W_{jk} (H^\dagger D_0 D_i H + D_0 D_i H^\dagger H)$
		& $\epsilon_{ijk}W^\smallIdx{$I$}_{jk} (H^\dagger W_0^\smallIdx{$I$} D_i H + H W_0^\smallIdx{$I$} D_i H^\dagger)$
		&\\
        &
		& $\epsilon_{ijk}W^\smallIdx{$I$}_{jk} (H^\dagger B_0\,\tau^\smallIdx{$I$} D_i H + H B_0\,\tau^\smallIdx{$I$} D_i H^\dagger)$
		&
		\\
        
		& $\epsilon_{ijk}(D_0 W_{jk}) (H^\dagger i\lrD_{\!\!i} H)$
		& $\epsilon_{ijk}(W_0^\smallIdx{$I$} W_{jk}^\smallIdx{$I$}) (H^\dagger i\lrD_{\!\!i} H)$
		& \\
		\hline
		\hline
		\multicolumn{1}{|c|}{
			Operators:
		} &
		\multicolumn{1}{l|}{12 + {\Green 4}}
		& 10 + {\magenta 10}&
		2\\
		\hline
	\end{tabular}}
	\caption{%
		Dimension-six: class $H^{2}XD^{2}$.
		See tabs.~\ref{tab:dim5H2XD} and~\ref{tab:dim5X2D}
		for the color coding.
	}
	\label{tab:dim6H2XD2}
\end{table}

%%%%%%%%%%%%%%%%%%%%%%%%%%%%%%%%%%%%%%%%%%%%%%%%%%%%%%%

\begin{table}[t]
\centering
	\scalebox{\classtablescale}{%
	\begin{tabular}{| c | p{0.35\textwidth}|p{0.36\textwidth}|}
		\hline
		\multicolumn{3}{|c|}{
		\textbf{Operator class $X^3$} ($T\neq 0$)}
		\\
		\hline
		HS ($T\neq 0$) &
		\multicolumn{1}{c|}{Inv.\ operators ($T\neq 0$)} &
		\multicolumn{1}{c|}{Static 3D form}
		\\
		\hline
		\hline
	
		$B_\rmii{E}\,G_\rmii{E}\,G_\rmii{M}$ &
		$
			\epsilon_{ijk}\,B_{0i}G^\smallIdx{$A$}_{0j}
			\big(\epsilon_{k l m}G^\smallIdx{$A$}_{l m}\big)$& \cellmag
		$
			\epsilon_{ijk}\,
			\bigl[D_i, B_{0}\bigr]
			\bigl[D_j, G^\smallIdx{$A$}_{0}\bigr]
			\bigl(\epsilon_{k l m}G^\smallIdx{$A$}_{l m}\bigr)$
		\\
		\hdashline[1pt/2pt]
		$B_\rmii{M}\,G_\rmii{E}\,G_\rmii{M}$
		&
		$
			\epsilon_{ijk}
			\big(\epsilon_{i l m}B_{l m}\big)
			G^\smallIdx{$A$}_{0j}
			\big(\epsilon_{krs}G^\smallIdx{$A$}_{rs}\big)$&
		\cellmag
		$
		\epsilon_{ijk}
		\big(\epsilon_{i l m}B_{l m}\big)
		\big[D_j, G^\smallIdx{$A$}_{0}\big]
		\big(\epsilon_{krs}G^\smallIdx{$A$}_{rs}\big)$  \\
		\hdashline[1pt/2pt]
		$B_\rmii{E}\,W_\rmii{E}\,W_\rmii{M}$ &
		$
			\epsilon_{ijk}\,B_{0i}W^\smallIdx{$I$}_{0j}
			\big(\epsilon_{k l m}W^\smallIdx{$I$}_{l m}\big)$&
		\cellmag
		$
		\epsilon_{ijk}\,
		\big[D_i, B_{0}\big]
		\big[D_j, W^\smallIdx{$I$}_{0}\big]
		\big(\epsilon_{k l m}W^\smallIdx{$I$}_{l m}\big)$
		\\
		\hdashline[1pt/2pt]
		$B_\rmii{M}\,W_\rmii{E}\,W_\rmii{M}$ &
		$
			\epsilon_{ijk}
			\big(\epsilon_{i l m}B_{l m}\big)
			W^\smallIdx{$I$}_{0j}\big(\epsilon_{krs}W^\smallIdx{$I$}_{rs}\big)$&
		\cellmag $
			\epsilon_{ijk}
			\big(\epsilon_{i l m}B_{l m}\big)
			\big[D_j, W^\smallIdx{$I$}_{0}\big]
			\big(\epsilon_{krs}W^\smallIdx{$I$}_{rs}\big)$ 
		\\
		\hline
		\hline
		\multicolumn{1}{|c|}{
			Operators:
		} & 
		\multicolumn{1}{l|}{
			4
		} & 
        \multicolumn{1}{l|}{
			{\magenta 4}
		}
        \\
		\hline
	\end{tabular}
	}
	\caption{%
		Dimension-six: class $X^3$ ($T\neq 0$ only).
		See~\cite{Chala:2025aiz,Bernardo:2026nyq}
        for such operators in top-down matching.
        See tabs.~\ref{tab:dim5H2XD} and~\ref{tab:dim5X2D}
		for the color coding.
		All operators in this class are CP-odd.
		Each is an admixture of
		electric ($X_\rmii{E}$) and
		magnetic ($X_\rmii{M}$) components,
		making them 3D analogs of $F\widetilde{F}$-type structures.
		They are discussed further in sec.~\ref{sec:pure:finiteT:operators}.
		The operators of class $X^3$ that also exist at $T=0$
		are listed in tab.~\ref{tab:dim6X3}.
	}
	\label{tab:dim6X3:finite_only_2}
\end{table}
%%%%%%%%%%%%%%%%%%%%%%%%%%%%%%%%%%%%%%%%%%%%%%%%%%%%%

\begin{table}[t]
	\centering
	\scalebox{\classtablescale}{%
	\begin{tabular}{| c | p{0.35\textwidth}|p{0.40\textwidth}|p{0.25\textwidth}|}
		\hline
		\multicolumn{4}{|c|}{\textbf{Operator class $X^2 D^2$ (with $X\equiv B$) }}\\
		\hline
		HS ($T\neq 0$) &
		\multicolumn{1}{c|}{Inv.\ operators ($T\neq 0$)} &
		\multicolumn{1}{c|}{Static 3D form} &
		\multicolumn{1}{c|}{Inv.\ operators ($T=0$)} \\
		\hline
		\hline
        
		\multirow{2}{*}{$2B_\rmii{E}^{2}D_0^{2}$} &
		$(D_0 B_{0i}) (D_0 B_{0i})$ \newline $(D_0^2 B_{0i})  B_{0i}$
		& \multirow{2}{*}{\cellmag{$(B_0 [D_i, B_{0}]) (B_0 [D_i, B_{0}])$}}
		&
		\\
        \noalign{\vskip -.3pt}
        \cdashline{1-3}[1pt/2pt]
        \noalign{\vskip .3pt}
		\multirow{3}{*}{$3B_\rmii{E}B_\rmii{M}D_0^{2}$}
		& $(D_0 B_{0i}) (\epsilon_{ijk}D_0 B_{jk})$ \newline $(D_0^2 B_{0i}) (\epsilon_{ijk} B_{jk})$ \newline $(B_{0i}) (\epsilon_{ijk}D_0^2 B_{jk})$
		& \multirow{3}{*}{\cellmag{$(B_0 [D_i, B_{0}]) (B_0 \epsilon_{ijk}B_{jk})$}}
		&
		\\
        \noalign{\vskip -.3pt}
        \cdashline{1-3}[1pt/2pt]
        \noalign{\vskip .3pt}
        
		\multirow{2}{*}{$2B_\rmii{M}^{2}D_0^{2}$}
		& $(\epsilon_{ijk}D_0 \,B_{jk}) (\epsilon_{ilm} D_0  B_{lm})$ \newline  $(\epsilon_{ijk}D_0^2 \,B_{jk}) (\epsilon_{ilm}  B_{lm})$ & \multirow{2}{*}{$B_0^2( \epsilon_{ijk}\, B_{jk}  \epsilon_{ilm}\,B_{lm})$} &
		\\
        \noalign{\vskip -.3pt}
        \cdashline{1-3}[1pt/2pt]
        \noalign{\vskip .3pt}

		\multirow{2}{*}{$2\,B_\rmii{E}B_\rmii{M}\,D_0D_i$}
		& \cellgreen{$(D_0\,B_{0i}) (\epsilon_{ijk}D_j \epsilon_{klm} B_{lm})$}
		\newline
    	$B_{0i} (\epsilon_{ijk}D_j \epsilon_{klm} D_0 B_{lm})$
		& \multirow{2}{*}{\cellmag{$(B_0\,[D_i, B_0]) (\epsilon_{ijk}D_j \epsilon_{klm} B_{lm})$}}
		& $(\partial_{\mu}B^{\mu\nu})(\partial^{\rho}B_{\rho\nu})$
		\\
        \noalign{\vskip -.3pt}
        \cdashline{1-3}[1pt/2pt]
        \noalign{\vskip .3pt}

      $B_\rmii{E}^{2}D_0 D_i$
		& \cellgreen{$(D_0 B_{0i}) (\epsilon_{ijk}D_j B_{0k})$}
		& \cellmag{$(B_0 [D_i, B_{0}]) (\epsilon_{ijk}\,D_j [D_k, B_{0}])$}
		&
		\\
			$B_\rmii{E}^{2}D_i^{2}$
		& \cellgreen{$(\epsilon_{ijk} D_j B_{0k}) (\epsilon_{ilm} D_l B_{0m})$}
		& \cellmag{($\epsilon_{ijk}D_j [D_k,B_{0}]) (\epsilon_{ilm}D_l [D_m, B_{0}])$}
		&
		\\
			$B_\rmii{E}B_\rmii{M}D_i^{2}$
		& \cellgreen{$(\epsilon_{ijk}\,D_j B_{0k})(\epsilon_{ilm} D_l \,\epsilon_{mpq}B_{pq})$}
		& \cellmag{$(\epsilon_{ijk}\,D_j [D_k,B_0])(\epsilon_{ilm} D_l \,\epsilon_{mpq}B_{pq})$}
		&
		\\
			$B_\rmii{M}^{2}D_i^{2}$ &
		\cellgreen{$(\epsilon_{ijk}\epsilon_{klm}D_{j} B_{lm})(\epsilon_{ipq}\epsilon_{qrs}D_{p}B_{rs})$}
		& $(\epsilon_{ijk}\epsilon_{klm}D_{j} B_{lm})(\epsilon_{ipq}\epsilon_{qrs}D_{p}B_{rs})$
		&
		\\
    	$B_\rmii{M}^{2}D_0D_i$
		& \cellgreen{$(\epsilon_{ijk}D_{0} B_{jk})(\epsilon_{ipq}\epsilon_{qrs}D_{p}B_{rs})$}
		& $(\epsilon_{ijk}B_0 B_{jk})(\epsilon_{ipq}\epsilon_{qrs}D_{p}B_{rs})$ &
		\\
		\hline
		\hline
		\multicolumn{1}{|c|}{
			Operators:
		} &
		\multicolumn{1}{l|}{
			{7 + {\Green 7}}
		 } & 3 + {\magenta 6 } &
		1\\
		\hline
	\end{tabular}%
	}
	\caption{%
		Dimension-six: class $X^2 D^2$, with $X\equiv B$.
		See tabs.~\ref{tab:dim5H2XD} and~\ref{tab:dim5X2D}
		for the color coding.
	}
    \label{tab:dim6X2D2B}
\end{table}
%%%%%%%%%%%%%%%%%%%%%%%%%%%%%%%%%%%%%%%%%%%%%%%%%%%%%

\begin{table}[t]
	\centering
	\scalebox{\classtablescale}{%
	\begin{tabular}{| c | p{0.35\textwidth}|p{0.40\textwidth}|p{0.25\textwidth}|}
		\hline
		\multicolumn{4}{|c|}{\textbf{Operator class $X^2 D^2$ (with $X\equiv W$) }}\\
		\hline
		HS ($T\neq 0$) &
		\multicolumn{1}{c|}{Inv.\ operators ($T\neq 0$)} &
		\multicolumn{1}{c|}{Static 3D form} &
		\multicolumn{1}{c|}{Inv.\ operators ($T=0$)} \\
		\hline
		\hline

		\multirow{2}{*}{$2\,W_\rmii{E}^{2}D_0^{2}$}
		& $(D_0 W_{0i}^\smallIdx{$I$}) (D_0 W_{0i}^\smallIdx{$I$})$ \newline $(D_0^2 W_{0i}^\smallIdx{$I$})  W_{0i}^\smallIdx{$I$}$
		& \cellmag{$(W_0^\smallIdx{$I$} [D_i, W_{0}^\smallIdx{$I$}]) (W_0^\smallIdx{$J$} [D_i, W_{0}^\smallIdx{$J$}])$ \newline 
		$(W_0^\smallIdx{$I$} [D_i, W_{0}^\smallIdx{$J$}]) (W_0^\smallIdx{$I$} [D_i, W_{0}^\smallIdx{$J$}])$}
		& \multirow{14}{*}{$(D_{\mu}W^{\smallIdx{$I$}\,\mu\nu})(D^{\rho}W^\smallIdx{$I$}_{\rho\nu})$}	
		\\
        \noalign{\vskip -.3pt}
        \cdashline{1-3}[1pt/2pt]
        \noalign{\vskip .3pt}

		\makebox[0pt][l]{\hypertarget{op:dim6-x2d2w-wewm-d02}{}}
        \multirow{3}{*}{$3W_\rmii{E} W_\rmii{M}D_0^{2}$}
		& $(D_0 W_{0i}^\smallIdx{$I$}) (\epsilon_{ijk}D_0 W_{jk}^\smallIdx{$I$})$ \newline $(D_0^2 W_{0i}^\smallIdx{$I$}) (\epsilon_{ijk}\,W_{jk}^\smallIdx{$I$})$
		 \newline $(W_{0i}^\smallIdx{$I$}) (\epsilon_{ijk}D_0^2 W_{jk}^\smallIdx{$I$})$
		& \multirow{3}{*}{\cellmag{\makecell{$\epsilon^\smallIdx{$IJK$}(W_0^\smallIdx{$I$} [D_i, W_{0}^\smallIdx{$I$}]) (W_0^\smallIdx{$J$} \epsilon_{ijk}W_{jk}^\smallIdx{$J$})$ \\
		$\epsilon^\smallIdx{$IJK$}(W_0^\smallIdx{$I$} [D_i, W_{0}^\smallIdx{$J$}]) (W_0^\smallIdx{$I$} \epsilon_{ijk}W_{jk}^\smallIdx{$J$})$}}}
		&
		\\
        \noalign{\vskip -.3pt}
        \cdashline{1-3}[1pt/2pt]
        \noalign{\vskip .3pt}
		\multirow{2}{*}{$2\,W_\rmii{M}^{2}D_0^{2}$}
		& $(\epsilon_{ijk}D_0 \,W_{jk}^\smallIdx{$I$}) (\epsilon_{ilm} D_0  W_{lm}^\smallIdx{$I$})$ \newline
		$(\epsilon_{ijk}D_0^2 \,W_{jk}^\smallIdx{$I$}) (\epsilon_{ilm}  W_{lm}^\smallIdx{$I$})$
		& $(W_0^\smallIdx{$I$} \epsilon_{ijk}\, W_{jk}^\smallIdx{$I$})(W_0^\smallIdx{$J$}  \epsilon_{ilm}\,W_{lm}^\smallIdx{$J$})$ \newline
		$(W_0^\smallIdx{$I$} \epsilon_{ijk}\, W_{jk}^\smallIdx{$J$})(W_0^\smallIdx{$I$}  \epsilon_{ilm}\,W_{lm}^\smallIdx{$J$})$
		&
		\\
        \noalign{\vskip -.3pt}
        \cdashline{1-3}[1pt/2pt]
        \noalign{\vskip .3pt}
		\multirow{2}{*}{$2\,W_\rmii{E} W_\rmii{M}\,D_0D_i$}
		& \cellgreen{
			$(D_0\,W_{0i}^\smallIdx{$I$}) (\epsilon_{ijk}D_j \epsilon_{klm} W_{lm}^\smallIdx{$I$})$}
			\newline
			$W_{0i}^\smallIdx{$I$} (\epsilon_{ijk}D_j \epsilon_{klm} D_0 W_{lm}^\smallIdx{$I$})$
		& \multirow{2}{*}{\cellmag{%
				$\epsilon^\smallIdx{$IJK$}(W_0^\smallIdx{$I$}\,[D_i, W_0^\smallIdx{$J$}]) (\epsilon_{ijk}D_j \epsilon_{klm} W_{lm}^\smallIdx{$K$})$
				}}
		&
		\\
        \noalign{\vskip -.3pt}
        \cdashline{1-3}[1pt/2pt]
        \noalign{\vskip .3pt}
     
		$W_\rmii{E}^{2}D_0 D_i$
		& \cellgreen{$(D_0 W_{0i}^\smallIdx{$I$}) (\epsilon_{ijk}D_j W_{0k}^\smallIdx{$I$})$}
		& \cellmag{$\epsilon^\smallIdx{$IJK$}(W_0^\smallIdx{$I$} [D_i, W_{0}^\smallIdx{$J$}]) (\epsilon_{ijk}\,D_j [D_k, W_{0}^\smallIdx{$K$}])$}
		&
		\\

		$W_\rmii{E}^{2}D_i^{2}$
		& \cellgreen{$(\epsilon_{ijk} D_j W_{0k}^\smallIdx{$I$}) (\epsilon_{ilm} D_l W_{0m}^\smallIdx{$I$})$}
		& \cellmag{$(\epsilon_{ijk}D_j [D_k,W_{0}^\smallIdx{$I$}]) (\epsilon_{ilm}D_l [D_m, W_{0}^\smallIdx{$I$}])$}
		&
		\\

		\makebox[0pt][l]{\hypertarget{op:dim6-x2d2w-wewm-di2}{}}$W_\rmii{E} W_\rmii{M}D_i^{2}$
		& \cellgreen{$(\epsilon_{ijk}\,D_j W_{0k}^\smallIdx{$I$})(\epsilon_{ilm} D_l \,\epsilon_{mpq}W_{pq}^\smallIdx{$I$})$}
		& \cellmag{$(\epsilon_{ijk}\,D_j [D_k,W_0^\smallIdx{$I$}])(\epsilon_{ilm} D_l \,\epsilon_{mpq}W_{pq}^\smallIdx{$I$})$}
		&
		\\

		$W_\rmii{M}^{2}D_i^{2}$
		& \cellgreen $(\epsilon_{ijk}\epsilon_{klm}D_{j} W_{lm}^\smallIdx{$I$})(\epsilon_{ipq}\epsilon_{qrs}D_{p}W_{rs}^\smallIdx{$I$})$
		& $(\epsilon_{ijk}\epsilon_{klm}D_{j} W_{lm}^\smallIdx{$I$})(\epsilon_{ipq}\epsilon_{qrs}D_{p}W_{rs}^\smallIdx{$I$})$
		&
		\\

		$W_\rmii{M}^{2}D_0D_i$
		& \cellgreen{$(\epsilon_{ijk}D_{0} W^\smallIdx{$I$}_{jk})(\epsilon_{ipq}\epsilon_{qrs}D_{p}W^\smallIdx{$I$}_{rs})$}
		& $\epsilon^\smallIdx{$IJK$}(\epsilon_{ijk}W^\smallIdx{$I$}_0 W^\smallIdx{$J$}_{jk})(\epsilon_{ipq}\epsilon_{qrs}D_{p}W^\smallIdx{$K$}_{rs})$
		&
		\\
		\hline
		\hline
		\multicolumn{1}{|c|}{
			Operators:
		} &
		\multicolumn{1}{l|}{
			{7 + {\Green 7}}
		 } &  4 + {\magenta 8} &
		1\\
		\hline
	\end{tabular}%
	}
	\caption{%
		Dimension-six: class $X^2 D^2$, with $X\equiv W$.
		See tabs.~\ref{tab:dim5H2XD} and~\ref{tab:dim5X2D}
		for the color coding.
	}
     \label{tab:dim6X2D2W}
\end{table}
%%%%%%%%%%%%%%%%%%%%%%%%%%%%%%%%%%%%%%%%%%%%%%%%%%%%%%

\begin{table}[t]
	\centering
	\scalebox{\classtablescale}{%
	\begin{tabular}{| c | p{0.35\textwidth}|p{0.40\textwidth}|p{0.25\textwidth}|}
		\hline
		\multicolumn{4}{|c|}{\textbf{Operator class $X^2 D^2$ (with $X\equiv G$) }}\\
		\hline
		HS ($T\neq 0$) &
		\multicolumn{1}{c|}{Inv.\ operators ($T\neq 0$)} &
		\multicolumn{1}{c|}{Static 3D form} &
		\multicolumn{1}{c|}{Inv.\ operators ($T=0$)} \\
		\hline
		\hline

		\multirow{2}{*}{$2 G_\rmii{E}^{2}D_0^{2}$}
		& $(D_0 G_{0i}^\smallIdx{$A$}) (D_0 G_{0i}^\smallIdx{$A$})$ \newline $(D_0^2 G_{0i}^\smallIdx{$A$})  G_{0i}^\smallIdx{$A$}$
		& \cellmag{$(G_0^\smallIdx{$A$} [D_i, G_{0}^\smallIdx{$A$}]) (G_0^\smallIdx{$B$} [D_i, G_{0}^\smallIdx{$B$}])$ \newline $(G_0^\smallIdx{$A$} [D_i, G_{0}^\smallIdx{$B$}]) (G_0^\smallIdx{$A$} [D_i, G_{0}^\smallIdx{$B$}])$}
		&
		\\
        \noalign{\vskip -.3pt}
        \cdashline{1-3}[1pt/2pt]
        \noalign{\vskip .3pt}
		\multirow{3}{*}{$3 G_\rmii{E}^{ } G_\rmii{M}^{ } D_0^{2}$}
		& $(D_0 G_{0i}^\smallIdx{$A$}) (\epsilon_{ijk}D_0 G_{jk}^\smallIdx{$A$})$ \newline  $(D_0^2 G_{0i}^\smallIdx{$A$}) (\epsilon_{ijk}\,  G_{jk}^\smallIdx{$A$})$ \newline $( G_{0i}^\smallIdx{$A$}) (\epsilon_{ijk}D_0^2\,G_{jk}^\smallIdx{$A$})$
		& \multirow{3}{*}{\cellmag{\makecell{$(G_0^\smallIdx{$A$} [D_i, G_{0}^\smallIdx{$B$}]) (G_0^\smallIdx{$A$}\epsilon_{ijk}G_{jk}^\smallIdx{$B$})$ \\ $(G_0^\smallIdx{$A$} [D_i, G_{0}^\smallIdx{$A$}]) (G_0^\smallIdx{$B$}\epsilon_{ijk}G_{jk}^\smallIdx{$B$})$}}}
		&
		\\
        \noalign{\vskip -.3pt}
        \cdashline{1-3}[1pt/2pt]
        \noalign{\vskip .3pt}

		\multirow{2}{*}{$2\,G_\rmii{M}^{2}D_0^{2}$}
		& $(\epsilon_{ijk}D_0 \,G_{jk}^\smallIdx{$A$}) (\epsilon_{ilm} D_0  G_{lm}^\smallIdx{$A$})$ \newline
		$(\epsilon_{ijk}D_0^2 \,G_{jk}^\smallIdx{$A$}) (\epsilon_{ilm}  G_{lm}^\smallIdx{$A$})$
		& $(G_0^\smallIdx{$A$} \epsilon_{ijk}\, G_{jk}^\smallIdx{$A$})(G_0^\smallIdx{$B$}  \epsilon_{ilm}\,G_{lm}^\smallIdx{$B$})$ \newline $(G_0^\smallIdx{$A$} \epsilon_{ijk}\, G_{jk}^\smallIdx{$B$})(G_0^\smallIdx{$A$}  \epsilon_{ilm}\,G_{lm}^\smallIdx{$B$})$
		&
		\\
        \noalign{\vskip -.3pt}
        \cdashline{1-3}[1pt/2pt]
        \noalign{\vskip .3pt}

		\multirow{2}{*}{$2\,G_\rmii{E} G_\rmii{M}\,D_0D_i$}
		& \cellgreen{$(D_0\,G_{0i}^\smallIdx{$A$}) (\epsilon_{ijk}D_j \epsilon_{klm} G_{lm}^\smallIdx{$A$})$}
		\newline
		$G_{0i}^\smallIdx{$A$} (\epsilon_{ijk}D_j \epsilon_{klm} D_0 G_{lm}^\smallIdx{$A$})$
		& \multirow{2}{*}{\cellmag{$f^\smallIdx{$ABC$}(G_0^\smallIdx{$A$}\,[D_i, G_0^\smallIdx{$B$}]) (\epsilon_{ijk}D_j \epsilon_{klm} G_{lm}^\smallIdx{$C$})$}}
		& $(D_{\mu}G^{\smallIdx{$A$}\,\mu\nu})(D^{\rho}G^\smallIdx{$A$}_{\rho\nu})$
		\\
        \noalign{\vskip -.3pt}
        \cdashline{1-3}[1pt/2pt]
        \noalign{\vskip .3pt}

		$G_\rmii{E}^{2}D_0^{ } D_i^{ }$
		& \cellgreen{$(D_0 G_{0i}^\smallIdx{$A$}) (\epsilon_{ijk}D_j G_{0k}^\smallIdx{$A$})$}
		& \cellmag{$f^\smallIdx{$ABC$}(G_0^\rmii{$A$} [D_i, G_{0}^\rmii{$B$}]) (\epsilon_{ijk}\,D_j [D_k, G_{0}^\smallIdx{$C$}])$}
		&
		\\

		$G_\rmii{E}^{2}D_i^{2}$
		& \cellgreen{$(\epsilon_{ijk} D_j G_{0k}^\smallIdx{$A$}) (\epsilon_{ilm} D_l G_{0m}^\smallIdx{$A$})$}
		& \cellmag{$(\epsilon_{ijk}D_j [D_k,G_{0}^\smallIdx{$A$}]) (\epsilon_{ilm}D_l [D_m, G_{0}^\smallIdx{$A$}])$}
		&
		\\

		$G_\rmii{E}^{ } G_\rmii{M}^{ } D_i^{2}$
		& \cellgreen{$(\epsilon_{ijk}\,D_j G_{0k}^\smallIdx{$A$})(\epsilon_{ilm} D_l \,\epsilon_{mpq}G_{pq}^\smallIdx{$A$})$}
		& \cellmag{$(\epsilon_{ijk}\,D_j [D_k,G_0^\smallIdx{$A$}])(\epsilon_{ilm} D_l \,\epsilon_{mpq}G_{pq}^\smallIdx{$A$})$}
		&
		\\

		$G_\rmii{M}^{2}D_i^{2}$
		& \cellgreen{$(\epsilon_{ijk}\epsilon_{klm}D_{j} G_{lm}^\smallIdx{$A$})(\epsilon_{ipq}\epsilon_{qrs}D_{p}G_{rs}^\smallIdx{$A$})$}
		& $(\epsilon_{ijk}\epsilon_{klm}D_{j} G_{lm}^\smallIdx{$A$})(\epsilon_{ipq}\epsilon_{qrs}D_{p}G_{rs}^\smallIdx{$A$})$
		&
		\\

		$G_\rmii{M}^{2}D_0^{ }D_i^{ }$
		& \cellgreen{$(\epsilon_{ijk}D_{0} G^\smallIdx{$A$}_{jk})(\epsilon_{ipq}\epsilon_{qrs}D_{p}G^\smallIdx{$A$}_{rs})$}
		& $f^\smallIdx{$ABC$}(\epsilon_{ijk}G^\smallIdx{$A$}_0 G^\smallIdx{$B$}_{jk})(\epsilon_{ipq}\epsilon_{qrs}D_{p}G^\smallIdx{$C$}_{rs})$
		&
		\\
		\hline
		\hline
		\multicolumn{1}{|c|}{
			Operators:
		} &
		\multicolumn{1}{l|}{
			{7 + {\Green 7}}
		 } &  4 + {\magenta 8} &
		1\\
		\hline
	\end{tabular}%
	}
	\caption{%
		Dimension-six: class $X^2 D^2$, with $X\equiv G$.
		See tabs.~\ref{tab:dim5H2XD} and~\ref{tab:dim5X2D}
		for the color coding.
	}
     \label{tab:dim6X2D2G}
\end{table}

%%%%%%%%%%%%%%%%%%%%%%%%% SECTION %%%%%%%%%%%%%%%%%%%%%%%%%%%%%%%%%%%%%%%%%
%
\subsection{Classification of operators}
\label{sec:classification_operators}

We list all the non-redundant dimension-five operators in
tabs.~\ref{tab:dim5H4D}--\ref{tab:dim5X2D}, and
dimension-six operators in
tabs.~\ref{tab:dim6X3}--\ref{tab:dim6X2D2G}.
In each table in this subsection,
the first column lists
the monomials resulting from the Hilbert series calculation.
In the second column, we write the invariant polynomials
as composite operators.
We also further reduce them to operators in 3D-EFT
by removing the temporal properties,
see sec.~\ref{sec:temporal-deriv-gf}, and
the final operators in their static 3D form
are listed in the third column of the respective tables. In the case of dimension-six operators, we also add a fourth column in each table,%
\footnote{%
	Except for tab.~\ref{tab:dim6X3:finite_only_2},
	whose operators have no representative at zero temperature.} containing their representative operators in zero temperature.
In the following,
we discuss
the presence of some operators only at $T\neq 0$
in sec.~\ref{sec:pure:finiteT:operators},
the roles of different constraints
in sec.~\ref{sec:constraints:impact}, and
the impact of gauge choices on the operator set
in sec.~\ref{sec:gauge}.

%%%%%%%%%%%%%%%%%%%%%%%%% SECTION %%%%%%%%%%%%%%%%%%%%%%%%%%%%%%%%%%%%%%%%%
%
\subsubsection{Role of constraints on operator basis}
\label{sec:constraints:impact}

We highlight the impact of different constraints,
discussed in sec.~\ref{sec:HSonR3S1}.
All the dimension-five and -six operators in each table are compatible with
\hyperref[sec:constraint-I:def]{Constraint~I}. 
The green shaded operators contain the following terms:
$\epsilon_{ijk} D_j X_{0k}\equiv \epsilon_{ijk} D_j X_{\rmii{E}\,k} $,
$\epsilon_{ijk} D_j \epsilon_{klm}X_{lm}\equiv \epsilon_{ijk} D_j X_{\rmii{M}\,k} $, where
$X_\rmii{E}$ and $X_\rmii{M}$ are electric and magnetic fields associated with field tensors  $X\in \{B,W,G \}$.
These are proportional to the curl of electric and(or) magnetic fields, and thus vanish when
the
\hyperref[sec:constraint-II:def]{Constraint~II}
is employed.
We display the
\hyperref[sec:constraint-I:def]{Constraint~I}
results separately to isolate the effect of the additional curl constraints
on the operator basis,
and allow for generalized setups with more relaxed constraints.

%%%%%%%%%%%%%%%%%%%%%%%%% SECTION %%%%%%%%%%%%%%%%%%%%%%%%%%%%%%%%%%%%%%%%%
%
\subsubsection{Impact of gauge choice on operator set}
\label{sec:gauge}

The EOM of the temporal gauge field
$X_0$ is $\frac{{\rm d} X_0}{{\rm d}\tau}=0$.
The generic choice of gauge is
$X_0 =\mathcal{F} (\mathbb{R}^3)$.
In that case, the electric component of the field tensor is $[D_i, X_0]= X_{i0} \neq 0$.
However, one can certainly choose a gauge where $X_0=\text{const}$.
Then, the electric component  $[D_i, X_0]=0$.
In each table,
we highlight the operators that vanish for $X_0=\text{const}$ 
(magenta).
In this gauge, the non-perturbative effects from
Polyakov loops are captured
by a constant shift of the Matsubara modes
by a non-integer real number~\cite{Chakrabortty:2024wto},
which subsequently affects the evaluation of thermal Wilson coefficients.
This also explains why operators built from temporal gauge fields alone
do not appear as independent entries in our basis.
They are absorbed into the Wilson coefficients via the Polyakov loop
and are therefore captured more systematically there.

Gauge fixing at finite temperature is more subtle than at zero temperature.
In particular, one cannot set $X_0=0$, since this is incompatible
with the periodic boundary conditions of the spatial gauge fields $X_i$
around the compact direction $S^1$~\cite{Weiss:1980rj}.

%%%%%%%%%%%%%%%%%%%%%%%%% SECTION %%%%%%%%%%%%%%%%%%%%%%%%%%%%%%%%%%%%%%%%%
%
\subsubsection{%
    Pure finite-temperature operators
}
\label{sec:pure:finiteT:operators}

The SMEFT at $T=0$ gives rise to a single dimension-five operator,
$(\ell^\rmii{T} i\tau_2 H) C (H^\rmii{T} i\tau_2 \ell)+\text{h.c.}$,
which is
the Weinberg operator.
Here, $\ell$ is the SM lepton doublet, the superscript $\rmii{T}$ denotes transposition, $C$ is the charge-conjugation operator, $(i\tau_2)_{jk} = \epsilon_{jk}$, and ``$+\text{h.c.}$'' means we are adding the Hermitian conjugate.
This operator violates the lepton number ($L$) by $\Delta L=2$.
At finite temperature,
the space-time manifold and its related symmetry change, thus the fate of this operator.
In our 3D-EFT,
the absence of fermion modes forbids the Weinberg operator.
However, it allows the emergence of different dimension-five operators;
see tabs.~\ref{tab:dim5H4D}--\ref{tab:dim5X2D}.
Since the Wilson coefficients of these operators are
analytic functions of $T$, they vanish in
the limit $T\to 0$,%
\footnote{%
	In the presence of chemical potential,
	some operators might also
	exist at $T=0$ with non-vanishing Wilson coefficients
	as discussed in~\cite{Kajantie:1997ky}.
}
justifying their absence in
the zero-temperature SMEFT operator basis.%
\footnote{%
		For $T > 0$
	the manifold remains $\mathbb{R}^3 \times S^1$.
	At $T = 0$, the thermal circle decompactifies ($\beta \to \infty$,
	$S^1 \to \mathbb{R}$),
	restoring the full Lorentz symmetry.
	Operator structures tied to the compact direction
	are therefore no longer independent at $T = 0$.
} 

In zero-temperature SM,
the tensors
\begin{align}
G_{\mu\rho}^\smallIdx{$A$} G^{\smallIdx{$A$}\,\rho}{}_{\nu}
\,,&&
G_{\mu\rho}^\smallIdx{$A$} \widetilde{G}^{\smallIdx{$A$}\,\rho}{}_{\nu}
\,,
\end{align}
are symmetric under \(\mu \leftrightarrow \nu\).
By contrast,
the field strengths
$B_{\mu\nu}$ and
$\widetilde{B}_{\mu\nu}$
are antisymmetric in their Lorentz indices.
Therefore, the contraction of
$B_{\mu\nu}$ or
$\widetilde{B}_{\mu\nu}$
with either of the above tensors vanishes identically.
As a result, mixed cubic structures, e.g., 
\begin{align}
    B_{\mu \nu} G^{\smallIdx{$A$}\,\nu \rho}\widetilde{G}^\smallIdx{$A$}_{\rho \mu}
    \;=\;
    \widetilde{B}_{\mu \nu} G^{\smallIdx{$A$}\,\nu \rho}G^\smallIdx{$A$}_{\rho \mu}
    \;=\;
    B_{\mu \nu} G^{\smallIdx{$A$}\,\nu \rho} G^\smallIdx{$A$}_{\rho \mu}
    \;=\; 0\,,
\end{align}
do not appear in the zero-temperature SMEFT.

This can also be understood from the symmetry structure of the theory. At zero temperature, full Lorentz invariance relates the electric and magnetic components of a field strength tensor, so they are not independent operator building blocks.
Instead, they are tied together as components of the same Lorentz tensor and can only appear in combinations consistent with four-dimensional Lorentz symmetry. At finite temperature, however, 
as discussed in sec.~\ref{sec:HSonR3S1}, the electric and magnetic components can be treated as independent building blocks, and structures that vanish in the zero-temperature theory need not vanish anymore.
Operators such as those listed in tab.~\ref{tab:dim6X3:finite_only_2}
can arise at finite temperature.

%%%%%%%%%%%%%%%%%%%%%%%%% SECTION %%%%%%%%%%%%%%%%%%%%%%%%%%%%%%%%%%%%%%%%%
%
\section{Comparison to standard dimensional reduction}
\label{sec:dim-reduc}

The operator bases tabulated in sec.~\ref{sec:classification_operators}
differ in structure and multiplicity from those obtained via
the standard dimensional reduction procedure~\cite{%
	Ginsparg:1980ef,Appelquist:1981vg,Susskind:1979up,Weiss:1980rj,
	Kajantie:1995dw}.
We compare our results with~\cite{Chapman:1994vk,Laine:2018lgj}
for QCD and with~\cite{%
	Kajantie:1995dw,Moore:1995jv,Kajantie:1997ky,
	Chala:2024xll,Bernardo:2025vkz,Chala:2025aiz,Bernardo:2026nyq}
for the SM and its subsectors.
Where needed, we extract the temporal gauge fields from $D_0$
as in eq.~\eqref{eq:temporal-and-gauge}.

%%%%%%%%%%%%%%%%%%%%%%%%% SECTION %%%%%%%%%%%%%%%%%%%%%%%%%%%%%%%%%%%%%%%%%
%
\subsection{Comparing operator bases}

One immediate difference between the HS-constructed basis
and the standard dimensionally reduced basis is
the absence of operators with
higher powers of the temporal gauge fields $X_0$,
which play the role of adjoint scalars in the dimensionally reduced theory.
In our construction, these degrees of freedom are encoded in
the covariant derivative $D_0$ (cf.\ sec.~\ref{sec:temporal-deriv-gf})
and the non-local Polyakov loop eq.~\eqref{eq:polyakov}.
A local expansion of the latter
produces a tower of higher-dimensional operators
with explicit $X_0$ insertions
(cf.\ sec.~\ref{sec:finiteTmanifold}).

A concrete illustration is QCD,
where such a local expansion of the effective action~\cite{Megias:2003ui,Bernardo:2026nyq}
reproduces known operators such as
$\tr\{G_0^6\}$ and
$\tr\{G_0^4\}$~\cite{Chapman:1994vk,Laine:2018lgj}.
We can make this comparison explicit by examining
the operator classes
$X^3$ in tab.~\ref{tab:dim6X3} and
$X^2 D^2$ in tab.~\ref{tab:dim6X2D2G},
and comparing them with the basis of~\cite{Laine:2018lgj}.

The class $X^3$ is an admixture of CP-even and CP-odd operators,
and contains two independent operators
that are even in $G_\rmii{E}$,
$G_\rmii{M}^3$ and
$G_\rmii{E}^2 G_\rmii{M}^{ }$,
both matching~\cite{Laine:2018lgj}.
The remaining two operators,
$G_\rmii{E}^3$ and $G_\rmii{E}^{ } G_\rmii{M}^2$,
are odd in $G_\rmii{E}$.
In the class $X^2 D^2$,
the operators
$D_i^2 G_\rmii{M}^2$ and
$D_i^2 G_\rmii{E}^2$
appear at the correct multiplicity
relative to the Lagrangian of~\cite{Chapman:1994vk,Laine:2018lgj}
and are removed in tab.~\ref{tab:dim6X2D2G}
by constraint~\ref{sec:constraint-II:def},
as in the field-redefined basis of QCD~\cite{Bernardo:2026nyq}.
Additional operators in this class are
$G_\rmii{E}^{ } G_\rmii{M}^{ } D_i^2$,
$2\,G_\rmii{E}^{ } G_\rmii{M}^{ } D_0 D_i$,
$G_\rmii{E}^2 D_0 D_i$, and
$G_\rmii{M}^2 D_0 D_i$,
of which some are present
in~\cite{Chapman:1994vk,Megias:2003ui,Laine:2018lgj}.

Beyond these, the class $X^2 D^2$
contains two independent operators with purely temporal derivatives,
$D_0^2 G_\rmii{E}^2$ and
$D_0^2 G_\rmii{M}^2$.
For the former, multiplicities agree between the HS and the standard basis.
For the latter, however, the HS construction undercounts by one operator.
For $SU(N)$ gauge theories,
the operator $\tr\{G_0^2\}\tr\{G_{ij}^2\}$
is independent when $G_0$
is treated as a fundamental adjoint scalar~\cite{Laine:2018lgj},
but it does not appear in the HS basis.
The two approaches simply differ in their input field content.
In the standard basis,
$G_0$ is treated as an independent adjoint scalar,
making $\tr\{G_0^2\}\tr\{G_{ij}^2\}$ a manifest entry.
In the HS construction, the fundamental building block is
the covariant derivative $D_0$,
which naturally encodes non-perturbative effects
such as the Polyakov loop eq.~\eqref{eq:polyakov}
through the thermal Wilson coefficients.
Operators such as $\tr\{G_0^2\}\tr\{G_{ij}^2\}$
can still be generated via a local, perturbative
expansion of these Wilson coefficients.%
\footnote{%
	One can alternatively build a HS in which $X_0$ enters
	as a fundamental field with its own character and plethystic exponential;
	such a construction would yield a basis in closer
	correspondence with the dimensionally reduced one
	and is a possible extension of the framework presented here.
	However, treating $X_0$ as a perturbative field
	would miss non-perturbative effects of the Polyakov loop,
	which are captured through the thermal Wilson coefficients.
}

The odd-in-$G_\rmii{E}$
operators
\begin{align}
	&
	G_\rmii{E}^{3}
	\,,&
	&
	G_\rmii{E}^{ } G_\rmii{M}^{2}	
	\,,&
	&
	3\,G_\rmii{E}^{ } G_\rmii{M}^{ } D_0^2
	\,,
\end{align}
are absent in the standard QCD operator basis because
they are only generated at finite chemical potential~\cite{Bochkarev:1989kp},
and are discussed further in
sec.~\ref{sec:CP-violating-operators}.
All remaining operators, absent from the HS basis
relative to~\cite{Laine:2018lgj},
are generated by the Polyakov loop.

The comparison of the HS operator basis to the dimensional reduction procedure is also apparent in the SM.
Here, we can directly compare the operator bases
with~\cite{Chala:2025aiz,Bernardo:2025vkz,Bernardo:2026nyq} and specifically
focus on the dimension-five and dimension-six operators
of the 3D-SMEFT.
As a first example, consider the class
$H^2 D^4$ in tab.~\ref{tab:dim6H2D4},
which contains one operator of the form ${\cal O}_{D_0^4 H^2} = (D_0D_0 H)^\dagger(D_0D_0 H)$.
\begin{comment}
   $\mathcal{O}_{H^2 D^4} = \bigl\{
    {\cal O}_{D_0^4 H^2},
    {\cal O}_{D_0^2 D_i^2 H^2}
\bigr\}$
with
${\cal O}_{D_0^4 H^2} = (D_0D_0 H)^\dagger(D_0D_0 H)$ and
${\cal O}_{H^2XD^2} = (D_0D_i H)^\dagger(D_0D_i H)$. 
\end{comment}
To make contact with the existing EFT operators~\cite{Chala:2025aiz,Bernardo:2025vkz,Bernardo:2026nyq},
it is useful to expand these representative structures
${\cal O}_{D_0^4 H^2}$ 
which makes explicit the temporal gauge fields hidden inside the covariant derivative $D_0$.
In this way, one finds, schematically,
\begin{align}
    {\cal O}_{D_0^4 H^2} &\supset
		\Bigl\{
			B_0^4 H^\dagger H,
		  (W_0^\smallIdx{$I$} W_0^\smallIdx{$I$} H^\dagger)(W_0^\smallIdx{$J$} W_0^\smallIdx{$J$} H)
		\Bigr\}
    \,.
\end{align}
As argued above,
standalone $X_0$ operators are
absent, as they are captured by the Polyakov loop.

As another example,
we investigate the operator class
$H^4D^2$ in tab.~\ref{tab:dim6H4D2}.
After expanding the corresponding operators,
\begin{align}
    (D_0(H^\dagger H))(D_0(H^\dagger H)) &\sim
    B_0^2 (H^\dagger H)^2
    + (W^\smallIdx{$I$}_0 H^\dagger \tau^\smallIdx{$I$} H)(W_0^\smallIdx{$J$} H^\dagger \tau^\smallIdx{$J$} H)
    \,,
    \\
    (H^\dagger D_0 H)\,(H^\dagger D_0 H)^\dagger &\sim B_0^2 (H^\dagger H)^2 + (H^\dagger W^\smallIdx{$I$} H)(H W^\smallIdx{$I$} H^\dagger)
    \,,\\
    (H^\dagger  H)\,D_0D_0(H^\dagger  H) &\sim  B_0^2 (H^\dagger H)^2 + W_0^\smallIdx{$I$} W_0^\smallIdx{$J$} (H^\dagger H) \text{tr}[\tau^\smallIdx{$I$}  \tau^\smallIdx{$J$}]  (H^\dagger  H)
    \,,
\end{align}
the temporal gauge-field contributions
that appear in the standard thermal EFT picture
are again generated naturally by
the temporal covariant derivatives $D_0$ in the HS operator basis.
This comparison
also agrees with
the dimension-six operators
induced by fermions,
as first derived
by evaluating the thermal fluctuation determinant in
the SM~\cite{Moore:1995jv}.

%%%%%%%%%%%%%%%%%%%%%%%%% SECTION %%%%%%%%%%%%%%%%%%%%%%%%%%%%%%%%%%%%%%%%%
%
\subsection{Parity violating and CP violating operators}
\label{sec:CP-violating-operators}

In principle,
not all symmetry-breaking effects are captured
by the super-renormalizable part of the dimensionally reduced theory;
instead, they enter
the effective theory through
symmetry-breaking higher-dimensional operators.
The HS classification automatically captures operators induced
by symmetry breaking in the parent theory to all orders
in the operator basis.
Specific examples to investigate are
parity (P) and
charge conjugation (C) violation in the SM
and how these effects are passed down to the thermal EFT
operators $O^{\rmii{PC}}$.
While parity violation is absent
in the dimension-four part of the 3D-EFT Lagrangian,
the emergent dimension-five and -six
locally gauge invariant operators
are investigated in~\cite{Kajantie:1997ky}.

In the super-renormalizable part of
the dimensionally reduced EFT
such charge-conjugation violating operators appear only
in the presence of a non-zero chemical potential~\cite{Bochkarev:1989kp}.
In correlator matching, these terms arise from
sum-integrals with shifted Matsubara frequencies, $p_0 \to p_0 + i\mu$,
which can generate Chern--Simons-like contributions~\cite{Gynther:2003za}.

At dimension five, the situation is similar, and
we focus specifically on the following CP violating operators.
Using that
$X_{\rmii{M}\,i}$ denotes the magnetic field
(cf.\ eq.~\eqref{eq:EB}), the corresponding expanded operators are~\cite{Kajantie:1997ky}
\begin{align}
	O_{1}^{-+}
	&= i\,H^\dagger\{D_i,W_{\rmii{M}\,i}\}H
	\nn &\sim
	(H^\dagger \tau^\smallIdx{$I$} i\lrD_{\!\!i} H)\,
	\epsilon_{ijk}^{ }W^\smallIdx{$I$}_{jk}
	\,,
	&\text{corresponding to }&
	\hyperlink{op:dim5-h2xd-wm-di}{HH^\dagger W_\rmii{M} D_i}
	\,,\\[2mm]
	O_{2}^{-+}
	&= i\,\tr [D_i,W_0][W_{\rmii{M}\,i},W_0]
	\nn &\sim
	\epsilon_{ijk}^{ } D_0^{ } W_{i0}^{\smallIdx{$I$}} W_{jk}^\smallIdx{$I$}
	\,,
	&\text{corresponding to }&
	\hyperlink{op:dim5-x2d-wewm-d0}{W_\rmii{E}W_\rmii{M}D_0}
	\,,\\[2mm]
	O_{3}^{-+}
	&= \epsilon_{ijk}\,\tr W_{\rmii{M}\,i} [D_j,W_{\rmii{M}\,k}]
	\nn &\sim
	\epsilon_{ijk}^{ }W^\smallIdx{$I$}_{jk}
	\bigl(\epsilon_{ipq}^{ }D_p^{ }\epsilon_{qlm}^{ }W^\smallIdx{$I$}_{lm}\bigr)
	\,,
	&\text{corresponding to }&
	\hyperlink{op:dim5-x2d-wm2-di}{W_\rmii{M}^2 D_i^{ }}
\,.
\end{align}
The corresponding dimension-five operators are found
in tab.~\ref{tab:dim5H2XD} for $O_{1}^{-+}$ and
in tab.~\ref{tab:dim5X2D} for $O_{2}^{-+}$ and $O_{3}^{-+}$.
At dimension five additional
parity-conserving but C-violating structures
appear, such as
\begin{align}
	O_{1}^{+-}
	&= i(H^\dagger W_0 H)(H^\dagger H)
	\nn &\sim
	W_0^\smallIdx{$I$}(H^\dagger \tau^\smallIdx{$I$} H)(H^\dagger H)
	\,,
	&\text{corresponding to }&
	\hyperlink{op:dim5-h4d-w0-h4}{W_0^\smallIdx{$I$}(H^\dagger \tau^\smallIdx{$I$} H)(H^\dagger H)}
	\,,\\[2mm]
	O_{2}^{+-}
	&= i(H^\dagger W_0 H)\tr(W_0W_0)
	\nn &\sim
	(H^\dagger \tau^\smallIdx{$I$} H)
	W_0^\smallIdx{$I$} W_0^\smallIdx{$J$} W_0^\smallIdx{$J$}
	\,,
	&\text{corresponding to }&
	\hyperlink{op:dim5-h2d3}{H H^\dagger D_0^3}
	\,,\\[2mm]
	O_{3}^{+-}
	&= iH^\dagger[D_i,[D_i,W_0]]H
	\nn
	&= iH^\dagger[D_i,W_{{\rmii E}i}]H
	\,,
	&\text{corresponding to }&
	\hyperlink{op:dim5-h2xd-we-di}{HH^\dagger W_\rmii{E} D_i}
	\,,
	\\[2mm]
	O_{4}^{+-}
	&= iH^\dagger\{D_i,\{D_i,W_0\}\}H
	\,,
	&\text{corresponding to }&
	\hyperlink{op:dim5-h2xd-we-di}{HH^\dagger W_\rmii{E} D_i}
	\,.
\end{align}
Also, a T-odd structure appears, which is CP-even,
\begin{align}
	O_{1}^{++}
	&= H^\dagger\{D_i,[D_i,W_0]\}H
	\,,
	&\text{corresponding to }&
	\hyperlink{op:dim5-h2xd-we-di}{HH^\dagger W_\rmii{E} D_i}
	\,.
\end{align}
Here, $O_{1}^{+-}$ maps directly to
the explicit $H^4D$ entry in
tab.~\ref{tab:dim5H4D}, while
$O_{2}^{+-}$ belongs to the same temporal
$H^2D^3$ sector collected in tab.~\ref{tab:dim5H2D3}.
The operators $O_{3}^{+-}$, $O_{4}^{+-}$, and $O_{1}^{++}$ involve explicit
$W_0$ or $W_{{\rmii E}\,i}=[D_i,W_0]$ insertions and are therefore represented,
in our basis, by
the $H^2W_\rmii{E}D_i$ structures collected in
tab.~\ref{tab:dim5H2XD}.

At dimension six, we encounter for the first time
CP-violating operators,
that are not induced by chemical potential.
Following~\cite{Kajantie:1997ky}, 
their original form and
the corresponding
expanded version is
\begin{align}
	O_{4}^{-+}
	&= i\,\partial_k(H^\dagger H)\,\tr W_0 W_{\rmii{M}\,k}
	\nn &\sim
	\epsilon_{ijk}
	\bigl(D_0W_{jk}^\smallIdx{$I$}\bigr)
	\bigl(H^\dagger i\lrD_{\!\!i} H\bigr)
	\,,
	&\text{corresponding to }&
	\hyperlink{op:dim6-h2xd2-hwm-d0di}{H W_\rmii{M} D_0 D_i H^{\dagger}}
	\,,\\[2mm]
	O_{5}^{-+}
	&= i\,\partial_k\tr W_0 W_0\,\tr W_0 W_{\rmii{M}\,k}
	\nn &\sim
	\bigl(D_0W_{0i}^\smallIdx{$I$}\bigr)
	\bigl(\epsilon_{ijk}D_0W_{jk}^\smallIdx{$I$}\bigr)
	\nn &
	+ \bigl(D_0^2W_{0i}^\smallIdx{$I$}\bigr)
		\bigl(\epsilon_{ijk}W_{jk}^\smallIdx{$I$}\bigr)
	+ W_{0i}^\smallIdx{$I$}\bigl(\epsilon_{ijk}D_0^2W_{jk}^\smallIdx{$I$}\bigr)
	\,,
	&\text{corresponding to }&
	\hyperlink{op:dim6-x2d2w-wewm-d02}{W_\rmii{E} W_\rmii{M}D_0^{2}}
	\,,\\[2mm]
	O_{6}^{-+}
	&= \epsilon_{ijk}\,\tr[D_i,W_0][D_j,W_0][D_k,W_0]
	\nn &\sim
	\epsilon^\smallIdx{$IJK$}_{ }
	\epsilon_{ijk}^{ }
	W^\smallIdx{$I$}_{0i}W^\smallIdx{$J$}_{0j}W^\smallIdx{$K$}_{0k}
	\,,
	&\text{corresponding to }&
	\hyperlink{op:dim6-x3-we3}{W_\rmii{E}^{3}}
	\,,\\[2mm]
	O_{7}^{-+}
	&= i\,\epsilon_{ijk}\,\tr[D_i,[D_j,W_{km}]][D_m,W_0]
	\nn &\sim
	\bigl(\epsilon_{ijk}^{ }D_j^{ }W_{0k}^\smallIdx{$I$}\bigr)
	\bigl(\epsilon_{ilm}^{ }D_l^{ }\epsilon_{mpq}^{ }W_{pq}^\smallIdx{$I$}\bigr)
	\,,
	&\text{corresponding to }&
	\hyperlink{op:dim6-x2d2w-wewm-di2}{W_\rmii{E} W_\rmii{M}D_i^{2}}
\,.
\end{align}
The corresponding dimension-six operators are found in
tab.~\ref{tab:dim6H2XD2} for $O_{4}^{-+}$,
tab.~\ref{tab:dim6X2D2W} for $O_{5}^{-+}$ and $O_{7}^{-+}$, and
tab.~\ref{tab:dim6X3} for $O_{6}^{-+}$.
Since the SM is CP-violating
in the fundamental 4D theory,
these operators are generated in the thermal EFT,
with Wilson coefficients that are suppressed but non-vanishing
in the limit of zero chemical potential.
Another two independent CP-even operators appear in
the 3D-SMEFT~\cite{Kajantie:1997ky},
which in our notation can be written as
\begin{align}
	O_{1}^{--}
	&= H^\dagger \{D_i,\{W_0,W_{\rmii{M}\,i}\}\}H
	\nn &\sim
	\epsilon_{ijk}^{ }W_{jk}^\smallIdx{$I$}
	\bigl(D_i H^\dagger D_0 H + D_0 H^\dagger D_i H\bigr)
	\,,
	&\text{corresponding to }&
	\hyperlink{op:dim6-h2xd2-hwm-d0di}{H W_\rmii{M} D_0 D_i H^{\dagger}}
	\,,\\[2mm]
	O_{2}^{--}
	&= H^\dagger [D_i,[W_0,W_{\rmii{M}\,i}]]H
	\nn &\sim
	H^\dagger \tau^\smallIdx{$I$} H\,
	\epsilon^\smallIdx{$IJK$}_{ }
	\epsilon_{ijk}^{ }
	W^\smallIdx{$J$}_{0i}W^\smallIdx{$K$}_{jk}
	\,,
	&\text{corresponding to }&
	\hyperlink{op:dim6-h2x2-hwewm-tau}{H W_\rmii{E} W_\rmii{M} H^{\dagger}}
	\,.
\end{align}

The corresponding CP-even dimension-six operators are found in
tab.~\ref{tab:dim6H2XD2} for $O_{1}^{--}$ and in
tab.~\ref{tab:dim6H2X2} for $O_{2}^{--}$.
For the $H^2 X D^2$ structures,
coefficients can be $\mathcal{O}(1)$
since ${\rm CPT}=+1$ for these operators.

There are additional CP-odd operators that appear
in the pure gauge sector at dimension six
cubic in the field strengths.
Such operators are e.g.
\begin{align}
		\label{eq:O8+-}
		O_{8}^{+-} &= i\,\epsilon_{ijk}\,\tr W_0 W_0 W_{jk}
		\sim
		\epsilon^\smallIdx{$IJK$}_{ }
		\epsilon_{ijk}^{ }
		W^\smallIdx{$I$}_{0i}
		W^\smallIdx{$J$}_{0j}
		W^\smallIdx{$K$}_{jk}
		 \,,
		 &\text{corresponding to }&
	\hyperlink{op:dim6-x3-we2wm}{W_\rmii{E}^{2} W_\rmii{M}^{ }}
		 \,,\\[2mm]
	\label{eq:O9+-}
	O_{9}^{+-} &= i\,\epsilon_{ijk}\,\tr W_0 W_{jl} W_{lk}
		\sim
		\epsilon^\smallIdx{$IJK$}_{ }
		\epsilon_{ijk}^{ }
		W^\smallIdx{$I$}_{0i}
		W^\smallIdx{$J$}_{jl}
		W^\smallIdx{$K$}_{lk}
		\,,
		&\text{corresponding to }&
	\hyperlink{op:dim6-x3-wewm2}{W_\rmii{E}^{ } W_\rmii{M}^{2}}
		 \,,
\end{align}
as listed in tab.~\ref{tab:dim6X3}
in agreement with~\cite{Chala:2025aiz,Bernardo:2026nyq}.

Equivalent CP-violating operators at zero temperature,
e.g., $(\widetilde{F}_{\mu \nu}F_{\nu \rho}F_{\rho \mu})$,
can only be generated at two-loop or
beyond~\cite{DasBakshi:2020ejz,Bakshi:2021ofj,Naskar:2022rpg,Guedes:2023azv}.
Thus, they are expected to be suppressed compared to
other CP-violating operators
containing scalar fields, e.g.,
$\phi^\dagger \phi \widetilde{F}_{\mu \nu}F_{\mu \nu}$.
We expect a similar hierarchy in the size of
the respective Wilson coefficients at finite temperatures
of tab.~\ref{tab:dim6X3:finite_only_2}.
A consequence of these operators
is that they can source
additional CP violation in the thermal plasma.
Although their Wilson coefficients are expected to be suppressed,
similarly to
$\mathcal{O}_8^{+-}$ and $\mathcal{O}_9^{+-}$,
this source goes beyond
the CP-violation in the fundamental theory~\cite{%
	vandeVis:2025efm,
	Prokopec:2003pj,Kainulainen:2001cn,Prokopec:2004ic,
	Cirigliano:2011di,Cirigliano:2009yt,Konstandin:2004gy},
and could have implications for
baryogenesis and the dynamics of the electroweak phase transition.

We close this comparison with two practical remarks.
First, our basis omits operators built purely out of the temporal scalar $X_0$,
because their effect is encoded
in the Polyakov loop and therefore enters
through the Wilson coefficients rather than as independent operators in the basis.
A local expansion of the Polyakov loop
then redistributes these contributions onto $X_0$-monomials
at the price of a tower of higher-dimensional operators~\cite{Chapman:1994vk};
this perspective is the easiest way to obtain a workable
standard EFT operator basis without requiring lattice input for the Polyakov loop.
Second, given any operator written
in a standard 3D-EFT basis with explicit $X_0$ insertions,
the corresponding HS entry is recovered by
absorbing $X_0$ into the temporal covariant derivative $D_0$ and,
where needed, applying spatial IBP and EOM to bring the result into
the non-redundant form tabulated in sec.~\ref{sec:classification_operators}.

%%%%%%%%%%%%%%%%%%%%%%%%% SECTION %%%%%%%%%%%%%%%%%%%%%%%%%%%%%%%%%%%%%%%%%
%
\section{Conclusions}
\label{sec:conclusions}

In this paper,
we study finite-temperature QFT in the imaginary-time formalism.
The space-time manifold is
$\mathbb{R}^3 \times S^1$, where bosonic fields satisfy periodic
boundary conditions along $S^1$, whose radius $\beta$ is set by the
inverse temperature.
The corresponding space-time symmetry is
$\widetilde{E}(3) \times SO(2)
\equiv SU(2)_\T \ltimes \mathbb{T}_3 \times SO(2)$.
Our construction mirrors a static 3D effective theory,
so there is no dynamics in the temporal direction.
Accordingly, full Lorentz invariance is absent because the thermal background
selects a preferred timelike direction and therefore breaks boost invariance.

We employ the Hilbert series method to compute the effective operators.
To this end, we calculate the characters for scalar and vector fields,
with the latter decomposed into the electric and magnetic components of the
field strengths, while incorporating the constraints from spatial IBP and
EOMs.
We then introduce the bosonic SM fields that transform under this space-time
symmetry and under the internal gauge symmetry
$SU(3)_c\times SU(2)_\rmii{$L$} \times U(1)_\rmii{$Y$}$.
Combining these ingredients,
we construct the full Hilbert series for the SM
at finite temperature.
Compared with strictly static constructions,
the finite-temperature Hilbert series on $\mathbb{R}^3\times S^1$ is better suited
to capture generalized thermal effects,
including Polyakov-loop-sensitive structures,
in a systematic and symmetry-controlled way.
This provides a broader and more robust framework for basis construction
beyond the static-$X_0$ limit.

We compute the SMEFT operators up to dimension-six.
In particular, we highlight the distinct role of the temporal derivative $D_0$
relative to the spatial derivatives $D_i$, and explain why operators built from
$D_0$ are more numerous than those built from $D_i$ at finite temperature.
We tabulate the dimension-five and -six operators in
tabs.~\ref{tab:dim5H4D}--\ref{tab:dim5X2D}, and
tabs.~\ref{tab:dim6X3}--\ref{tab:dim6X2D2G}, respectively.
We reduce these operators to the form appropriate for a static 3D-EFT.
For completeness, we also list representative covariant operators at $T=0$.

In each table, we identify the operators that vanish when
(i) the curl of the electric and magnetic components of the gauge field strengths
is null-valued (green shade), and
(ii) the temporal gauge fields are constant, so that the electric components
are identically zero (magenta shade).
We further point out operators that are present only at finite temperature,
including the dimension-six operators in tab.~\ref{tab:dim6X3:finite_only_2}
and all dimension-five operators.
The Wilson coefficients of these intrinsically thermal operators carry positive powers of $T$ and
therefore vanish as $T \to 0$, which is why they are absent from
the standard zero-temperature SMEFT basis.
The presence of such operators is a consequence of
the reduced symmetry at finite temperature,
which allows for new operator structures that are not invariant under
the full Lorentz group but are consistent with the symmetries of the thermal background.
Some of these operators are CP-odd and
can source additional CP violation in the thermal plasma,
with potential implications for
baryogenesis and
the dynamics of cosmological phase transitions.

Finally,
we plan to extend the Hilbert series construction
to other theories of interest,
such as scalar extensions of the SM,
and to incorporate fermionic fields,
which satisfy anti-periodic boundary conditions on $S^1$
and therefore enter the character computation differently from bosons.
Scalar extensions are particularly interesting
in the context of cosmological phase transitions
that may be affected by the presence of
previously unconsidered operators.

%%%%%%%%%%%%%%%%%%%%%%%%% SECTION %%%%%%%%%%%%%%%%%%%%%%%%%%%%%%%%%%%%%%%%%
%
\section*{Acknowledgments}

We thank
Fabio Bernardo,
Debmalya Dey,
Romain Guillermo Reinle,
and
Andreas Helset
for enlightening discussions.
JC acknowledges the hospitality of HRI, Allahabad, India, where part of the research was done.
This work is supported by the Core Research Grant (CRG/2023/003200), SERB, India.
BSE was supported by ``Fundação de Amparo à Pesquisa do Estado de São Paulo'' (FAPESP) under contracts 2024/01764-9 and 2025/05822-6, and thanks Francesco Riva for the hospitality at the University of Geneva.
PS was supported by
the Swiss National Science Foundation (SNSF) under grant
\href{https://data.snf.ch/grants/grant/215997}{\tt PZ00P2-215997}.

%%%%%%%%%%%%%%%%%%%%%%%%% SECTION %%%%%%%%%%%%%%%%%%%%%%%%%%%%%%%%%%%%%%%%%
%
\section*{Data availability statement}
No Data associated with the manuscript.

% %%%%%%%%%%%%%%%%%%%%%%%%%%%%%%%%%%%%%%%%%%%%%%%%%%%%%%%%%%%%%%%%%%%%%%%%%%%%%%%%%%%%%%%%%%%%%%%%%%%%
\appendix
\renewcommand{\thesection}{\Alph{section}}
\renewcommand{\thesubsection}{\Alph{section}.\arabic{subsection}}
\renewcommand{\theequation}{\Alph{section}.\arabic{equation}}

%%%%%%%%%%%%%%%%%%%%%%%%% SECTION %%%%%%%%%%%%%%%%%%%%%%%%%%%%%%%%%%%%%%%%%
%
\section{Hilbert series output}
\label{app:HS:details}
%%%%%%%%%%%%%%%%%%%%%%%%% SECTION %%%%%%%%%%%%%%%%%%%%%%%%%%%%%%%%%%%%%%%%%
%
\subsection{
			\hyperref[sec:constraint-I:def]{Constraint~I}$\;\oplus\;$Spatial IBP:
            }
\label{sec:constraint-I}

The Hilbert series eq.~\eqref{eq:HSspatial} with
\begin{align}
	S_\text{scalar}&=\{H,H^\dagger\}
	\,,&
	S_\text{vector}^{\rmii{\ref{sec:constraint-I:def}}}&=\{B_\rmii{E},B_\rmii{M},W_\rmii{E},W_\rmii{M},G_\rmii{E},G_\rmii{M}\}
	\,,&
	S_\text{vector}^{\rmii{\ref{sec:constraint-II:def}}}&=\emptyset
	\,,
\end{align}
at mass dimensions five and six read:
\begin{align}
 \label{app:eq:I-dim5}
 \mathcal{H}_\rmi{dim5}^{\rmii{\ref{sec:constraint-I:def}}} &=
          H H^{\dagger } B_\rmii{E} D_i
        + B_\rmii{E} B_\rmii{M} D_i
        + B_\rmii{E}^2 D_i
        + 2 B_\rmii{E} B_\rmii{M} D_0
        + B_\rmii{E}^2 D_0+H H^{\dagger } B_\rmii{M} D_i
        + B_\rmii{M}^2 D_i
        \nn &
        + B_\rmii{M}^2 D_0
        + G_\rmii{E} G_\rmii{M} D_i
        + G_\rmii{E}^2 D_i
        + 2 G_\rmii{E} G_\rmii{M} D_0
        + G_\rmii{E}^2 D_0
        + H H^{\dagger } W_\rmii{E} D_i
        \nn &
        + W_\rmii{E} W_\rmii{M} D_i
        + W_\rmii{E}^2 D_i
        + 2 W_\rmii{E} W_\rmii{M} D_0
        + W_\rmii{E}^2 D_0
        + G_\rmii{M}^2 D_i
        + G_\rmii{M}^2 D_0
        \nn &
        + H H^{\dagger } W_\rmii{M} D_i
        + 4 H H^{\dagger } D_0^3
        + 2 H^2 \bigl(H^{\dagger }\bigr)^2 D_0
        + W_\rmii{M}^2 D_i
        + W_\rmii{M}^2 D_0
    \,,
    \\[2mm]
\label{app:eq:I-dim6}
    \mathcal{H}_\rmi{dim6}^{\rmii{\ref{sec:constraint-I:def}}} &=
          H H^{\dagger } B_\rmii{E} B_\rmii{M}
        + H H^{\dagger } B_\rmii{E}^2
        + H H^{\dagger } B_\rmii{M}^2
        + 3 B_\rmii{E} B_\rmii{M} D_0^2
        + 2 B_\rmii{E}^2 D_0^2+2 B_\rmii{M}^2 D_0^2
        \nn &
        + 3 G_\rmii{E} G_\rmii{M} D_0^2
        + 2 G_\rmii{E}^2 D_0^2
        + 2 G_\rmii{M}^2 D_0^2
        + 3 W_\rmii{E} W_\rmii{M} D_0^2
        + 2 W_\rmii{E}^2 D_0^2
        + 2 W_\rmii{M}^2 D_0^2
        \nn &
        + 6 H^2 \bigl(H^{\dagger }\bigr)^2 D_0^2
        + 5 H H^{\dagger } D_0^4
        + 2 B_\rmii{E} B_\rmii{M} D_0 D_i
        + 3 H H^{\dagger } B_\rmii{E} D_0 D_i
        + B_\rmii{E}^2 D_0 D_i
        \nn &
        + 3 H H^{\dagger } B_\rmii{M} D_0 D_i
        + B_\rmii{M}^2 D_0 D_i
        + 2 G_\rmii{E} G_\rmii{M} D_0 D_i
        + G_\rmii{E}^2 D_0 D_i
        + G_\rmii{M}^2 D_0 D_i
        \nn &
        + 2 W_\rmii{E} W_\rmii{M} D_0 D_i
        + 3 H H^{\dagger } W_\rmii{E} D_0 D_i
        + W_\rmii{E}^2 D_0 D_i
        + 3 H H^{\dagger } W_\rmii{M} D_0 D_i
        + W_\rmii{M}^2 D_0 D_i
        \nn &
        + B_\rmii{E} B_\rmii{M} D_i^2
        + H H^{\dagger } B_\rmii{E} D_i^2
        + B_\rmii{E}^2 D_i^2
        + H H^{\dagger } B_\rmii{M} D_i^2
        + B_\rmii{M}^2 D_i^2
        + G_\rmii{E} G_\rmii{M} D_i^2
        \nn &
        + G_\rmii{E}^2 D_i^2
        + G_\rmii{M}^2 D_i^2
        + W_\rmii{E} W_\rmii{M} D_i^2
        + H H^{\dagger } W_\rmii{E} D_i^2
        + W_\rmii{E}^2 D_i^2
        + H H^{\dagger } W_\rmii{M} D_i^2
        + W_\rmii{M}^2 D_i^2
        \nn &
        + 2 H^2 \bigl(H^{\dagger }\bigr)^2 D_i^2
        + B_\rmii{E} G_\rmii{E} G_\rmii{M}
        + B_\rmii{M} G_\rmii{E} G_\rmii{M}
        + H H^{\dagger } G_\rmii{E} G_\rmii{M}
        + G_\rmii{E} G_\rmii{M}^2
        + G_\rmii{E}^2 G_\rmii{M}
        \nn &
        + H H^{\dagger } G_\rmii{E}^2
        + H H^{\dagger } G_\rmii{M}^2
        + H H^{\dagger } B_\rmii{E} W_\rmii{E}
        + H H^{\dagger } B_\rmii{M} W_\rmii{E}
        + B_\rmii{E} W_\rmii{E} W_\rmii{M}
        + B_\rmii{M} W_\rmii{E} W_\rmii{M}
        \nn &
        + 2 H H^{\dagger } W_\rmii{E} W_\rmii{M}
        + W_\rmii{E} W_\rmii{M}^2
        + W_\rmii{E}^2 W_\rmii{M}
        + H H^{\dagger } W_\rmii{E}^2
        + H H^{\dagger } B_\rmii{E} W_\rmii{M}
        + H H^{\dagger } B_\rmii{M} W_\rmii{M}
        \nn* &
        + H H^{\dagger } W_\rmii{M}^2
        + H^3 \bigl(H^{\dagger }\bigr)^3
        + G_\rmii{E}^3
        + G_\rmii{M}^3
        + W_\rmii{E}^3
        + W_\rmii{M}^3
    \,.
\end{align}

%%%%%%%%%%%%%%%%%%%%%%%%% SECTION %%%%%%%%%%%%%%%%%%%%%%%%%%%%%%%%%%%%%%%%%
%
\subsection{
			\hyperref[sec:constraint-II:def]{Constraint~II}$\;\oplus\;$Spatial IBP:
            }
\label{app:constraint-II}

The Hilbert series eq.~\eqref{eq:HSspatial} with
\begin{align}
	S_\text{scalar}&=\{H,H^\dagger\}
	\,,&
	S_\text{vector}^{\rmii{\ref{sec:constraint-I:def}}}&=\emptyset
	\,,&
	S_\text{vector}^{\rmii{\ref{sec:constraint-II:def}}}&=
	\{B_\rmii{E},B_\rmii{M},W_\rmii{E},W_\rmii{M},G_\rmii{E},G_\rmii{M}\}
	\,,
\end{align}
at mass dimensions five read:
\begin{align}
\label{app:eq:II-dim5}
	\mathcal{H}_\rmi{dim5}^{\rmii{\ref{sec:constraint-II:def}}}&=
			H H^{\dagger } B_\rmii{E} D_i
		+ 2 B_\rmii{E} B_\rmii{M} D_0
		+ B_\rmii{E}^2 D_0
		+ H H^{\dagger } B_\rmii{M} D_i
		+ B_\rmii{M}^2 D_0
		+ 2 G_\rmii{E} G_\rmii{M} D_0
    \nn &
    + G_\rmii{E}^2 D_0
		+ H H^{\dagger } W_\rmii{E} D_i
		+ 2 W_\rmii{E} W_\rmii{M} D_0
		+ W_\rmii{E}^2 D_0
		+ G_\rmii{M}^2 D_0
		+ H H^{\dagger } W_\rmii{M} D_i
    \nn &
    + 4 H H^{\dagger } D_0^3
		+ 2 H^2 \bigl(H^{\dagger }\bigr)^2 D_0
		+ W_\rmii{M}^2 D_0
		\,.
\end{align}
The number of dimension-five operators
that vanish exclusively because of
null curl constraint in eq.~\eqref{eq:curlSM}
is given by subtracting eq.~\eqref{app:eq:II-dim5}
from eq.~\eqref{app:eq:I-dim5},
{\em viz.}
\begin{align}
\label{app:eq:difII-dim5}
		\mathcal{H}_\rmi{dim5}^{\rmii{\ref{sec:constraint-I:def}}}
	- \mathcal{H}_\rmi{dim5}^{\rmii{\ref{sec:constraint-II:def}}} &=
    	B_\rmii{E} B_\rmii{M} D_i
		+ B_\rmii{E}^2 D_i
		+ B_\rmii{M}^2 D_i
		+ G_\rmii{E} G_\rmii{M} D_i
		+ G_\rmii{E}^2 D_i
    \nn &
    + W_\rmii{E} W_\rmii{M} D_i
		+ W_\rmii{E}^2 D_i
		+ G_\rmii{M}^2 D_i
		+ W_\rmii{M}^2 D_i
		\,.
\end{align}

At mass dimension six,
the series reads:
\begin{align}
\label{app:eq:II-dim6}
	\mathcal{H}_\rmi{dim6}^{\rmii{\ref{sec:constraint-II:def}}} &=
	3 H H^{\dagger } B_\rmii{E} D_0 D_i
	+ 3 B_\rmii{E} B_\rmii{M} D_0^2
	+ 2 B_\rmii{E}^2 D_0^2
	+ 3 H H^{\dagger } B_\rmii{M} D_0 D_i
	+ 2 B_\rmii{M}^2 D_0^2
	+ B_\rmii{E} G_\rmii{E} G_\rmii{M}
	\nn &
	+ B_\rmii{M} G_\rmii{E} G_\rmii{M}
	+ H H^{\dagger } B_\rmii{M} W_\rmii{E}
	+ H H^{\dagger } B_\rmii{E} W_\rmii{M}
	+ H H^{\dagger } B_\rmii{E} B_\rmii{M}
	+ H H^{\dagger } B_\rmii{E} W_\rmii{E}
	+ H H^{\dagger } B_\rmii{E}^2
	\nn &
	+ B_\rmii{E} W_\rmii{E} W_\rmii{M}
	+ B_\rmii{M} W_\rmii{E} W_\rmii{M}
	+ H H^{\dagger } B_\rmii{M} W_\rmii{M}
	+ H H^{\dagger } B_\rmii{M}^2
	+ 3 G_\rmii{E} G_\rmii{M} D_0^2
	+ 2 G_\rmii{E}^2 D_0^2
	\nn &
	+ 3 H H^{\dagger } W_\rmii{E} D_0 D_i
	+ 3 W_\rmii{E} W_\rmii{M} D_0^2
	+ 2 W_\rmii{E}^2 D_0^2
	+ 2 G_\rmii{M}^2 D_0^2
	+ 3 H H^{\dagger } W_\rmii{M} D_0 D_i
	+ 5 H H^{\dagger } D_0^4
	\nn &
	+ 2 H^2 \bigl(H^{\dagger }\bigr)^2 D_i^2
	+ 6 H^2 \bigl(H^{\dagger }\bigr)^2 D_0^2
	+ 2 W_\rmii{M}^2 D_0^2
	+ H H^{\dagger } G_\rmii{M} G_\rmii{E}
	+ H H^{\dagger } G_\rmii{E}^2
	+ G_\rmii{E} G_\rmii{M}^2
	\nn &
	+ G_\rmii{E}^2 G_\rmii{M}
	+ G_\rmii{E}^3
	+ 2 H H^{\dagger } W_\rmii{E} W_\rmii{M}
	+ H H^{\dagger } W_\rmii{E}^2
	+ W_\rmii{E} W_\rmii{M}^2
	+ W_\rmii{E}^2 W_\rmii{M}
	\nn &
	+ W_\rmii{E}^3
	+ H H^{\dagger } G_\rmii{M}^2
	+ G_\rmii{M}^3
	+ H H^{\dagger } W_\rmii{M}^2
	+ H^3 \bigl(H^{\dagger }\bigr)^3
	+ W_\rmii{M}^3
	\,.
\end{align}
The number of dimension-six operators that vanish exclusively
because of null curl constraint in eq.~\eqref{eq:curlSM} is given
by subtracting eq.~\eqref{app:eq:II-dim6} from eq.~\eqref{app:eq:I-dim6},
{\em viz.}
\begin{align}
\label{app:eq:difII-dim6}
   	\mathcal{H}_\rmi{dim6}^{\rmii{\ref{sec:constraint-I:def}}}
	-	\mathcal{H}_\rmi{dim6}^{\rmii{\ref{sec:constraint-II:def}}} &=
	   H H^{\dagger } B_\rmii{E} D_i^2
	   + 2 B_\rmii{E} B_\rmii{M} D_0 D_i
	   + B_\rmii{E} B_\rmii{M} D_i^2
	   + B_\rmii{E}^2 D_0 D_i
	   + B_\rmii{E}^2 D_i^2
	   + H H^{\dagger } B_\rmii{M} D_i^2
	   \nn &
	   + B_\rmii{M}^2 D_0 D_i
	   + B_\rmii{M}^2 D_i^2
	   + 2 G_\rmii{E} G_\rmii{M} D_0 D_i
	   + G_\rmii{E} G_\rmii{M} D_i^2
	   + G_\rmii{E}^2 D_0 D_i
	   + G_\rmii{E}^2 D_i^2
	   \nn &
	   + H H^{\dagger } W_\rmii{E} D_i^2
	   + 2 W_\rmii{E} W_\rmii{M} D_0 D_i
	   + W_\rmii{E} W_\rmii{M} D_i^2
	   + W_\rmii{E}^2 D_0 D_i
	   + W_\rmii{E}^2 D_i^2
	   \nn &
	   + G_\rmii{M}^2 D_0 D_i
	   + G_\rmii{M}^2 D_i^2
	   + H H^{\dagger } W_\rmii{M} D_i^2
	   + W_\rmii{M}^2 D_0 D_i
	   + W_\rmii{M}^2 D_i^2
	   \,.
\end{align}

%%%%%%%%%%%%%%%%%%%%%%%%% SECTION %%%%%%%%%%%%%%%%%%%%%%%%%%%%%%%%%%%%%%%%%
%
\section{Growth of the number of independent operators}
\label{app:OperatorGrowth}
%%%%%%%%%%%%%%%%%%%%%%%%% SECTION %%%%%%%%%%%%%%%%%%%%%%%%%%%%%%%%%%%%%%%%%
%
Here, we display the total number of independent SMEFT operators,
up to mass dimension ten,
at finite temperature, and
compare them with the zero-temperature
SMEFT operator basis~\cite{Henning:2015alf}.
To be on the same footing, 
we consider only the bosonic degrees of freedom for both scenarios. 

\begin{figure}[t]
    \centering
    \includegraphics[width=0.6\textwidth]{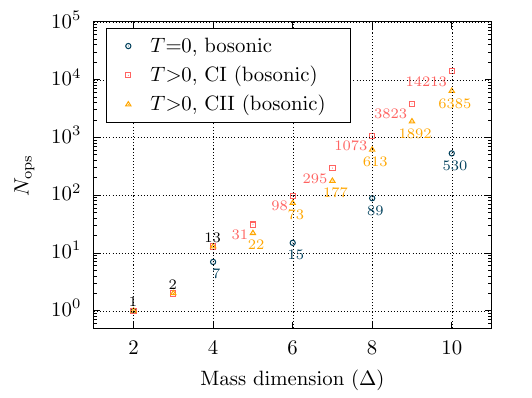}%
    \caption{%
			Total number of independent operators
			$N_\rmi{ops}$ for bosonic SMEFT
			at different mass dimensions:
			a comparison among
			(i) $T=0$,
			(ii) $T > 0$ with~\ref{sec:constraint-I:def},
			and
			(iii) $T > 0$ with~\ref{sec:constraint-II:def}.
		}
		\label{fig:count-total}
\end{figure}
In fig.~\ref{fig:count-total},
we find that the number of effective operators for
$T > 0$ with~\ref{sec:constraint-I:def},
at each mass dimension,
is greater than or equal to that for the $T=0$ case.
That number is significantly reduced once
we employ~\ref{sec:constraint-II:def}.
It is worth noting that, in the absence of fermion fields,
there is no operator of odd mass dimension
at zero temperature~\cite{Kobach:2016ami},
while such operators emerge in the finite-temperature SMEFT.
The symmetry of the modified space-time manifold and
the temporal derivative $D_0$ play a crucial role in this outcome.

%%%%%%%%%%%%%%%%%%%%%%%%% SECTION %%%%%%%%%%%%%%%%%%%%%%%%%%%%%%%%%%%%%%%%%
%
{\small
%%%%%%%%%%%%%%%%%%%%%%%%% BIBLIO %%%%%%%%%%%%%%%%%%%%%%%%%%%%%%%%%%%%%%%%%
%
\bibliographystyle{utphys}
\bibliography{HSFT}

@article{Dolan:2005wy,
    author = "Dolan, F. A.",
    title = "{Character formulae and partition functions in higher dimensional conformal field theory}",
    eprint = "hep-th/0508031",
    archivePrefix = "arXiv",
    reportNumber = "DAMTP-05-69",
    doi = "10.1063/1.2196241",
    journal = "J. Math. Phys.",
    volume = "47",
    pages = "062303",
    year = "2006"
}

@article{Beckers:1981yq,
    author = "Beckers, J. and Hussin, V.",
    title = "{SPINORS AND RELATED TENSORS INVARIANT UNDER E(3) AND ITS SUBGROUPS}",
    doi = "10.1088/0305-4470/14/2/008",
    journal = "J. Phys. A",
    volume = "14",
    pages = "317--326",
    year = "1981"
}

@article{Sitenko:2023rex,
    author = "Sitenko, Yurii A.",
    title = "{Path integral formalism for finite-temperature field theory and generation of chiral currents}",
    eprint = "2303.02145",
    archivePrefix = "arXiv",
    primaryClass = "hep-th",
    month = "3",
    year = "2023"
}

@article{aguilar1999,
      title= "{The Universal Covering Group of U(n) and Projective Representations}", 
      author={M. A. Aguilar and M. Socolovsky},
      year={1999},
      eprint={math-ph/9911028},
      archivePrefix={arXiv},
      primaryClass={math-ph},
      url={https://arxiv.org/abs/math-ph/9911028}, 
}

@article{Bacry1970,
  author="Bacry H, Combe Ph and Richard J L",
  title="{Group-theoretical analysis of elementary particles in an external electromagnetic field: I. The relativistic particle in a constant and uniform field}",
  journal={Nuovo Cimento},
  issn={0369-3546},
  volume={67},
  year={1970},
  number={2},
  pages={267--99},
  url={https://doi.org/10.1007/BF02725178},
  DOI={10.1007/BF02725178}
}

@article{Douglas_2011,
  author = {Douglas, Michael J. and Repka, Joe},
  title = "{Embeddings of the Euclidean algebra $\mathfrak{{e}}(3)$ into $\mathfrak{{sl}}(4, \mathbb{{C}})$ and restrictions of irreducible representations of $\mathfrak{{sl}}(4, \mathbb{{C}})$}",
  journal = {Journal of Mathematical Physics},
  volume = {52},
  issue = {1},
  pages = {013504},
  year = {2011},
  publisher = {AIP Publishing},
  doi = {10.1063/1.3531984},
  url = {https://doi.org/10.1063/1.3531984}
}

@article{Balui:2025kat,
    author = "Balui, Debanjan and Chakrabortty, Joydeep and Dey, Debmalya and Mohanty, Subhendra",
    title = "{Gauge invariant effective potential}",
    eprint = "2502.17156",
    archivePrefix = "arXiv",
    primaryClass = "hep-th",
    doi = "10.1103/PhysRevD.111.085032",
    journal = "Phys. Rev. D",
    volume = "111",
    number = "8",
    pages = "085032",
    year = "2025"
}

@article{Balui:2025yvd,
    author = "Balui, Debanjan and Biswas, Tisa and Chakrabortty, Joydeep and Dey, Debmalya and Englert, Christoph and Mohanty, Subhendra",
    title = "{Gauge choices, infrared pitfalls, and thermal effects in effective potentials}",
    eprint = "2507.22706",
    archivePrefix = "arXiv",
    primaryClass = "hep-th",
    doi = "10.1103/drsd-wfns",
    journal = "Phys. Rev. D",
    volume = "112",
    number = "5",
    pages = "056022",
    year = "2025"
}

@article{Chakrabortty:2024wto,
    author = "Chakrabortty, Joydeep and Mohanty, Subhendra",
    title = "{One Loop Thermal Effective Action}",
    eprint = "2411.14146",
    archivePrefix = "arXiv",
    primaryClass = "hep-th",
    doi = "10.1016/j.nuclphysb.2025.117165",
    journal = "Nucl. Phys. B",
    volume = "1020",
    pages = "117165",
    year = "2025"
}

@article{Laine:2018lgj,
    author = {Laine, M. and Schicho, P. and Schr{\"o}der, Y.},
    title = "{Soft thermal contributions to 3-loop gauge coupling}",
    eprint = "1803.08689",
    archivePrefix = "arXiv",
    primaryClass = "hep-ph",
    doi = "10.1007/JHEP05(2018)037",
    journal = "JHEP",
    volume = "05",
    pages = "037",
    year = "2018"
}

@article{Gynther:2003za,
    author = "Gynther, A.",
    title = "{Electroweak phase diagram at finite lepton number density}",
    eprint = "hep-ph/0303019",
    archivePrefix = "arXiv",
    reportNumber = "HIP-2003-12-TH",
    doi = "10.1103/PhysRevD.68.016001",
    journal = "Phys. Rev. D",
    volume = "68",
    pages = "016001",
    year = "2003"
}

@article{Gray:2008yu,
  author        = "Gray, James and Hanany, Amihay and He, Yang-Hui and Jejjala, Vishnu and Mekareeya, Noppadol",
  title         = "{SQCD: A Geometric Apercu}",
  eprint        = "0803.4257",
  archivePrefix = "arXiv",
  primaryClass  = "hep-th",
  journal       = "JHEP",
  volume        = "05",
  pages         = "099",
  year          = "2008",
  doi           = "10.1088/1126-6708/2008/05/099"
}

@article{Hanany:2014dia,
    author = "Hanany, Amihay and Kalveks, Rudolph",
    title = "{Highest Weight Generating Functions for Hilbert Series}",
    eprint = "1408.4690",
    archivePrefix = "arXiv",
    primaryClass = "hep-th",
    reportNumber = "IMPERIAL-TP-14-AH-07",
    doi = "10.1007/JHEP10(2014)152",
    journal = "JHEP",
    volume = "10",
    pages = "152",
    year = "2014"
}

@article{Henning:2017fpj,
  author        = "Henning, Brian and Lu, Xiaochuan and Melia, Tom and Murayama, Hitoshi",
  title         = "{Operator bases, $S$-matrices, and their partition functions}",
  eprint        = "1706.08520",
  archivePrefix = "arXiv",
  primaryClass  = "hep-th",
  journal       = "JHEP",
  volume        = "10",
  pages         = "199",
  year          = "2017",
  doi           = "10.1007/JHEP10(2017)199"
}

@article{Henning:2015alf,
    author = "Henning, Brian and Lu, Xiaochuan and Melia, Tom and Murayama, Hitoshi",
    title = "{2, 84, 30, 993, 560, 15456, 11962, 261485, ...: Higher dimension operators in the SM EFT}",
    eprint = "1512.03433",
    archivePrefix = "arXiv",
    primaryClass = "hep-ph",
    reportNumber = "UCB-PTH-15-14, IPMU15-0207",
    doi = "10.1007/JHEP08(2017)016",
    journal = "JHEP",
    volume = "08",
    pages = "016",
    year = "2017",
    note = "[Erratum: JHEP 09, 019 (2019)]"
}

@article{Ruhdorfer:2019qmk,
    author = "Ruhdorfer, Maximilian and Serra, Javi and Weiler, Andreas",
    title = "{Effective Field Theory of Gravity to All Orders}",
    eprint = "1908.08050",
    archivePrefix = "arXiv",
    primaryClass = "hep-ph",
    reportNumber = "TUM-HEP-1205-19",
    doi = "10.1007/JHEP05(2020)083",
    journal = "JHEP",
    volume = "05",
    pages = "083",
    year = "2020"
}

@article{Graf:2020yxt,
    author = "Graf, Lukas and Henning, Brian and Lu, Xiaochuan and Melia, Tom and Murayama, Hitoshi",
    title = "{2, 12, 117, 1959, 45171, 1170086, {\textellipsis}: a Hilbert series for the QCD chiral Lagrangian}",
    eprint = "2009.01239",
    archivePrefix = "arXiv",
    primaryClass = "hep-ph",
    doi = "10.1007/JHEP01(2021)142",
    journal = "JHEP",
    volume = "01",
    pages = "142",
    year = "2021"
}

@mastersthesis{Dujava:2022vqz,
    author = "Dujava, Jon{\'a}{\v{s}}",
    title = "{Counting operators in Effective Field Theories}",
    eprint = "2211.05759",
    archivePrefix = "arXiv",
    primaryClass = "hep-th",
    type = "Bachelor thesis",
    school = "Charles U.",
    year = "2022"
}

@article{Graf:2022rco,
    author = "Gr{\'a}f, Luk{\'a}{\v{s}} and Henning, Brian and Lu, Xiaochuan and Melia, Tom and Murayama, Hitoshi",
    title = "{Hilbert series, the Higgs mechanism, and HEFT}",
    eprint = "2211.06275",
    archivePrefix = "arXiv",
    primaryClass = "hep-ph",
    doi = "10.1007/JHEP02(2023)064",
    journal = "JHEP",
    volume = "02",
    pages = "064",
    year = "2023"
}

@article{Bijnens:2022zqo,
    author = "Bijnens, Johan and Gudnason, Sven Bjarke and Yu, Jiahui and Zhang, Tiantian",
    title = "{Hilbert series and higher-order Lagrangians for the O(N) model}",
    eprint = "2212.07901",
    archivePrefix = "arXiv",
    primaryClass = "hep-th",
    reportNumber = "LU TP 22-65",
    doi = "10.1007/JHEP05(2023)061",
    journal = "JHEP",
    volume = "05",
    pages = "061",
    year = "2023"
}

@article{Grojean:2023tsd,
    author = "Grojean, Christophe and Kley, Jonathan and Yao, Chang-Yuan",
    title = "{Hilbert series for ALP EFTs}",
    eprint = "2307.08563",
    archivePrefix = "arXiv",
    primaryClass = "hep-ph",
    reportNumber = "DESY-23-098, HU-EP-23/39",
    doi = "10.1007/JHEP11(2023)196",
    journal = "JHEP",
    volume = "11",
    pages = "196",
    year = "2023"
}

@article{Alonso:2024usj,
    author = "Alonso, Rodrigo and Rahaman, Shakeel Ur",
    title = "{Counting and building operators in theories with hidden symmetries and application to HEFT}",
    eprint = "2412.09463",
    archivePrefix = "arXiv",
    primaryClass = "hep-ph",
    reportNumber = "IPPP/24/80",
    doi = "10.1007/JHEP08(2025)071",
    journal = "JHEP",
    volume = "08",
    pages = "071",
    year = "2025"
}

@article{Banerjee:2020bym,
    author = "Banerjee, Upalaparna and Chakrabortty, Joydeep and Prakash, Suraj and Rahaman, Shakeel Ur",
    title = "{Characters and group invariant polynomials of (super)fields: road to {\textquotedblleft}Lagrangian{\textquotedblright}}",
    eprint = "2004.12830",
    archivePrefix = "arXiv",
    primaryClass = "hep-ph",
    doi = "10.1140/epjc/s10052-020-8392-x",
    journal = "Eur. Phys. J. C",
    volume = "80",
    number = "10",
    pages = "938",
    year = "2020"
}

@article{Anisha:2019nzx,
    author = "Anisha and Das Bakshi, Supratim and Chakrabortty, Joydeep and Prakash, Suraj",
    title = "{Hilbert Series and Plethystics: Paving the path towards 2HDM- and MLRSM-EFT}",
    eprint = "1905.11047",
    archivePrefix = "arXiv",
    primaryClass = "hep-ph",
    doi = "10.1007/JHEP09(2019)035",
    journal = "JHEP",
    volume = "09",
    pages = "035",
    year = "2019"
}

@article{Balantekin:2001id,
  author        = "Balantekin, A. B. and Cassak, P.",
  title         = "{Character expansions for the orthogonal and symplectic groups}",
  eprint        = "hep-th/0108130",
  archivePrefix = "arXiv",
  primaryClass  = "hep-th",
  journal       = "J. Math. Phys.",
  volume        = "43",
  pages         = "604--620",
  year          = "2002",
  doi           = "10.1063/1.1418014"
}

@article{Plymen:1976,
  author        = "Plymen, R. J.",
  title         = "{On the Weyl Character Formula for SU(N)}",
  journal       = "Int. J. Theor. Phys.",
  volume        = "15",
  pages         = "201--206",
  year          = "1976",
  doi           = "10.1007/BF01807092"
}

@article{koike1987,
  author        = "Koike, Kazuhiko and Terada, Itaru",
  title         = "{Young-diagrammatic methods for the representation theory of the classical groups of type $B_n$, $C_n$, $D_n$}",
  journal       = "J. Algebra",
  volume        = "107",
  pages         = "466--511",
  year          = "1987"
}

@book{littlewood1977theory,
  author        = "Littlewood, Dudley E.",
  title         = "{The Theory of Group Characters and Matrix Representations of Groups}",
  publisher     = "Clarendon Press",
  address       = "Oxford",
  edition       = "2",
  year          = "1977"
}

@book{rossmann2006lie,
  author        = "Rossmann, Wulf",
  title         = "{Lie Groups: An Introduction Through Linear Groups}",
  publisher     = "Oxford University Press",
  address       = "Oxford",
  year          = "2006",
  isbn          = "9780199202515"
}

@article{Kajantie:1995dw,
    author = "Kajantie, K. and Laine, M. and Rummukainen, K. and Shaposhnikov, Mikhail E.",
    title = "{Generic rules for high temperature dimensional reduction and their application to the standard model}",
    eprint = "hep-ph/9508379",
    archivePrefix = "arXiv",
    reportNumber = "CERN-TH-95-226, HU-TFT-95-50, IUHET-312",
    doi = "10.1016/0550-3213(95)00549-8",
    journal = "Nucl. Phys. B",
    volume = "458",
    pages = "90--136",
    year = "1996"
}

@article{Bernardo:2025vkz,
    author = "Bernardo, Fabio and Klose, Philipp and Schicho, Philipp and Tenkanen, Tuomas V. I.",
    title = "{Higher-dimensional operators at finite temperature affect gravitational-wave predictions}",
    eprint = "2503.18904",
    archivePrefix = "arXiv",
    primaryClass = "hep-ph",
    reportNumber = "HIP-2025-6/TH",
    doi = "10.1007/JHEP08(2025)109",
    journal = "JHEP",
    volume = "08",
    pages = "109",
    year = "2025"
}

@article{Ginsparg:1980ef,
    author = "Ginsparg, Paul H.",
    title = "{First Order and Second Order Phase Transitions in Gauge Theories at Finite Temperature}",
    reportNumber = "SACLAY-DPh-T 80/27",
    doi = "10.1016/0550-3213(80)90418-6",
    journal = "Nucl. Phys. B",
    volume = "170",
    pages = "388--408",
    year = "1980"
}

@article{Appelquist:1981vg,
    author = "Appelquist, Thomas and Pisarski, Robert D.",
    title = "{High-Temperature Yang-Mills Theories and Three-Dimensional Quantum Chromodynamics}",
    reportNumber = "Print-81-0020 (YALE), YTP-81-01, COO-3075-203",
    doi = "10.1103/PhysRevD.23.2305",
    journal = "Phys. Rev. D",
    volume = "23",
    pages = "2305",
    year = "1981"
}

@article{Matsubara:1955ws,
    author = "Matsubara, Takeo",
    title = "{A New approach to quantum statistical mechanics}",
    doi = "10.1143/PTP.14.351",
    journal = "Prog. Theor. Phys.",
    volume = "14",
    pages = "351--378",
    year = "1955"
}

@article{Ekstedt:2024etx,
    author = "Ekstedt, Andreas and Schicho, Philipp and Tenkanen, Tuomas V. I.",
    title = "{Cosmological phase transitions at three loops: The final verdict on perturbation theory}",
    eprint = "2405.18349",
    archivePrefix = "arXiv",
    primaryClass = "hep-ph",
    reportNumber = "HIP-2024-15/TH",
    doi = "10.1103/PhysRevD.110.096006",
    journal = "Phys. Rev. D",
    volume = "110",
    number = "9",
    pages = "096006",
    year = "2024"
}

@article{Croon:2020cgk,
    author = "Croon, Djuna and Gould, Oliver and Schicho, Philipp and Tenkanen, Tuomas V. I. and White, Graham",
    title = "{Theoretical uncertainties for cosmological first-order phase transitions}",
    eprint = "2009.10080",
    archivePrefix = "arXiv",
    primaryClass = "hep-ph",
    reportNumber = "HIP-2020-26/TH",
    doi = "10.1007/JHEP04(2021)055",
    journal = "JHEP",
    volume = "04",
    pages = "055",
    year = "2021"
}

@article{Ekstedt:2022bff,
    author = "Ekstedt, Andreas and Schicho, Philipp and Tenkanen, Tuomas V. I.",
    title = "{DRalgo: A package for effective field theory approach for thermal phase transitions}",
    eprint = "2205.08815",
    archivePrefix = "arXiv",
    primaryClass = "hep-ph",
    reportNumber = "HIP-2022-11/TH, NORDITA 2022-030",
    doi = "10.1016/j.cpc.2023.108725",
    journal = "Comput. Phys. Commun.",
    volume = "288",
    pages = "108725",
    year = "2023"
}

@article{Laine:2017hdk,
    author = "Laine, M. and Meyer, M. and Nardini, G.",
    title = "{Thermal phase transition with full 2-loop effective potential}",
    eprint = "1702.07479",
    archivePrefix = "arXiv",
    primaryClass = "hep-ph",
    doi = "10.1016/j.nuclphysb.2017.04.023",
    journal = "Nucl. Phys. B",
    volume = "920",
    pages = "565--600",
    year = "2017"
}

@article{Caron-Huot:2008zna,
    author = "Caron-Huot, Simon",
    title = "{O(g) plasma effects in jet quenching}",
    eprint = "0811.1603",
    archivePrefix = "arXiv",
    primaryClass = "hep-ph",
    doi = "10.1103/PhysRevD.79.065039",
    journal = "Phys. Rev. D",
    volume = "79",
    pages = "065039",
    year = "2009"
}

@article{Panero:2013pla,
    author = {Panero, Marco and Rummukainen, Kari and Sch{\"a}fer, Andreas},
    title = "{Lattice Study of the Jet Quenching Parameter}",
    eprint = "1307.5850",
    archivePrefix = "arXiv",
    primaryClass = "hep-ph",
    reportNumber = "HIP-2013-12-TH, IFT-UAM-CSIC-13-098",
    doi = "10.1103/PhysRevLett.112.162001",
    journal = "Phys. Rev. Lett.",
    volume = "112",
    number = "16",
    pages = "162001",
    year = "2014"
}

@article{DOnofrio:2014mld,
    author = "D'Onofrio, Michela and Kurkela, Aleksi and Moore, Guy D.",
    title = "{Renormalization of Null Wilson Lines in EQCD}",
    eprint = "1401.7951",
    archivePrefix = "arXiv",
    primaryClass = "hep-lat",
    reportNumber = "CERN-PH-TH-2014-031",
    doi = "10.1007/JHEP03(2014)125",
    journal = "JHEP",
    volume = "03",
    pages = "125",
    year = "2014"
}

@article{Ghiglieri:2013gia,
    author = "Ghiglieri, Jacopo and Hong, Juhee and Kurkela, Aleksi and Lu, Egang and Moore, Guy D. and Teaney, Derek",
    title = "{Next-to-leading order thermal photon production in a weakly coupled quark-gluon plasma}",
    eprint = "1302.5970",
    archivePrefix = "arXiv",
    primaryClass = "hep-ph",
    doi = "10.1007/JHEP05(2013)010",
    journal = "JHEP",
    volume = "05",
    pages = "010",
    year = "2013"
}

@article{Ghiglieri:2014kma,
    author = "Ghiglieri, Jacopo and Moore, Guy D.",
    title = "{Low Mass Thermal Dilepton Production at NLO in a Weakly Coupled Quark-Gluon Plasma}",
    eprint = "1410.4203",
    archivePrefix = "arXiv",
    primaryClass = "hep-ph",
    doi = "10.1007/JHEP12(2014)029",
    journal = "JHEP",
    volume = "12",
    pages = "029",
    year = "2014"
}

@article{Ghiglieri:2016xye,
    author = "Ghiglieri, J. and Laine, M.",
    title = "{Neutrino dynamics below the electroweak crossover}",
    eprint = "1605.07720",
    archivePrefix = "arXiv",
    primaryClass = "hep-ph",
    doi = "10.1088/1475-7516/2016/07/015",
    journal = "JCAP",
    volume = "07",
    pages = "015",
    year = "2016"
}

@article{Grzadkowski:2010es,
    author = "Grzadkowski, B. and Iskrzynski, M. and Misiak, M. and Rosiek, J.",
    title = "{Dimension-Six Terms in the Standard Model Lagrangian}",
    eprint = "1008.4884",
    archivePrefix = "arXiv",
    primaryClass = "hep-ph",
    reportNumber = "IFT-9-2010, TTP10-35",
    doi = "10.1007/JHEP10(2010)085",
    journal = "JHEP",
    volume = "10",
    pages = "085",
    year = "2010"
}

@article{Chala:2025aiz,
    author = "Chala, Mikael and Guedes, Guilherme",
    title = "{The high-temperature limit of the SM(EFT)}",
    eprint = "2503.20016",
    archivePrefix = "arXiv",
    primaryClass = "hep-ph",
    doi = "10.1007/JHEP07(2025)085",
    journal = "JHEP",
    volume = "07",
    pages = "085",
    year = "2025"
}

@article{Chala:2024xll,
    author = "Chala, Mikael and Criado, Juan Carlos and Gil, Luis and Miras, Javier L{\'o}pez",
    title = "{Higher-order-operator corrections to phase-transition parameters in dimensional reduction}",
    eprint = "2406.02667",
    archivePrefix = "arXiv",
    primaryClass = "hep-ph",
    doi = "10.1007/JHEP10(2024)025",
    journal = "JHEP",
    volume = "10",
    pages = "025",
    year = "2024"
}

@article{Kajantie:1997ky,
    author = "Kajantie, K. and Laine, M. and Rummukainen, K. and Shaposhnikov, Mikhail E.",
    title = "{High temperature dimensional reduction and parity violation}",
    eprint = "hep-ph/9710538",
    archivePrefix = "arXiv",
    reportNumber = "CERN-TH-97-298, NORDITA-97-78-P",
    doi = "10.1016/S0370-2693(97)01584-0",
    journal = "Phys. Lett. B",
    volume = "423",
    pages = "137--144",
    year = "1998"
}

@article{Weinberg:1979sa,
    author = "Weinberg, Steven",
    title = "{Baryon and Lepton Nonconserving Processes}",
    reportNumber = "HUTP-79-A050",
    doi = "10.1103/PhysRevLett.43.1566",
    journal = "Phys. Rev. Lett.",
    volume = "43",
    pages = "1566--1570",
    year = "1979"
}

@article{Buchmuller:1985jz,
    author = "Buchmuller, W. and Wyler, D.",
    title = "{Effective Lagrangian Analysis of New Interactions and Flavor Conservation}",
    reportNumber = "CERN-TH-4254/85",
    doi = "10.1016/0550-3213(86)90262-2",
    journal = "Nucl. Phys. B",
    volume = "268",
    pages = "621--653",
    year = "1986"
}

@article{Aebischer:2025qhh,
    author = "Aebischer, Jason and Buras, Andrzej J. and Kumar, Jacky",
    title = "{SMEFT ATLAS: The landscape beyond the Standard Model}",
    eprint = "2507.05926",
    archivePrefix = "arXiv",
    primaryClass = "hep-ph",
    reportNumber = "AJB-25-1, CERN-TH-2025-129, LA-UR-24-24665",
    doi = "10.1016/j.physrep.2026.07.003",
    journal = "Phys. Rept.",
    volume = "1198",
    pages = "1--181",
    year = "2026"
}

@article{Chapman:1994vk,
    author = "Chapman, Scott",
    title = "{A New dimensionally reduced effective action for QCD at high temperature}",
    eprint = "hep-ph/9407313",
    archivePrefix = "arXiv",
    reportNumber = "TPR-94-8",
    doi = "10.1103/PhysRevD.50.5308",
    journal = "Phys. Rev. D",
    volume = "50",
    pages = "5308--5313",
    year = "1994"
}

@article{Schwinger:1951nm,
    author = "Schwinger, Julian S.",
    editor = "Milton, K. A.",
    title = "{On gauge invariance and vacuum polarization}",
    doi = "10.1103/PhysRev.82.664",
    journal = "Phys. Rev.",
    volume = "82",
    pages = "664--679",
    year = "1951"
}

@article{DeWitt:1975ys,
    author = "DeWitt, Bryce S.",
    title = "{Quantum Field Theory in Curved Space-Time}",
    doi = "10.1016/0370-1573(75)90051-4",
    journal = "Phys. Rept.",
    volume = "19",
    pages = "295--357",
    year = "1975"
}

@article{Dyakonov:1984,
  author       = "D'yakonov, D I and Petrov, V Y and Yung, A V",
  title        = "{Quasiclassical expansion in an external Yang-Mills field and the approximate calculation of functional determinants}",
  url          = "https://www.osti.gov/biblio/6299831",
  journal      = "Sov. J. Nucl. Phys.",
  issn         = "ISSN SJNCA",
  volume       = "39:1",
  year         = "1984",
  month        = "01"}

@article{Henning:2015daa,
    author = "Henning, Brian and Lu, Xiaochuan and Melia, Tom and Murayama, Hitoshi",
    title = "{Hilbert series and operator bases with derivatives in effective field theories}",
    eprint = "1507.07240",
    archivePrefix = "arXiv",
    primaryClass = "hep-th",
    doi = "10.1007/s00220-015-2518-2",
    journal = "Commun. Math. Phys.",
    volume = "347",
    number = "2",
    pages = "363--388",
    year = "2016"
}

@article{Lehman:2015via,
    author = "Lehman, Landon and Martin, Adam",
    title = "{Hilbert Series for Constructing Lagrangians: expanding the phenomenologist's toolbox}",
    eprint = "1503.07537",
    archivePrefix = "arXiv",
    primaryClass = "hep-ph",
    doi = "10.1103/PhysRevD.91.105014",
    journal = "Phys. Rev. D",
    volume = "91",
    pages = "105014",
    year = "2015"
}

@article{Lehman:2015coa,
    author = "Lehman, Landon and Martin, Adam",
    title = "{Low-derivative operators of the Standard Model effective field theory via Hilbert series methods}",
    eprint = "1510.00372",
    archivePrefix = "arXiv",
    primaryClass = "hep-ph",
    doi = "10.1007/JHEP02(2016)081",
    journal = "JHEP",
    volume = "02",
    pages = "081",
    year = "2016"
}

@article{Marinissen:2020jmb,
    author = "Marinissen, Coenraad B. and Rahn, Rudi and Waalewijn, Wouter J.",
    title = "{..., 83106786, 114382724, 1509048322, 2343463290, 27410087742, ... efficient Hilbert series for effective theories}",
    eprint = "2004.09521",
    archivePrefix = "arXiv",
    primaryClass = "hep-ph",
    reportNumber = "NIKHEF 20-010",
    doi = "10.1016/j.physletb.2020.135632",
    journal = "Phys. Lett. B",
    volume = "808",
    pages = "135632",
    year = "2020"
}

@article{Melia:2020pzd,
    author = "Melia, Tom and Pal, Sridip",
    title = "{EFT Asymptotics: the Growth of Operator Degeneracy}",
    eprint = "2010.08560",
    archivePrefix = "arXiv",
    primaryClass = "hep-th",
    doi = "10.21468/SciPostPhys.10.5.104",
    journal = "SciPost Phys.",
    volume = "10",
    number = "5",
    pages = "104",
    year = "2021"
}

@article{Kondo:2022wcw,
    author = "Kondo, Dan and Murayama, Hitoshi and Okabe, Risshin",
    title = "{23, 381, 6242, 103268, 1743183, {\textellipsis} : Hilbert series for CP-violating operators in SMEFT}",
    eprint = "2212.02413",
    archivePrefix = "arXiv",
    primaryClass = "hep-ph",
    doi = "10.1007/JHEP03(2023)107",
    journal = "JHEP",
    volume = "03",
    pages = "107",
    year = "2023"
}

@article{Delgado:2022bho,
    author = "Delgado, Antonio and Martin, Adam and Wang, Runqing",
    title = "{Constructing operator basis in supersymmetry: a Hilbert series approach}",
    eprint = "2212.02551",
    archivePrefix = "arXiv",
    primaryClass = "hep-th",
    doi = "10.1007/JHEP04(2023)097",
    journal = "JHEP",
    volume = "04",
    pages = "097",
    year = "2023"
}

@article{Delgado:2023ivp,
    author = "Delgado, Antonio and Martin, Adam and Wang, Runqing",
    title = "{Counting operators in N = 1 supersymmetric gauge theories}",
    eprint = "2305.01736",
    archivePrefix = "arXiv",
    primaryClass = "hep-th",
    doi = "10.1007/JHEP07(2023)081",
    journal = "JHEP",
    volume = "07",
    pages = "081",
    year = "2023"
}

@article{Sun:2025zuk,
    author = "Sun, Hao and Wang, Yi-Ning and Yu, Jiang-Hao",
    title = "{Chiral effective field theories for strong and weak dynamics}",
    eprint = "2501.14018",
    archivePrefix = "arXiv",
    primaryClass = "hep-ph",
    doi = "10.1007/JHEP08(2025)185",
    journal = "JHEP",
    volume = "08",
    pages = "185",
    year = "2025"
}

@article{Grinstein:2023njq,
    author = "Grinstein, Benjam{\'\i}n and Lu, Xiaochuan and Merlo, Luca and Qu{\'\i}lez, Pablo",
    title = "{Hilbert series for covariants and their applications to minimal flavor violation}",
    eprint = "2312.13349",
    archivePrefix = "arXiv",
    primaryClass = "hep-ph",
    reportNumber = "IFT-UAM-CSIC-23-67",
    doi = "10.1007/JHEP06(2024)154",
    journal = "JHEP",
    volume = "2024",
    pages = "154",
    year = "2024",
    note = "[Erratum: JHEP 03, 072 (2025)]"
}

@article{Grinstein:2024jqt,
    author = "Grinstein, Benjam{\'\i}n and Lu, Xiaochuan and Mir{\'o}, Carlos and Qu{\'\i}lez, Pablo",
    title = "{Accidental symmetries, Hilbert series, and friends}",
    eprint = "2412.05359",
    archivePrefix = "arXiv",
    primaryClass = "hep-ph",
    doi = "10.1007/JHEP03(2025)172",
    journal = "JHEP",
    volume = "03",
    pages = "172",
    year = "2025"
}

@article{Grinstein:2024iyf,
    author = "Grinstein, Benjam{\'\i}n and Lu, Xiaochuan and Mir{\'o}, Carlos and Qu{\'\i}lez, Pablo",
    title = "{Most general EFTs from spurion analysis Hilbert series and minimal lepton flavor violation}",
    eprint = "2412.16285",
    archivePrefix = "arXiv",
    primaryClass = "hep-ph",
    doi = "10.1007/JHEP07(2025)259",
    journal = "JHEP",
    volume = "07",
    pages = "259",
    year = "2025"
}

@article{Kobach:2016ami,
    author = "Kobach, Andrew",
    title = "{Baryon Number, Lepton Number, and Operator Dimension in the Standard Model}",
    eprint = "1604.05726",
    archivePrefix = "arXiv",
    primaryClass = "hep-ph",
    reportNumber = "PHYS.LETT.-B758-(2016)-455-457",
    doi = "10.1016/j.physletb.2016.05.050",
    journal = "Phys. Lett. B",
    volume = "758",
    pages = "455--457",
    year = "2016"
}

@article{Weldon:1982aq,
    author = "Weldon, H. Arthur",
    title = "{Covariant Calculations at Finite Temperature: The Relativistic Plasma}",
    reportNumber = "PRINT-82-0313 (PENN)",
    doi = "10.1103/PhysRevD.26.1394",
    journal = "Phys. Rev. D",
    volume = "26",
    pages = "1394",
    year = "1982"
}

@article{Weiss:1980rj,
    author = "Weiss, Nathan",
    title = "{The Effective Potential for the Order Parameter of Gauge Theories at Finite Temperature}",
    reportNumber = "UBC-81",
    doi = "10.1103/PhysRevD.24.475",
    journal = "Phys. Rev. D",
    volume = "24",
    pages = "475",
    year = "1981"
}

@article{Susskind:1979up,
    author = "Susskind, Leonard",
    title = "{Lattice Models of Quark Confinement at High Temperature}",
    doi = "10.1103/PhysRevD.20.2610",
    journal = "Phys. Rev. D",
    volume = "20",
    pages = "2610--2618",
    year = "1979"
}

@article{Megias:2003ui,
    author = "Megias, E. and Ruiz Arriola, E. and Salcedo, L. L.",
    title = "{The Thermal heat kernel expansion and the one loop effective action of QCD at finite temperature}",
    eprint = "hep-ph/0312133",
    archivePrefix = "arXiv",
    doi = "10.1103/PhysRevD.69.116003",
    journal = "Phys. Rev. D",
    volume = "69",
    pages = "116003",
    year = "2004"
}

@article{Moral-Gamez:2011wcb,
    author = "Moral-Gamez, F. J. and Salcedo, L. L.",
    title = "{Derivative expansion of the heat kernel at finite temperature}",
    eprint = "1110.6300",
    archivePrefix = "arXiv",
    primaryClass = "hep-th",
    doi = "10.1103/PhysRevD.85.045019",
    journal = "Phys. Rev. D",
    volume = "85",
    pages = "045019",
    year = "2012"
}

@article{Popov:1988fdi,
    author = "Popov, V. N. and Fedotov, S. A.",
    title = "{The functional-integration method and diagram technique for spin systems}",
    journal = "Sov. Phys. JETP",
    volume = "67",
    number = "3",
    pages = "535--541",
    year = "1988"
}

@article{Prokofev:2011pof,
    author = "Prokof'ev, Nikolay and Svistunov, Boris",
    title = "{From Popov-Fedotov trick to universal fermionization}",
    eprint = "1103.3730",
    archivePrefix = "arXiv",
    primaryClass = "cond-mat.str-el",
    doi = "10.1103/PhysRevB.84.073102",
    journal = "Phys. Rev. B",
    volume = "84",
    pages = "073102",
    year = "2011"
}

@article{Pisarski:2002ji,
    author = "Pisarski, Robert D.",
    title = "{Notes on the deconfining phase transition}",
    booktitle = "{Cargese Summer School on QCD Perspectives on Hot and Dense Matter}",
    eprint = "hep-ph/0203271",
    archivePrefix = "arXiv",
    pages = "353--384",
    month = "3",
    year = "2002"
}

@article{Gross:1980br,
    author = "Gross, David J. and Pisarski, Robert D. and Yaffe, Laurence G.",
    title = "{QCD and Instantons at Finite Temperature}",
    reportNumber = "PRINT-80-0538 (PRINCETON)",
    doi = "10.1103/RevModPhys.53.43",
    journal = "Rev. Mod. Phys.",
    volume = "53",
    pages = "43",
    year = "1981"
}

@article{Pisarski:2000eq,
    author = "Pisarski, Robert D.",
    title = "{Quark gluon plasma as a condensate of SU(3) Wilson lines}",
    eprint = "hep-ph/0006205",
    archivePrefix = "arXiv",
    doi = "10.1103/PhysRevD.62.111501",
    journal = "Phys. Rev. D",
    volume = "62",
    pages = "111501",
    year = "2000"
}

@article{Banerjee:2020jun,
    author = "Banerjee, Upalaparna and Chakrabortty, Joydeep and Prakash, Suraj and Rahaman, Shakeel Ur and Spannowsky, Michael",
    title = "{Effective Operator Bases for Beyond Standard Model Scenarios: An EFT compendium for discoveries}",
    eprint = "2008.11512",
    archivePrefix = "arXiv",
    primaryClass = "hep-ph",
    reportNumber = "IPPP/20/37",
    doi = "10.1007/JHEP01(2021)028",
    journal = "JHEP",
    volume = "01",
    pages = "028",
    year = "2021"
}

@article{Bernardo:2026whs,
    author = "Bernardo, Fabio and Chala, Mikael and Gil, Luis and Schicho, Philipp",
    title = "{Hard thermal contributions to phase transition observables at NNLO}",
    eprint = "2602.06962",
    archivePrefix = "arXiv",
    primaryClass = "hep-ph",
    doi = "10.1007/JHEP06(2026)281",
    journal = "JHEP",
    volume = "06",
    pages = "281",
    year = "2026"
}

@article{Laine:1999rv,
    author = "Laine, M.",
    title = "{The Renormalized gauge coupling and nonperturbative tests of dimensional reduction}",
    eprint = "hep-ph/9903513",
    archivePrefix = "arXiv",
    reportNumber = "CERN-TH-99-62",
    doi = "10.1088/1126-6708/1999/06/020",
    journal = "JHEP",
    volume = "06",
    pages = "020",
    year = "1999"
}

@article{Laine:1996nz,
    author = "Laine, M.",
    title = "{Comparison of 4-D and 3-D lattice results for the electroweak phase transition}",
    eprint = "hep-lat/9604011",
    archivePrefix = "arXiv",
    reportNumber = "HD-THEP-96-08",
    doi = "10.1016/0370-2693(96)00854-4",
    journal = "Phys. Lett. B",
    volume = "385",
    pages = "249--253",
    year = "1996"
}

@article{Chala:2025gif,
    author = "Chala, Mikael and Fiore, Maria Cristina and Gil, Luis",
    title = "{Phase diagram of the standard model effective field theory}",
    eprint = "2507.16905",
    archivePrefix = "arXiv",
    primaryClass = "hep-ph",
    doi = "10.1103/zxzs-d6zt",
    journal = "Phys. Rev. D",
    volume = "113",
    number = "11",
    pages = "115061",
    year = "2026"
}

@article{Fuentes-Martin:2026bhr,
    author = "Fuentes-Mart{\'\i}n, Javier and L{\'o}pez Miras, Javier and Moreno-S{\'a}nchez, Adri{\'a}n",
    title = "{Matchotter: An Automated Tool for Dimensional Reduction at Finite Temperature}",
    eprint = "2604.21972",
    archivePrefix = "arXiv",
    primaryClass = "hep-ph",
    month = "4",
    year = "2026"
}

@article{Bochkarev:1989kp,
    author = "Bochkarev, A. I. and Khlebnikov, S. Yu. and Shaposhnikov, M. E.",
    title = "{Sphalerons and Baryogenesis: Electroweak {CP} Violation at High Temperatures}",
    reportNumber = "TPI-MINN-89/3-T",
    doi = "10.1016/0550-3213(90)90153-5",
    journal = "Nucl. Phys. B",
    volume = "329",
    pages = "493--518",
    year = "1990"
}

@article{Kobach:2018nmt,
    author = "Kobach, Andrew and Pal, Sridip",
    title = "{Conformal Structure of the Heavy Particle EFT Operator Basis}",
    eprint = "1804.01534",
    archivePrefix = "arXiv",
    primaryClass = "hep-ph",
    doi = "10.1016/j.physletb.2018.06.060",
    journal = "Phys. Lett. B",
    volume = "783",
    pages = "311--319",
    year = "2018"
}

@article{Kobach:2017xkw,
    author = "Kobach, Andrew and Pal, Sridip",
    title = "{Hilbert Series and Operator Basis for NRQED and NRQCD/HQET}",
    eprint = "1704.00008",
    archivePrefix = "arXiv",
    primaryClass = "hep-ph",
    doi = "10.1016/j.physletb.2017.06.026",
    journal = "Phys. Lett. B",
    volume = "772",
    pages = "225--231",
    year = "2017"
}

@article{Li:2026eym,
    author = "Li, Yong-Kang and Wang, Yi-Ning and Yu, Jiang-Hao",
    title = "{Systematic Operator Construction for Non-relativistic Effective Field Theories: Hilbert Series versus Young Tensor}",
    eprint = "2602.12263",
    archivePrefix = "arXiv",
    primaryClass = "hep-ph",
    month = "2",
    year = "2026"
}

@article{Moore:1995jv,
    author = "Moore, Guy D.",
    title = "{Fermion determinant and the sphaleron bound}",
    eprint = "hep-ph/9508405",
    archivePrefix = "arXiv",
    reportNumber = "PUPT-1557, PUP-TH-1557",
    doi = "10.1103/PhysRevD.53.5906",
    journal = "Phys. Rev. D",
    volume = "53",
    pages = "5906--5917",
    year = "1996"
}

@article{vandeVis:2025efm,
    author = "van de Vis, Jorinde and de Vries, Jordy and Postma, Marieke",
    title = "{Bubble trouble: A review on electroweak baryogenesis}",
    eprint = "2508.09989",
    archivePrefix = "arXiv",
    primaryClass = "hep-ph",
    reportNumber = "CERN-TH-2025-161, Nikhef 2025-012",
    doi = "10.1016/j.ppnp.2026.104244",
    journal = "Prog. Part. Nucl. Phys.",
    volume = "150",
    pages = "104244",
    year = "2026"
}

@article{Prokopec:2003pj,
    author = "Prokopec, Tomislav and Schmidt, Michael G. and Weinstock, Steffen",
    title = "{Transport equations for chiral fermions to order h bar and electroweak baryogenesis. Part 1}",
    eprint = "hep-ph/0312110",
    archivePrefix = "arXiv",
    reportNumber = "BNL-72343-2004-JA, HD-THEP-03-62",
    doi = "10.1016/j.aop.2004.06.002",
    journal = "Annals Phys.",
    volume = "314",
    pages = "208--265",
    year = "2004"
}

@article{Kainulainen:2001cn,
    author = "Kainulainen, Kimmo and Prokopec, Tomislav and Schmidt, Michael G. and Weinstock, Steffen",
    title = "{First principle derivation of semiclassical force for electroweak baryogenesis}",
    eprint = "hep-ph/0105295",
    archivePrefix = "arXiv",
    reportNumber = "HD-THEP-01-23, NORDITA-2001-9-HE",
    doi = "10.1088/1126-6708/2001/06/031",
    journal = "JHEP",
    volume = "06",
    pages = "031",
    year = "2001"
}

@article{Prokopec:2004ic,
    author = "Prokopec, Tomislav and Schmidt, Michael G. and Weinstock, Steffen",
    title = "{Transport equations for chiral fermions to order h-bar and electroweak baryogenesis. Part II}",
    eprint = "hep-ph/0406140",
    archivePrefix = "arXiv",
    reportNumber = "BNL-72342-2004-JA, HD-THEP-04-22",
    doi = "10.1016/j.aop.2004.06.001",
    journal = "Annals Phys.",
    volume = "314",
    pages = "267--320",
    year = "2004"
}

@article{Cirigliano:2011di,
    author = "Cirigliano, Vincenzo and Lee, Christopher and Tulin, Sean",
    title = "{Resonant Flavor Oscillations in Electroweak Baryogenesis}",
    eprint = "1106.0747",
    archivePrefix = "arXiv",
    primaryClass = "hep-ph",
    reportNumber = "MIT-CTP-4269",
    doi = "10.1103/PhysRevD.84.056006",
    journal = "Phys. Rev. D",
    volume = "84",
    pages = "056006",
    year = "2011"
}

@article{Cirigliano:2009yt,
    author = "Cirigliano, Vincenzo and Lee, Christopher and Ramsey-Musolf, Michael J. and Tulin, Sean",
    title = "{Flavored Quantum Boltzmann Equations}",
    eprint = "0912.3523",
    archivePrefix = "arXiv",
    primaryClass = "hep-ph",
    reportNumber = "NPAC-09-16, UCB-PTH-09-37",
    doi = "10.1103/PhysRevD.81.103503",
    journal = "Phys. Rev. D",
    volume = "81",
    pages = "103503",
    year = "2010"
}

@article{Konstandin:2004gy,
    author = "Konstandin, Thomas and Prokopec, Tomislav and Schmidt, Michael G.",
    title = "{Kinetic description of fermion flavor mixing and CP-violating sources for baryogenesis}",
    eprint = "hep-ph/0410135",
    archivePrefix = "arXiv",
    reportNumber = "HD-THEP-04-36, ITP-UU-04-21, SPIN-UU-04-12",
    doi = "10.1016/j.nuclphysb.2005.03.013",
    journal = "Nucl. Phys. B",
    volume = "716",
    pages = "373--400",
    year = "2005"
}

@article{DasBakshi:2020ejz,
    author = "Das Bakshi, Supratim and Chakrabortty, Joydeep and Englert, Christoph and Spannowsky, Michael and Stylianou, Panagiotis",
    title = "{$CP$ violation at ATLAS in effective field theory}",
    eprint = "2009.13394",
    archivePrefix = "arXiv",
    primaryClass = "hep-ph",
    reportNumber = "IPPP/20/43",
    doi = "10.1103/PhysRevD.103.055008",
    journal = "Phys. Rev. D",
    volume = "103",
    number = "5",
    pages = "055008",
    year = "2021"
}

@article{Bakshi:2021ofj,
    author = "Bakshi, Supratim Das and Chakrabortty, Joydeep and Englert, Christoph and Spannowsky, Michael and Stylianou, Panagiotis",
    title = "{Landscaping CP-violating BSM scenarios}",
    eprint = "2103.15861",
    archivePrefix = "arXiv",
    primaryClass = "hep-ph",
    reportNumber = "IPPP/20/90",
    doi = "10.1016/j.nuclphysb.2022.115676",
    journal = "Nucl. Phys. B",
    volume = "975",
    pages = "115676",
    year = "2022"
}

@article{Naskar:2022rpg,
    author = "Naskar, Wrishik and Prakash, Suraj and Rahaman, Shakeel Ur",
    title = "{EFT Diagrammatica. Part II. Tracing the UV origin of bosonic D6 CPV and D8 SMEFT operators}",
    eprint = "2205.00910",
    archivePrefix = "arXiv",
    primaryClass = "hep-ph",
    doi = "10.1007/JHEP08(2022)190",
    journal = "JHEP",
    volume = "08",
    pages = "190",
    year = "2022"
}

@article{Vuorinen:2006nz,
    author = "Vuorinen, A. and Yaffe, Laurence G.",
    title = "{Z(3)-symmetric effective theory for SU(3) Yang-Mills theory at high temperature}",
    eprint = "hep-ph/0604100",
    archivePrefix = "arXiv",
    doi = "10.1103/PhysRevD.74.025011",
    journal = "Phys. Rev. D",
    volume = "74",
    pages = "025011",
    year = "2006"
}

@article{Kurkela:2007dh,
    author = "Kurkela, A.",
    title = "{Framework for non-perturbative analysis of a Z(3)-symmetric effective theory of finite temperature QCD}",
    eprint = "0704.1416",
    archivePrefix = "arXiv",
    primaryClass = "hep-lat",
    reportNumber = "HIP-2007-18-TH",
    doi = "10.1103/PhysRevD.76.094507",
    journal = "Phys. Rev. D",
    volume = "76",
    pages = "094507",
    year = "2007"
}

@article{Polyakov:1975rs,
    author = "Polyakov, Alexander M.",
    editor = "Taylor, J. C.",
    title = "{Compact Gauge Fields and the Infrared Catastrophe}",
    doi = "10.1016/0370-2693(75)90162-8",
    journal = "Phys. Lett. B",
    volume = "59",
    pages = "82--84",
    year = "1975"
}

@article{Fukushima:2017csk,
    author = "Fukushima, Kenji and Skokov, Vladimir",
    title = "{Polyakov loop modeling for hot QCD}",
    eprint = "1705.00718",
    archivePrefix = "arXiv",
    primaryClass = "hep-ph",
    doi = "10.1016/j.ppnp.2017.05.002",
    journal = "Prog. Part. Nucl. Phys.",
    volume = "96",
    pages = "154--199",
    year = "2017"
}

@article{Ghiglieri:2020dpq,
    author = "Ghiglieri, Jacopo and Kurkela, Aleksi and Strickland, Michael and Vuorinen, Aleksi",
    title = "{Perturbative Thermal QCD: Formalism and Applications}",
    eprint = "2002.10188",
    archivePrefix = "arXiv",
    primaryClass = "hep-ph",
    reportNumber = "CERN-TH-2020-029, HIP-2020-6/TH",
    doi = "10.1016/j.physrep.2020.07.004",
    journal = "Phys. Rept.",
    volume = "880",
    pages = "1--73",
    year = "2020"
}

@article{Bergner:2013qaa,
    author = "Bergner, Georg and Langelage, Jens and Philipsen, Owe",
    title = "{Effective lattice Polyakov loop theory vs. full SU(3) Yang-Mills at finite temperature}",
    eprint = "1312.7823",
    archivePrefix = "arXiv",
    primaryClass = "hep-lat",
    doi = "10.1007/JHEP03(2014)039",
    journal = "JHEP",
    volume = "03",
    pages = "039",
    year = "2014"
}

@article{Langelage:2010yr,
    author = "Langelage, Jens and Lottini, Stefano and Philipsen, Owe",
    title = "{Centre symmetric 3d effective actions for thermal SU(N) Yang-Mills from strong coupling series}",
    eprint = "1010.0951",
    archivePrefix = "arXiv",
    primaryClass = "hep-lat",
    reportNumber = "BI-TP-2010-32",
    doi = "10.1007/JHEP07(2011)014",
    journal = "JHEP",
    volume = "02",
    pages = "057",
    year = "2011",
    note = "[Erratum: JHEP 07, 014 (2011)]"
}

@article{Kajantie:1998yc,
    author = "Kajantie, K. and Laine, M. and Rajantie, A. and Rummukainen, K. and Tsypin, M.",
    title = "{The Phase diagram of three-dimensional SU(3) + adjoint Higgs theory}",
    eprint = "hep-lat/9811004",
    archivePrefix = "arXiv",
    reportNumber = "CERN-TH-98-350, NORDITA-98-66-HE",
    doi = "10.1088/1126-6708/1998/11/011",
    journal = "JHEP",
    volume = "11",
    pages = "011",
    year = "1998"
}

@article{Bernardo:2026nyq,
    author = "Bernardo, Fabio and Reinle, Romain Guillermo and Schicho, Philipp",
    title = "{Matching higher-dimensional operators at finite temperature for general models}",
    eprint = "2605.15176",
    archivePrefix = "arXiv",
    primaryClass = "hep-ph",
    month = "5",
    year = "2026"
}

@article{Chala:2025cya,
    author = "Chala, Mikael and Dashko, Andrii and Guedes, Guilherme",
    title = "{Running couplings in high-temperature effective field theory}",
    eprint = "2510.26878",
    archivePrefix = "arXiv",
    primaryClass = "hep-ph",
    reportNumber = "CERN-TH-2025-220",
    doi = "10.1103/ptqg-5g38",
    journal = "Phys. Rev. D",
    volume = "113",
    number = "5",
    pages = "055026",
    year = "2026"
}

@article{Guedes:2023azv,
    author = "Guedes, Guilherme and Olgoso, Pablo and Santiago, Jos{\'e}",
    title = "{Towards the one loop IR/UV dictionary in the SMEFT: One loop generated operators from new scalars and fermions}",
    eprint = "2303.16965",
    archivePrefix = "arXiv",
    primaryClass = "hep-ph",
    reportNumber = "DESY-23-040",
    doi = "10.21468/SciPostPhys.15.4.143",
    journal = "SciPost Phys.",
    volume = "15",
    number = "4",
    pages = "143",
    year = "2023"
}

@article{Bandyopadhyay:2026nrv,
    author = "Bandyopadhyay, Siddhartha and Chakrabortty, Joydeep and Dey, Debmalya and Schicho, Philipp and Tushar",
    title = "{Higher-dimensional operators and Polyakov loop in hot Scalar QED from the heat kernel}",
    eprint = "2606.09779",
    archivePrefix = "arXiv",
    primaryClass = "hep-ph",
    month = "6",
    year = "2026"
}
}

\end{document}